\documentclass[mnsc]{informs3}

\usepackage[T1]{fontenc}
\usepackage{lmodern}
\usepackage[dvipsnames]{xcolor}
\usepackage{mathtools,amssymb,amsmath,array,multirow,graphicx,booktabs,mathrsfs}
\usepackage{bm}
\usepackage{natbib}
\usepackage{algorithm}
\usepackage{algpseudocode}
\usepackage{tabularx}
\usepackage{tikz}
\usepackage{makecell}
\usepackage{placeins}
\usetikzlibrary{positioning,calc,arrows.meta,fit}
\definecolor{obsfill}{RGB}{246,250,248}
\definecolor{obsline}{RGB}{0,121,107}
\definecolor{repeatline}{RGB}{150,150,150}
\definecolor{latentfill}{RGB}{245,248,252}
\definecolor{latentline}{RGB}{0,114,178}
\definecolor{modelfill}{RGB}{253,248,242}
\definecolor{modelline}{RGB}{213,94,0}
\definecolor{softredcolor}{RGB}{180,45,45}
\definecolor{softbluecolor}{RGB}{35,85,170}
\definecolor{prodA}{HTML}{2A78D6}
\definecolor{prodB}{HTML}{EB6834}
\definecolor{prodC}{HTML}{1BAF7A}
\usepackage[hmargin=1in,vmargin=1in]{geometry}
\usepackage[hidelinks]{hyperref}
\hypersetup{
    colorlinks,
    breaklinks,
    linkcolor=BrickRed,
    urlcolor=black,
    anchorcolor=black,
    citecolor=black
}

\allowdisplaybreaks
\OneAndAHalfSpacedXI
\TheoremsNumberedThrough
\ECRepeatTheorems
\EquationsNumberedThrough

\bibpunct[, ]{(}{)}{,}{a}{}{,}%
\def\bibfont{\small}%

\newcommand{\myqed}{\hfill$\square$}
\renewenvironment{proof}[1][Proof]{%
    \par\addvspace{\medskipamount}%
    \noindent\textbf{#1:}\quad\ignorespaces
}{%
    \myqed
    \par\addvspace{\medskipamount}%
}

\newcommand{\R}{\mathbb R}
\newcommand{\Z}{\mathbb Z}

\newcommand{\E}{\mathbb E}
\newcommand{\PP}{\mathbb P}

\newcommand{\I}[1]{\mathbb I_{#1}}

\newcommand{\bv}{\bm v}
\newcommand{\btheta}{\bm\theta}
\newcommand{\bbeta}{\bm\beta}
\newcommand{\by}{\bm y}
\newcommand{\bx}{\bm x}
\newcommand{\bc}{\bm{c}}

\newcommand{\bzero}{\bm 0}
\newcommand{\cA}{\mathcal A}
\newcommand{\cB}{\mathcal B}
\newcommand{\cC}{\mathcal C}
\newcommand{\cD}{\mathcal D}

\newcommand{\Poi}{\operatorname{Poi}}
\newcommand{\GammaDist}{\operatorname{Gamma}}

\newcommand{\NMNL}{\texttt{Na\"ive MNL} }
\newcommand{\SAA}{\texttt{SK--SAA} }
\newcommand{\DAMC}{\texttt{DA--MCMC} }
\newcommand{\EMNL}{\texttt{Exact--MLE} }

\renewcommand{\varrho}{\rho}

\newcolumntype{Y}{>{\raggedright\arraybackslash}X}

\begin{document}

\RUNAUTHOR{Wang, Jiang, and Chen}

\TITLE{Demand Estimation with Unobserved Stockout Substitution: Exact Likelihood Maximization}

\ARTICLEAUTHORS{%
\AUTHOR{Chenhao Wang}
\AFF{School of Economics and Management, Tongji University, Shanghai, China, \EMAIL{chenhaowang@tongji.edu.cn}}
\AUTHOR{Hansheng Jiang}
\AFF{Rotman School of Management, University of Toronto, Toronto, Ontario, Canada, \EMAIL{hansheng.jiang@rotman.utoronto.ca}}
\AUTHOR{Ningyuan Chen}
\AFF{Department of Management, University of Toronto Mississauga; Rotman School of Management, University of Toronto, Ontario, Canada, \EMAIL{ningyuan.chen@utoronto.ca}}
}

\ABSTRACT{When a product is sold out, customers who prefer it may buy a substitute or leave without purchasing. Efficient demand estimation in this setting is challenging when inventory is observed only at replenishments, because the stockout times, and hence the products available to each customer, are unobserved.  Estimating demand efficiently under stockout substitution has been a challenging problem since \citet{anupindi1998estimation}, and it has become more pressing as the scale of such problem instances grows thanks to innovations in retailing such as unmanned stores. We provide a complete solution to this problem for the multinomial logit (MNL) model, the most widely used model of customer choice. Using a continuous-time representation of the purchase process developed in \cite{zhou2025optimal}, we show that the exact likelihood and its gradient are one-dimensional integrals, no matter how many products, customers, or stockouts there are. Both can be evaluated at a cost linear in the numbers of products and periods. Our method estimates demand for $2{,}000$ products from $2{,}000$ selling periods in under ten minutes.
}

\maketitle

\section{Introduction}
\label{sec:introduction}
\begin{figure}[t]
\centering
\begin{tikzpicture}[
  x=1cm, y=1cm, >=Latex, font=\footnotesize,
  mk/.style={circle, line width=0.8pt, minimum size=4.6mm, inner sep=0pt, font=\scriptsize\bfseries},
  c0/.style={mk, draw=black!50, fill=white, text=black!75},
  c1/.style={mk, draw=prodA, fill=prodA!22},
  c2/.style={mk, draw=prodB, fill=prodB!22},
  c3/.style={mk, draw=prodC, fill=prodC!22},
  obsbox/.style={draw=black!70, line width=0.6pt, rounded corners=3pt, fill=black!2},
  latbox/.style={draw=black!50, densely dashed, line width=0.7pt, rounded corners=3pt},
  lab/.style={font=\scriptsize, text=black!85},
  head/.style={font=\scriptsize\bfseries, text=black!85},
  tag/.style={font=\scriptsize\itshape, text=black!55},
  sep/.style={draw=black!55, densely dashed, line width=0.6pt},
  brk/.style={draw=black!45, line width=0.5pt},
  flow/.style={-{Latex[length=2mm]}, line width=0.7pt, draw=black!60},
  soldunit/.style={draw=black!45, densely dashed, line width=0.5pt, fill=white},
]

\draw[obsbox] (0,-0.8) rectangle (2.55,2.2);
\node[head] at (1.275,1.95) {Initial inventory};
\node[tag] at (1.275,1.63) {observed};
\foreach \k in {0,1}   {\fill[prodA] (0.38,0.10+0.36*\k) rectangle ++(0.44,0.32);}
\foreach \k in {0,1,2} {\fill[prodB] (1.055,0.10+0.36*\k) rectangle ++(0.44,0.32);}
\foreach \k in {0,1}   {\fill[prodC] (1.73,0.10+0.36*\k) rectangle ++(0.44,0.32);}
\node[lab] at (0.60,-0.12) {1}; \node[lab] at (1.275,-0.12) {2}; \node[lab] at (1.95,-0.12) {3};

\draw[latbox] (3.05,-0.8) rectangle (12.75,2.2);
\node at (7.9,1.95) {{\scriptsize\bfseries\color{black!85}Customers in order of arrival}\ {\scriptsize\itshape\color{black!55}(unobserved)}};
\def\xs{{4.05,5.2,6.35,7.5,8.65,9.8,10.95,12.1}}
\node[head] at (3.45,1.0) {A};
\foreach \c [count=\t from 0] in {1,1,3,0,2,3,0,0} {\node[c\c] at ({\xs[\t]},1.0) {\c};}
\draw[sep] (5.775,0.72) -- (5.775,1.62);  \draw[sep] (10.375,0.72) -- (10.375,1.62);
\draw[brk] (3.80,1.30) -- ++(0,0.06) -- (5.45,1.36) -- ++(0,-0.06);
\draw[brk] (6.10,1.30) -- ++(0,0.06) -- (10.05,1.36) -- ++(0,-0.06);
\draw[brk] (10.70,1.30) -- ++(0,0.06) -- (12.35,1.36) -- ++(0,-0.06);
\node[lab] at (4.625,1.52) {$\{1,2,3\}$};
\node[lab] at (8.075,1.52) {$\{2,3\}$};
\node[lab] at (11.525,1.52) {$\{2\}$};
\node[head] at (3.45,-0.2) {B};
\foreach \c [count=\t from 0] in {3,0,2,3,1,0,1,0} {\node[c\c] at ({\xs[\t]},-0.2) {\c};}
\draw[sep] (8.075,-0.48) -- (8.075,0.42);  \draw[sep] (11.525,-0.48) -- (11.525,0.42);
\draw[brk] (3.80,0.10) -- ++(0,0.06) -- (7.75,0.16) -- ++(0,-0.06);
\draw[brk] (8.40,0.10) -- ++(0,0.06) -- (11.20,0.16) -- ++(0,-0.06);
\draw[brk] (11.85,0.10) -- ++(0,0.06) -- (12.35,0.16) -- ++(0,-0.06);
\node[lab] at (5.775,0.32) {$\{1,2,3\}$};
\node[lab] at (9.8,0.32) {$\{1,2\}$};
\node[lab] at (12.1,0.32) {$\{2\}$};
\node[tag] at (7.9,-0.62) {two of the $1{,}680$ purchase sequences consistent with the data};

\draw[obsbox] (13.25,-0.8) rectangle (15.8,2.2);
\node[head] at (14.525,1.95) {Terminal sales};
\node[tag] at (14.525,1.63) {observed};
\foreach \k in {0,1}   {\draw[soldunit] (13.63,0.10+0.36*\k) rectangle ++(0.44,0.32);}
\foreach \k in {0,1}   {\fill[prodB] (14.305,0.10+0.36*\k) rectangle ++(0.44,0.32);}
\draw[soldunit] (14.305,0.82) rectangle ++(0.44,0.32);
\foreach \k in {0,1}   {\draw[soldunit] (14.98,0.10+0.36*\k) rectangle ++(0.44,0.32);}
\node[lab] at (13.85,-0.12) {1}; \node[lab] at (14.525,-0.12) {2}; \node[lab] at (15.2,-0.12) {3};

\draw[flow] (2.57,0.7) -- (3.03,0.7);
\draw[flow] (12.77,0.7) -- (13.23,0.7);
\end{tikzpicture}
\caption{One example selling period}
\label{fig:schematic}
\smallskip
\begin{minipage}{\textwidth}
\footnotesize
\textit{Notes.} The initial inventory is $(2,3,2)$, and eight customers arrive. Sequences A and B both produce the terminal sales $(2,1,2)$, in which products 1 and 3 are sold out. Each circle shows a customer's choice, with $0$ denoting no purchase. The brackets indicate the products available to each customer.
\end{minipage}
\end{figure}
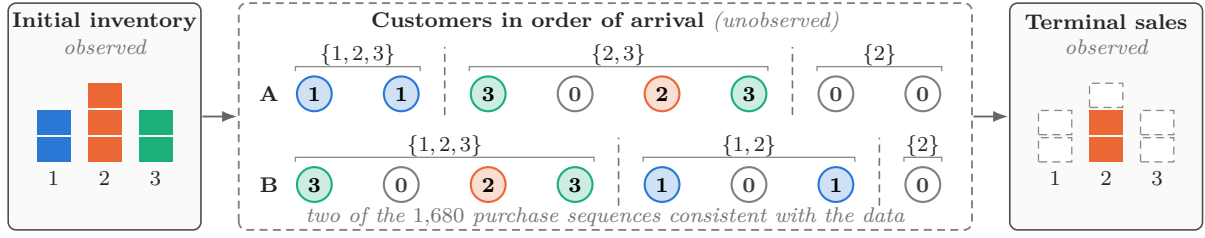

Consider a retailer who periodically replenishes an array of products in a vending machine.
The inventory of each product is observed only at replenishments, immediately before and after each one.
Between two replenishments, which we call a \emph{selling period}, customers arrive sequentially and choose from the products that are still available; a customer whose preferred product has sold out may choose a substitute or leave without purchasing.
The retailer observes none of these choices and, in particular, does not know when each product sold out.
The question is whether the retailer can estimate the demand for each product from the initial inventories and the number of units of each product sold in each period, which we call the terminal sales.
The same data structure (illustrated in Figure~\ref{fig:schematic} and formally introduced in Section~\ref{sec:data}) arises in traditional and unmanned retail stores, where inventory is typically counted only at replenishments.

Stockouts cannot be ignored: after a product sells out, its sales stop and part of its demand shifts to the remaining products.
Accounting for this shift is difficult because stockouts occur at unobserved times.
The terminal sales reveal which products sold out, but not when, and therefore not which products were available to each customer.
The same terminal sales can result from many purchase sequences, i.e., orders in which customers bought products or left, and their number grows exponentially with the number of customers.
In the small example of Figure~\ref{fig:schematic}, with three products and eight customers, $1{,}680$ purchase sequences are consistent with the observed initial inventory and terminal inventory/sales.
These sequences do not share the same likelihoods, because a product that sells out early changes the choices of all later customers.
For instance, product~1 sells out after the second customer in sequence~A but only after the seventh in sequence~B, so customers~3 to~7 can choose product~1 in sequence~B but not in sequence~A.
The likelihood of the observed data is therefore a sum over all the purchase sequences consistent with the initial inventory and terminal sales.

This estimation problem dates back at least to \citet{anupindi1998estimation} and remains an active area of research (see Section~\ref{sec:literature}).
Existing likelihood-based methods enumerate, sample, or approximate the purchase sequences, and each approach breaks down as the numbers of products and stockouts grow.
Expectation--maximization (EM) methods average over the possible times and orders of stockouts within each period, which is tractable only when few products have stockouts: \citet{anupindi1998estimation} carry out the EM calculations for up to two stockouts per period, and \citet{conlon2013demand} report that direct calculation is feasible for up to three.
Bayesian data augmentation \citep{musalem2010structural,wan2018demand} samples the purchase sequences; in our experiments, it takes several minutes even with ten products.
\citet{shao2020joint} group the purchase sequences by the order in which products sell out and sample these orders.
Their sampling and approximation method reduces the computational cost, but introduces simulation error or approximation bias that can affect parameter estimates, and controlling these errors requires additional computation.
This tradeoff between computational cost and estimation accuracy is compounded because the likelihood must be evaluated for every selling period at every iteration of the optimization.
To our knowledge, the exact likelihood has not been computed efficiently at a realistic scale, even under the multinomial logit (MNL) model, the most widely used model of customer choice.
The gap has widened as data have grown: assortments now often contain hundreds or thousands of products.

In this paper, we close this gap for the MNL model.
We build on a recent result of \citet{zhou2025optimal}, who represent the purchase sequence under MNL by a continuous-time system of independent Poisson processes and use it to study inventory allocation under dynamic substitution.
We use this representation for a different purpose: to compute the probability or likelihood of an entire vector of terminal sales.
Our finding is surprising: the exact likelihood of each selling period is a \emph{one-dimensional} integral, no matter how many products, customers, or stockouts there are.
The gradient of the log-likelihood is also a one-dimensional integral, and both the likelihood and its gradient can be computed efficiently through the integral expression.
These results amount to a complete solution to the problem: the purchase sequences are integrated out \emph{analytically} rather than enumerated, sampled, or approximated, and the cost of evaluating the likelihood and its gradient is linear in both the number of products and the number of selling periods.
In the numerical experiments,
our method estimates demand for $2{,}000$ products from $2{,}000$ selling periods within ten minutes on a single GPU, whereas no other benchmarks complete much smaller instances within the time limit.

{
\subsection{Literature Review}
\label{sec:literature}
A stream of literature studies demand estimation with stockouts, typically in setups that differ from ours.
We compare these studies with ours below.
\citet{anupindi1998estimation} and \citet{conlon2013demand} both estimate demand from periodic inventory records, as we do.
\citet{anupindi1998estimation} study Poisson demand with a particular substitution pattern, and \citet{conlon2013demand} consider more flexible choice models, including nested and random-coefficients logit.
Both use EM algorithms to handle the unobserved stockout times.
However, the expectation step becomes complicated as the number of stockouts in a period grows as illustrated in Figure~\ref{fig:schematic}, and both studies carry out the exact calculations only when few products sell out in a period.

\citet{musalem2010structural} and \citet{wan2018demand} tackle the problem by Bayesian data augmentation, sampling the unobserved purchase sequences that are consistent with the data.
Their setups differ slightly: \citet{musalem2010structural} allow heterogeneous customer preferences, and \citet{wan2018demand} consider nested logit models with substitution across both products and stores.
Both treat the total number of customers in each period, including those who leave without purchasing, as known, obtained from total transactions or estimated externally (see Section~\ref{sec:data} for more details).
However, sampling the purchase sequences is computationally intensive, and in our experiments it is practical only for the smallest instances (Section~\ref{sec:numerical-experiment}).

\citet{shao2020joint} is closest to ours, because they also attempt to formulate the likelihood exactly and maximize it directly.
They jointly estimate a Poisson arrival rate and the parameters of a general discrete-choice model from initial inventories and terminal sales, without observing individual choices, no-purchases, or stockout times.
They group choice sequences by their stockout orders, i.e., the order in which products sell out.
This allows them to simplify the calculation of the likelihood significantly.
However, the likelihood still requires enumerating or sampling stockout orders, which remains prohibitive for large instances in our experiments.
Our computation of the likelihood avoids enumerating or sampling high-dimensional latent spaces altogether, and this is our most important contribution to the literature.
\citet{shao2020joint} also leave identification open, suggesting that because stockouts create more variety in choice sets, the number of potential customers may be identifiable even when the initial assortment is fixed; we prove this under MNL (Section~\ref{sec:estimation-theory}). Table~\ref{tab:literature-comparison} summarizes the studies above.

\begin{table}[htbp]
\begingroup
\renewcommand{\baselinestretch}{1}
\small
\selectfont
\setlength{\tabcolsep}{4pt}
\renewcommand{\arraystretch}{1.2}
\begin{tabularx}{\textwidth}{@{}
>{\raggedright\arraybackslash}p{0.24\textwidth}
>{\raggedright\arraybackslash}X
>{\centering\arraybackslash}p{0.16\textwidth}
>{\raggedright\arraybackslash}X
@{}}
\toprule
& Choice model & Number of customers & Latent stockout history \\
\midrule
\citet{anupindi1998estimation} & Poisson demand with one-stage substitution & Estimated & EM; up to two stockouts per period \\
\addlinespace
\citet{musalem2010structural} & Random-coefficients logit & Given & Sampled by MCMC \\
\addlinespace
\citet{conlon2013demand} & Nested and random-coefficients logit & Given & EM; up to three stockouts per period \\
\addlinespace
\citet{wan2018demand} & Nested logit; substitution across stores & Given & Sampled by MCMC \\
\addlinespace
\citet{shao2020joint} & General discrete choice & Estimated & Enumerated or sampled \\
\addlinespace
This paper & MNL & Given or estimated & Integrated out analytically in the exact likelihood function\\
\bottomrule
\end{tabularx}
\centering
\caption{Closely related studies}
\label{tab:literature-comparison}
\smallskip
\endgroup
\end{table}

When individual transaction times, and hence stockout times, are observed or can be inferred, the problem becomes much more tractable, even though no-purchases and potential customer arrivals remain unobserved.
\citet{letham2016bayesian} exploit this information in a Bayesian model with nonhomogeneous Poisson arrivals and heterogeneous preferences.
They construct a transaction-time likelihood from purchase rates and note that the MNL no-purchase parameter is not identifiable when the arrival function is unknown.
\citet{chen2026bayesian} develop a related model with covariate-dependent arrivals and stockout-based substitution and derive an analytic transaction-time likelihood.
\citet{karabati2009method} group point-of-sale intervals by availability scenario and infer substitution from differences in purchase rates.
\citet{zheng2026offline} estimate base demand from observations unlikely to contain substitution and recover substitution probabilities by comparing demand across stockout conditions.
Even when all choices and stockout times are observed, \citet{liu2026lessdata} show that maximum likelihood estimates of MNL choice probabilities are biased by stockout-induced endogeneity, and they remove the bias by dropping the purchases that trigger stockouts.

A broader literature jointly estimates customer arrivals and preferences when no-purchases are unobserved but the offered assortment is recorded for each period or transaction.
Under MNL, the arrival rate and the no-purchase probability are separated either through a common arrival rate \citep{talluri2004revenue,newman2014estimation} or by combining period-specific arrival rates with external market-share information \citep{vulcano2012estimating,abdallah2021demand}.
Related methods accommodate rank-based, Markov chain, nested logit, and mixture-of-logit choice models \citep{vanryzin2017rank,simsek2018estimation, cho2022estimating,wang2025joint}, incorporate competitor information \citep{talluri2025estimation}, guard against errors in market-share information \citep{cho2024robust}, or develop integer-programming and moment-based estimators \citep{subramanian2021demand,li2025estimating}.
In all of these studies, the assortment faced by each customer is known.
In ours, it changes within the period at unobserved times, and we show that these unobserved changes are nevertheless enough to separate a common arrival rate from the MNL preferences.

A separate line of literature studies demand estimation in inventory control with censoring.
Classical work treats a single product \citep{nahmias1994demand,agrawal1996estimating}, and later work exploits the timing of stockouts \citep{jain2015demand} or corrects for inaccurate inventory records \citep{mersereau2015demand}; \citet{steeneck2022estimating} estimate lost sales for substitutable products when on-shelf availability is uncertain, using external market-share and availability information.
Our setting differs in that inventory records are accurate and no external information is used, but the times at which products sell out, and hence the choice sets of later customers, are unobserved.

Finally, assortment and inventory decisions under stockout substitution have attracted considerable attention recently.
Several studies estimate substitution from retail sales and then optimize assortments or inventories, typically treating availability as observed or constant within a period \citep{kok2007demand,fisher2014demand,akchen2026size}; \citet{lee2016stockout} do so for new and used textbooks from season sales with unobserved stockout times.
With known demand parameters, a large literature optimizes assortments and inventories under stockout-based substitution \citep{smith2000management,mahajan2001stocking,honhon2010assortment}, often through tractable approximations \citep{hopp2008static,liang2021assortment,aouad2023stability,zhang2025leveraging}, and related work learns demand while making these decisions \citep{chen2020dynamic,liang2026online}.
The continuous-time representations of MNL substitution developed for such problems \citep{zhou2025optimal,zhou2026optimal} are the starting point of our analysis.
}

\section{Data and Model}
\label{sec:data}
A retailer sells $J$ products, indexed by $j\in[J]=\{1,\ldots,J\}$, across $N$ selling periods, indexed by $n=1,\ldots,N$.
In period $n$, the retailer starts with the observed inventory vector $\by_n=(y_{1,n},\ldots,y_{J,n})\in\Z_+^J$, where $\Z_+=\{0,1,\ldots\}$. Customers then arrive sequentially and choose among the products that remain in stock or leave without a purchase (choosing the \emph{outside option}). 
To describe the dynamics of the inventory process, let $\bm I_{t,n}=(I_{1,t,n},\ldots,I_{J,t,n})$ denote the inventory immediately after customer $t$ has chosen, with $\bm I_{0,n}=\by_n$. Customer $t$ sees the available set $S_{t,n}=\{j\in[J]:I_{j,t-1,n}>0\}$. Product $j$ has attraction $\theta_j=e^{v_j}>0$, and the outside option, product $0$, 
is normalized to have attraction $\theta_0=1$, or equivalently $v_0=0$.
Observing $S_{t,n}$, the choice of customer $t$ in period $n$, $C_{t,n}\in\{0,1,\ldots,J\}$, satisfies the classic MNL model:

\begin{equation}
\PP(C_{t,n}=j\mid S_{t,n})=
\begin{cases}
\dfrac{\theta_j}{1+\sum_{k\in S_{t,n}}\theta_k},&j\in S_{t,n},\\[8pt]
\dfrac{1}{1+\sum_{k\in S_{t,n}}\theta_k},&j=0,\\[8pt]
0,&j\in[J]\setminus S_{t,n},
\end{cases}
\label{eq:mnl}
\end{equation}
where  $0$ denotes the outside option or no-purchase. 
Inventory evolves according to
\begin{equation}
I_{j,t,n}=I_{j,t-1,n}-\I{\{C_{t,n}=j\}},
\qquad j\in[J],
\label{eq:inventory-update}
\end{equation}
where $\I{\{\cdot\}}$ denotes the indicator function. Thus each arrival either leaves inventory unchanged or consumes one unit of an available product. 
If customer $t$ buys the last unit of product $j$, then $j$ is removed from $S_{t+1,n}$ and all subsequent available sets.
At the end of the period, the retailer observes terminal sales $\bx_n=(x_{1,n},\ldots,x_{J,n})$, a realization of the random terminal-sales vector $\bm X_n$.

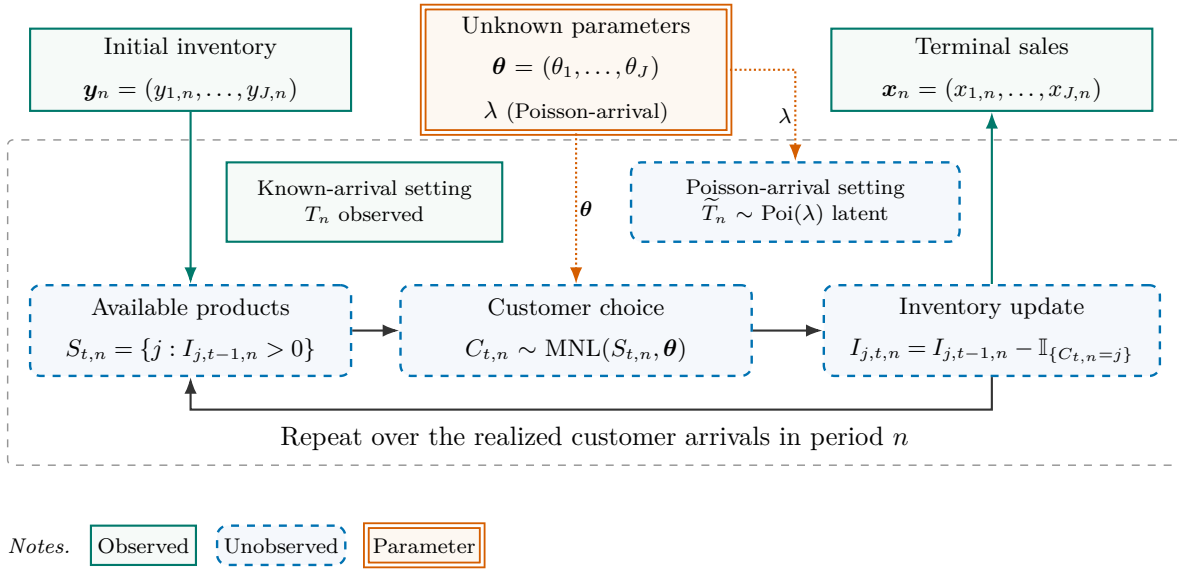
\begin{figure}[htbp]
\centering
\vspace{0.8em}
\begin{tikzpicture}[ >=Latex, block/.style={ draw, align=center, inner sep=3.5pt, font=\footnotesize }, observed/.style={ block, fill=obsfill, draw=obsline, line width=0.8pt }, latent/.style={ block, fill=latentfill, draw=latentline, dashed, rounded corners=4pt, line width=0.9pt }, model/.style={ block, fill=modelfill, draw=modelline, double, double distance=1.1pt, line width=0.7pt }, loopbox/.style={ draw=repeatline, dashed, line width=0.55pt, rounded corners=2pt, inner sep=8pt }, dataflow/.style={ -{Latex[length=2.2mm]}, thick, draw=black!80 }, obsflow/.style={ -{Latex[length=2.2mm]}, thick, draw=obsline }, paramflow/.style={ -{Latex[length=2.2mm]}, thick, densely dotted, draw=modelline }, legenditem/.style={ minimum height=0.52cm, inner sep=2.5pt, font=\scriptsize } ]
\node[ observed, text width=4.0cm, minimum height=1.05cm ] (inputs) at (-5.1,0) {Initial inventory\\ $\by_n=(y_{1,n},\ldots,y_{J,n})$};

\node[ model, text width=3.8cm, minimum height=1.35cm ] (param) at (0,0) {Unknown parameters\\ $\btheta=(\theta_1,\ldots,\theta_J)$\\[1pt] $\lambda$ {\scriptsize(Poisson-arrival)}};

\node[ observed, text width=4.0cm, minimum height=1.05cm ] (outcome) at (5.5,0) {Terminal sales\\ $\bx_n=(x_{1,n},\ldots,x_{J,n})$};

\node[ observed, text width=3.4cm, minimum height=1.05cm, font=\scriptsize ] (known) at (-2.8,-1.75) {Known-arrival setting\\ $T_n$ observed};

\node[ latent, text width=4.1cm, minimum height=1.05cm, font=\scriptsize ] (poisson) at (2.9,-1.75) {Poisson-arrival setting\\ $\widetilde T_n\sim\Poi(\lambda)$ latent};

\node[ latent, text width=4.0cm, minimum height=1.2cm ] (state) at (-5.1,-3.45) {Available products\\ $S_{t,n}=\{j:I_{j,t-1,n}>0\}$};

\node[ latent, text width=4.4cm, minimum height=1.2cm ] (choice) at (0,-3.45) {Customer choice\\ $C_{t,n}\sim\mathrm{MNL}(S_{t,n},\btheta)$};

\node[ latent, text width=4.2cm, minimum height=1.2cm ] (inventory) at (5.5,-3.45) {Inventory update\\ $I_{j,t,n}=I_{j,t-1,n}-\I{\{C_{t,n}=j\}}$};

\path (0,-4.95) coordinate (repeatfloor); \node[ loopbox, fit=(known)(poisson)(state)(choice)(inventory)(repeatfloor) ] (loop) {};

\node[ font=\small, fill=white, inner sep=1.5pt, anchor=south, yshift=1.5mm ] at (loop.south) {Repeat over the realized customer arrivals in period \(n\)};

\draw[obsflow] (inputs.south) -- (state.north);
\draw[obsflow] (inventory.north) -- (outcome.south);

\draw[paramflow]
    (param.south) --
    node[right, font=\scriptsize, inner sep=1pt] {$\btheta$}
    (choice.north);

\draw[paramflow]
    (param.east)
    -- (param.east -| poisson.north)
    -- node[midway, left, font=\scriptsize, inner sep=1pt] {$\lambda$}
    (poisson.north);

\draw[dataflow] (state.east) -- (choice.west);
\draw[dataflow] (choice.east) -- (inventory.west);
\draw[dataflow]
    (inventory.south) -- ++(0,-0.43) -| (state.south);

\node[ font=\scriptsize, anchor=west, minimum height=0.55cm, inner sep=0pt ] (notekey) at ([yshift=-1.05cm]loop.south west) {\it Notes. };

\node[ observed, right=0.25cm of notekey, legenditem ] (obskey) {Observed};

\node[ latent, right=0.25cm of obskey, legenditem ] (latkey) {Unobserved};

\node[ model, right=0.25cm of latkey, legenditem ] (modkey) {Parameter};

\end{tikzpicture}
\caption{Data-generating process for one selling period}
\label{fig:data_generation}
\end{figure}

We consider two data settings. In the \emph{known-arrival setting}, the observed data are $\cD_N^{\mathrm K}=\{(\by_n,\bx_n,T_n)\}_{n=1}^N$, and the unknown parameter is $\btheta$, or equivalently $\bv=\log\btheta$. In the \emph{Poisson-arrival setting}, the observed data are $\cD_N^{\mathrm P}=\{(\by_n,\bx_n)\}_{n=1}^N$, while $\widetilde T_n\sim\Poi(\lambda)$ independently across periods (we use $\widetilde T_n$ to distinguish from $T_n$ because $\widetilde T_n$ are unobserved latent variables). Here $\lambda>0$ is the expected number of arrivals per selling period. Representing each period by the interval $[0,1]$, we refer to $\lambda$ as the arrival rate. In addition to $\bv$, this rate is unknown and to be estimated. 
Figure~\ref{fig:data_generation} summarizes the inventory process and the  information structure.

In the rest of the paper, when we discuss a particular period and there is no confusion, we suppress the index $n$ and use $(\by,\bx)$ to denote the observed inventories and sales, with $0\le x_j\le y_j$ for every product. Let $\cA=\{j:y_j>0\}$ be the initially available products, $\cB=\{j\in\cA:x_j=y_j\}$ the products that sell out by the end of the period, and $\cC=\cA\setminus\cB$ the products that remain in stock. Adding the subscript $n$ denotes these sets for period $n$.

\section{Continuous-Time Representation and Exact Likelihood}
\label{sec:embedding}
\label{sec:likelihoods}

A natural choice to estimate the parameters in Section~\ref{sec:data} is a likelihood-based method. Consider the known-arrival setting for simplicity. For a trajectory of purchases $\bc=(c_1,\ldots,c_T)$, including no-purchase outcomes, let $\bm C=(C_1,\ldots,C_T)$ and let $S_t(\bc)$ be the available set obtained by applying \eqref{eq:inventory-update}. Its \emph{complete-data likelihood} is
\begin{equation*}
\PP_{\bv}(\bm C=\bc\mid\by,T) = \prod_{t=1}^{T} \frac{e^{v_{c_t}}}{1+\sum_{k\in S_t(\bc)}e^{v_k}}.
\end{equation*}
However, the seller does not observe the choice trajectory, which must be consistent with the terminal sales:
\[
\mathcal P(\bx,T) = \left\{ \bc\in\{0,1,\ldots,J\}^{T}: \sum_{t=1}^{T}\I{\{c_t=j\}}=x_j \ \text{for every }j\in[J] \right\}.
\]
Because $x_j\le y_j$, every trajectory in $\mathcal P(\bx,T)$ is consistent with the initial inventory and the terminal sales: before any purchase of $j$, fewer than $x_j\le y_j$ units of $j$ have been sold. The number of such trajectories can be exponentially large in $T$, as illustrated in Figure~\ref{fig:schematic}. The \emph{observed-data} likelihood aggregates the complete-data likelihood over possible choice trajectories:
\begin{equation}
L(\bv\mid\by,\bx,T)=\sum_{\bc\in \mathcal P(\bx,T)}
\PP_{\bv}(\bm C=\bc\mid\by,T).
\label{eq:path-sum-likelihood}
\end{equation}
Note that we treat $\by$ and $T$ as given and $\bx$ as the random sample being generated when expressing the likelihood in this way.
Directly enumerating $\mathcal P(\bx,T)$ can therefore be computationally prohibitive, particularly when likelihood evaluation is repeated during optimization. Applying EM does not by itself remove this difficulty: its expectation step must still average over latent histories conditional on terminal sales. This latent-history calculation is a central challenge in \cite{musalem2010structural}, \cite{conlon2013demand}, \cite{shao2020joint}, and \cite{steeneck2022estimating}.

We next represent the process in a different way to provide a tractable characterization of the likelihood. We use the \emph{continuous-time representation} developed for stockout-based substitution in \citet{zhou2025optimal}. The key idea is to embed the discrete-time events in Section~\ref{sec:data}, such as $C_{t,n}$ and $I_{j,t,n}$, in a continuous-time system with \emph{Poisson processes}. More precisely, consider $J+1$ independent Poisson processes $(\Pi_j(\tau))_{\tau\ge0}$, $j=0,\ldots,J$, with rates $\theta_j$ and $\theta_0=1$. We use $\tau$ for continuous time and retain $t$ for the customer index. For $j\ge1$, only the first $y_j$ counts can represent sales; process $0$ records no-purchase outcomes and remains uncapped. Define the capped processes by
\begin{equation}
N_0(\tau)=\Pi_0(\tau),\qquad
N_j(\tau)=\min\{y_j,\Pi_j(\tau)\},\qquad j\in[J].
\label{eq:capped-counts}
\end{equation}
Note that the capped processes are still independent. This differs from the discrete-time model, where products share the same sequence of customers. To connect the two systems, define $N_{\mathrm{tot}}(\tau)\coloneqq N_0(\tau)+\sum_{j=1}^J N_j(\tau)$ as the total number of customer choices by (continuous) time $\tau$, including no purchases. For each integer $t\ge1$, define two quantities
\[
\sigma_t=\inf\{\tau\ge0:N_{\mathrm{tot}}(\tau)\ge t\},
\qquad
D_t=j
\iff
N_j(\sigma_t)-N_j(\sigma_t^-)=1,\qquad j=0,\ldots,J,
\]
where \(N_j(\sigma_t^-)=\lim_{\tau\uparrow\sigma_t}N_j(\tau)\). Here, \(\sigma_t\) is the first time at which \(N_{\mathrm{tot}}(\tau)\) reaches \(t\), while \(D_t\) records which process contributes the \(t\)-th choice.
In this definition, \(D_1,D_2,\ldots\) in the continuous-time system represents sequence of choices and can be mapped to the customer choices \(C_1,C_2,\ldots\) in the discrete-time system. \citet[Section~4.1]{zhou2025optimal} give the following equivalence in distribution:
\begin{lemma}[Continuous-time representation]
\label{lem:clock-embedding}
For every fixed \(t\in\Z_+\),
\begin{equation}
(C_1,\ldots,C_t)\overset{d}{=}(D_1,\ldots,D_t),
\label{eq:embedding-law}
\end{equation}
where recall that $C_t$ and $D_t$ are the $t$-th purchased product (including the outside option) in the discrete-time and continuous-time systems, respectively.
\end{lemma}

We provide a brief introduction to the intuition behind Lemma~\ref{lem:clock-embedding} while leaving the details to \citet{zhou2025optimal}. The result hinges on the Independence of Irrelevant Alternatives (IIA) property of the MNL choice probability, which keeps the choice-probability ratios among available alternatives unchanged as other products sell out. 
Under the exponential interarrival times of independent Poisson processes,  these choice probabilities become competing clocks, with the memoryless property preserving the same choice dynamics after each event. 
These properties (IIA and memoryless) guarantee that although the inventory processes of different products in the discrete-time system seem to be more intertwined, they are essentially ``independent'' under the continuous-time view.

One implication of Lemma~\ref{lem:clock-embedding} is that, if we plug in $t=T_n$, the number of customers in period $n$, then $(N_1(\sigma_{T_n}),\dots,N_J(\sigma_{T_n}))$ have the same joint distribution as the terminal-sales vector $\bm X_n$, conditional on $(\by_n,T_n)$. We note that the underlying Poisson processes are independent as stochastic processes, so $N_j(\tau)$s are independent at any fixed $\tau$. 
However, $(N_1(\sigma_{T_n}),\dots,N_J(\sigma_{T_n}))$ are dependent because the common random time $\sigma_{T_n}$ is determined by the total arrivals reaching $T_n$.

Using Lemma~\ref{lem:clock-embedding}, we next analyze the likelihood of terminal sales $(x_{1,n},\dots,x_{J,n})$, or equivalently the joint distribution of $(N_1(\sigma_{T_n}),\dots,N_J(\sigma_{T_n}))$, through the continuous-time system. To do that, we first study the joint probability at a fixed continuous time, without conditioning on a stopping time. Given the terminal sales $\bx$, one can recover $\cB$ and $\cC$, the subsets of products that are sold out and in stock at the end of the period, respectively. For total no-purchase $m$ and a fixed time $\tau>0$, by independence of the Poisson processes, we have 
\begin{equation}
G_m(\tau;\btheta)
\coloneqq \PP(N_0(\tau)=m, N_1(\tau)=x_1,\dots,N_J(\tau)=x_J)=
\pi_m(\tau)
\prod_{j\in\cC}\pi_{x_j}(\theta_j\tau)
\prod_{j\in\cB}\overline F_{y_j}(\theta_j\tau).
\label{eq:Gm}
\end{equation}
Here, \(\pi_k(\mu)\coloneqq \PP(\Poi(\mu)=k)=e^{-\mu}\mu^k/k!\) is the probability mass function (PMF) of the Poisson distribution, and \(\overline F_k(\mu)\coloneqq \PP\{\Poi(\mu)\ge k\}=\sum_{l=k}^{\infty}e^{-\mu}\mu^l/l!\) is the tail probability. The joint distribution reflects products in $\cB$ and $\cC$ differently: for $j\in\cC$, $N_j(\tau)=x_j<y_j$ is equivalent to $\Pi_j(\tau)=x_j$; for $j\in\cB$, $N_j(\tau)=y_j$ is equivalent to $\Pi_j(\tau)\ge y_j$. The number of no purchases $N_0(\tau)$ remains uncapped.

The expression \eqref{eq:Gm} gives the probability of the state $(m,\bx)$ at a fixed continuous time $\tau$. To relate it to the likelihood of the discrete-time system, we need to integrate out $\tau$.
More precisely, we are interested in the probability of the event $(N_0(\tau),N_1(\tau),\dots,N_J(\tau))=(m,x_1,\dots,x_J)$ for some $\tau>0$.
Once the continuous-time system ever reaches the state $(m,x_1,\dots,x_J)$, in the coupled discrete-time system, what we observe is the terminal sales being $\bx =(x_1,x_2,\dots,x_J)$ given $T=m+x_1+\dots+x_J$ customers and initial inventory $\by$.
Vice versa, if the discrete-time system observes the terminal sales observation $\bx$ under $T$ and $\bm y$, then the continuous-time system $(N_0(\tau),N_1(\tau),\dots,N_J(\tau))$ ends in $(m,x_1,\dots,x_J)$.

\begin{lemma}[Terminal-sales probability representation]
\label{lem:fixed-count-occupation}
The probability of $(N_0,N_1,\ldots,N_J)$ hitting $(m,x_1,\ldots,x_J)$ is
\begin{equation}
\PP((N_0(\tau),N_1(\tau),\dots,N_J(\tau))=(m,x_1,\dots,x_J)\text{ for some }\tau>0)
=
(1+\sum_{j\in\cC}\theta_j ) \int_0^\infty G_m(\tau;\btheta)\,d\tau.
\label{eq:fixed-count-occupation}
\end{equation}
\end{lemma}

To understand \eqref{eq:fixed-count-occupation}, it is helpful to view $(N_0(\tau),N_1(\tau),\dots,N_J(\tau))$ as a continuous-time Markov chain. Because $G_m(\tau;\btheta)$ is the probability of being in the state $(m,\bx)$ at time $\tau$,
\[
\int_{0}^{\infty} G_m(\tau;\btheta)\,d\tau
=
\E \left[\int_{0}^{\infty} \I{\{(N_0(\tau),\dots,N_J(\tau))=(m,\bx)\}}\,d\tau \right]
\]
is the expected occupation time of that state. The process $(N_0(\tau),N_1(\tau),\dots,N_J(\tau))$ is nondecreasing in each coordinate, so after leaving $(m,\bx)$ it cannot return. In this state, the outside-option process has rate \(1\), the products \(j\in\cC\) that remain in stock have rates \(\theta_j\), and the sold-out products cannot change any more. Hence the mean holding time in the state is \(1/(1+\sum_{j\in\cC}\theta_j)\), which yields \eqref{eq:fixed-count-occupation}. By Lemma~\ref{lem:clock-embedding}, reaching $(m,\bx)$ is equivalent in distribution to observing terminal sales $\bx$ after \(T=m+\sum_{j=1}^Jx_j\) customer choices.

Compared with the likelihood representation in \eqref{eq:path-sum-likelihood}, Lemma~\ref{lem:fixed-count-occupation} already removes the summation over feasible choice trajectories. We next use this representation to derive tractable likelihood expressions under the two data settings.

\subsection{The Known-Arrival Setting}
\label{subsec:known-likelihood}

For a feasible observation with $\sum_jx_j\le T$, Lemma~\ref{lem:fixed-count-occupation} gives the likelihood
\begin{equation}
L^{\mathrm K}(\bv\mid\by,\bx,T)
\coloneqq
(1+\sum_{j\in\cC}\theta_j) \int_0^\infty
\pi_m(\tau)
\prod_{j\in\cC}\pi_{x_j}(\theta_j\tau)
\prod_{j\in\cB}\overline F_{y_j}(\theta_j\tau)
\,d\tau,
\label{eq:known-occupation}
\end{equation}
where $s=\sum_{j=1}^Jx_j$ is total product sales and $m=T-s$ is the observed number of no-purchase outcomes. The notation \(L^{\mathrm K}(\bv\mid\by,\bx,T)\) emphasizes that the likelihood is a function of \(\bv\) (equivalently, \(\btheta\)), whereas \(\by\), \(\bx\), and \(T\) are observed.

We next move the terms unrelated to $\tau$ out of the integral. Let $c=\sum_{j\in\cC}x_j$ be the total sales of the products that remain in stock, and set $\rho_j=\theta_j/(1+\sum_{k\in\cC}\theta_k)$ for $j\in\cA$. Expanding the Poisson masses in \eqref{eq:known-occupation} and changing variables to $u=(1+\sum_{j\in\cC}\theta_j)\tau$ gives
\begin{align}
L^{\mathrm K}
&=\frac{(1+\sum_{j\in\cC}\theta_j)\prod_{j\in\cC}\theta_j^{x_j}}{m!\prod_{j\in\cC}x_j!}
\int_0^\infty e^{-(1+\sum_{j\in\cC}\theta_j)\tau}\tau^{m+c}
\prod_{j\in\cB}\overline F_{y_j}(\theta_j\tau)\,d\tau
\label{eq:thm1-derivation}\\
&=\frac{(1+\sum_{j\in\cC}\theta_j)^{-m}\prod_{j\in\cC}\rho_j^{x_j}}{m!\prod_{j\in\cC}x_j!}
\int_0^\infty u^{m+c}e^{-u}
\prod_{j\in\cB}\overline F_{y_j}(\rho_ju)\,du\notag\\
&=\underbrace{\frac{(m+c)!}{m!\prod_{j\in\cC}x_j!}
(1+\sum_{j\in\cC}\theta_j)^{-m}\prod_{j\in\cC}\rho_j^{x_j}}_{C^{\mathrm K}(\bv)}
\times\underbrace{\frac1{(m+c)!}\int_0^\infty u^{m+c}e^{-u}
\prod_{j\in\cB}\overline F_{y_j}(\rho_ju)\,du}_{R^{\mathrm K}(\bv)}.\notag
\end{align}

In the remaining integral, $u^{m+c}e^{-u}/(m+c)!$, $u>0$, is the density of $U\sim\GammaDist(m+c+1,1)$. Thus
\[
R^{\mathrm K}(\bv)
=
\E\left[
\prod_{j\in\cB}
\overline F_{y_j}(\rho_jU)
\right].
\]
This leads to the following result.

\begin{theorem}[Known-arrival likelihood]
\label{thm:known-gamma}
The likelihood of the known-arrival setting \eqref{eq:known-occupation} can be written as
\begin{equation}
L^{\mathrm K}(\bv\mid\by,\bx,T) = C^{\mathrm K}(\bv)R^{\mathrm K}(\bv),
\label{eq:known-factorization}
\end{equation}
where
\begin{align}
C^{\mathrm K}(\bv)
&=
\frac{{(m+c)!}}
{{m!}\prod_{j\in\cC}{x_j!}}
\frac{\prod_{j\in\cC}{\rho}_j^{x_j}}{(1+\sum_{j\in\cC}\theta_j)^m},
\label{eq:known-prefactor}\\
R^{\mathrm K}(\bv)
&=
\E_{U\sim\GammaDist(m+c+1,1)}
\left[
\prod_{j\in\cB}
\overline F_{y_j}({\rho}_jU)
\right].
\label{eq:known-gamma}
\end{align}
\end{theorem}

This factorization is helpful because the terms are treated separately in the log-likelihood $\log L^{\mathrm K}=\log C^{\mathrm K}+\log R^{\mathrm K}$. 
When no product sells out, $\cB=\varnothing$, $c=s$, $m+s=T$, and $R^{\mathrm K}=1$, so the likelihood reduces to $C^{\mathrm K}$
\[
L^{\mathrm K}(\bv\mid\by,\bx,T)
=
\frac{T!}{m!\prod_{j\in\cA}x_j!}
\left(\frac{1}{1+\sum_{k\in\cA}\theta_k}\right)^{m}
\prod_{j\in\cA}\left(\frac{\theta_j}{1+\sum_{k\in\cA}\theta_k}\right)^{x_j}.
\]
This is the usual multinomial likelihood under the constant assortment $\cA$, considering the possible permutations. 
When $\cB\ne\varnothing$, the product $\prod_{j\in\cB}\overline F_{y_j}(\rho_ju)$ is the probability that independent Poisson random variables with means $(\rho_ju)_{j\in\cB}$ reach their respective inventory, and $R^{\mathrm K}$ averages this probability over the Gamma variable \(U\).
The factor $R^{\mathrm K}$ is therefore interpreted as the correction for the stockout-induced changes in the assortment.

\begin{example}
Consider the period in Figure~\ref{fig:schematic}, with $\by=(2,3,2)$, $T=8$, and $\bx=(2,1,2)$. Products~1 and~3 sell out, so $\cB=\{1,3\}$, $\cC=\{2\}$, $m=3$, and $c=1$, and Theorem~\ref{thm:known-gamma} gives
\[
C^{\mathrm K}(\bv)=\frac{4\theta_2}{(1+\theta_2)^4},
\qquad
R^{\mathrm K}(\bv)=\E_{U\sim\GammaDist(5,1)}\left[\overline F_2\left(\frac{\theta_1U}{1+\theta_2}\right)\overline F_2\left(\frac{\theta_3U}{1+\theta_2}\right)\right].
\]
The likelihood is thus a single one-dimensional expectation rather than a sum over the $1{,}680$ purchase sequences consistent with the data.
\end{example}

\subsection{Poisson-Arrival Setting}
\label{subsec:poisson-likelihood}

When the total number of arrivals is unobserved, the number of no-purchase customers is also latent, so the likelihood sums over its possible values:
\begin{align}
L^{\mathrm P}(\lambda,\bv\mid\by,\bx) &= \sum_{m=0}^\infty \pi_{s+m}(\lambda)\,\PP_{\bv}(\bm X=\bx\mid\by,\widetilde T=s+m)\notag\\
&= \sum_{m=0}^\infty \pi_{s+m}(\lambda) L^{\mathrm K}(\bv\mid\by,\bx,T=s+m).
\label{eq:poisson-mixture}
\end{align}
Substituting \eqref{eq:known-occupation} into \eqref{eq:poisson-mixture} and interchanging the sum and integral lead to
\begin{align*}
L^{\mathrm P}(\lambda,\bv\mid\by,\bx) &=(1+\sum_{j\in\cC}\theta_j)\int_0^\infty \sum_{m=0}^\infty \pi_{s+m}(\lambda) \pi_m(\tau) \prod_{j\in\cC}\pi_{x_j}(\theta_j\tau) \prod_{j\in\cB}\overline F_{y_j}(\theta_j\tau) \,d\tau.
\end{align*}
The only term in the integral that involves $m$ is \(\sum_{m=0}^\infty\pi_{s+m}(\lambda)\pi_m(\tau)\). Expanding the Poisson PMF, we rewrite it as
\[
\sum_{m=0}^\infty \pi_{s+m}(\lambda) \pi_m(\tau) = e^{-(\lambda+\tau)} \frac{\lambda^s}{s!} \Phi_s(\lambda\tau),
\]
where
\begin{equation}
\Phi_s(z) \coloneqq s! \sum_{m=0}^\infty \frac{z^m}{m!(s+m)!},\qquad z\ge0.
\label{eq:phi_s-def}
\end{equation}
The factor $e^{-\lambda}\lambda^s/s!$ does not depend on $\tau$, while $\Phi_s(\lambda\tau)$ retains the effect of summing over the latent no purchases.

Expanding the remaining Poisson masses and applying the same change of variables $u=(1+\sum_{j\in\cC}\theta_j)\tau$ gives
\begin{align*}
L^{\mathrm P}(\lambda,\bv\mid\by,\bx)
&=
\frac{e^{-\lambda}\lambda^s
(1+\sum_{j\in\cC}\theta_j)\prod_{j\in\cC}\theta_j^{x_j}}
{{s!}\prod_{j\in\cC}{x_j!}}
\int_0^\infty
e^{-(1+\sum_{j\in\cC}\theta_j)\tau}\tau^c
\Phi_s(\lambda\tau)
\prod_{j\in\cB}\overline F_{y_j}(\theta_j\tau)
\,d\tau\\
&=
\frac{e^{-\lambda}\lambda^s
\prod_{j\in\cC}{\rho}_j^{x_j}}
{{s!}\prod_{j\in\cC}{x_j!}}
\int_0^\infty
u^ce^{-u}
\Phi_s\!\left({\frac{\lambda u}{1+\sum_{j\in\cC}\theta_j}}\right)
\prod_{j\in\cB}\overline F_{y_j}({\rho}_j u)
\,du\\
&=
\underbrace{
e^{-\lambda}\lambda^s
\frac{{c!}}{{s!}}
\prod_{j\in\cC}
\frac{{\rho}_j^{x_j}}{{x_j!}}
}_{C^{\mathrm P}(\lambda,\bv)} \times
\underbrace{
\frac{1}{{c!}}
\int_0^\infty
u^ce^{-u}
\Phi_s\!\left({\frac{\lambda u}{1+\sum_{j\in\cC}\theta_j}}\right)
\prod_{j\in\cB}\overline F_{y_j}({\rho}_j u)
\,du
}_{R^{\mathrm P}(\lambda,\bv)}.
\end{align*}
In the remaining integral, \(u^ce^{-u}/{c!}\) is the density of \(U\sim\GammaDist(c+1,1)\). Therefore,
\[
R^{\mathrm P}(\lambda,\bv)
=
\E\left[
\Phi_s\!\left({\frac{\lambda U}{1+\sum_{j\in\cC}\theta_j}}\right)
\prod_{j\in\cB}\overline F_{y_j}({\rho}_jU)
\right].
\]
Compared with \eqref{eq:thm1-derivation}, the additional factor $\Phi_s(\lambda u/(1+\sum_{j\in\cC}\theta_j))$ remains inside the integral because it depends on $u$. Thus, it preserves the one-dimensional expectation, giving the following result.

\begin{theorem}[Poisson-arrival likelihood]
\label{thm:poisson-gamma}
The likelihood of the Poisson-arrival setting \eqref{eq:poisson-mixture} can be written as
\begin{equation}
L^{\mathrm P}(\lambda,\bv\mid\by,\bx)
=
C^{\mathrm P}(\lambda,\bv)R^{\mathrm P}(\lambda,\bv),
\label{eq:poisson-factorization}
\end{equation}
where
\begin{align}
C^{\mathrm P}(\lambda,\bv) &= e^{-\lambda}\lambda^s \frac{{c!}}{{s!}} \prod_{j\in\cC} \frac{{\rho}_j^{x_j}}{{x_j!}},
\label{eq:poisson-prefactor}\\
R^{\mathrm P}(\lambda,\bv) &= \E_{U\sim\GammaDist(c+1,1)} \left[ \Phi_s\!\left({\frac{\lambda U}{1+\sum_{j\in\cC}\theta_j}}\right) \prod_{j\in\cB} \overline F_{y_j}({\rho}_jU) \right].
\label{eq:poisson-gamma}
\end{align}
\end{theorem}

The main difference from the known-arrival setting is that \(R^{\mathrm P}\) contains an additional factor associated with the latent no purchases. The tail factors \(\overline F_{y_j}(\rho_jU)\) account for sold-out products as in \(R^{\mathrm K}\), whereas \(\Phi_s(\lambda U/(1+\sum_{j\in\cC}\theta_j))\) accounts for the unobserved total number of arrivals. For an observed sales vector $\bx$ with no sold-out products ($\cB=\varnothing$), every customer faced the same assortment $\cA$ throughout the period. The likelihood therefore reduces to $L^{\mathrm P}=\prod_{j\in\cA}\pi_{x_j}(\lambda\theta_j/(1+\sum_{k\in\cA}\theta_k))$, the same probability of observing $\bx$ as in a model without inventory constraints.

\section{The Gradient of the Log-Likelihood and Its Computation}
\label{sec:score-structure}
The exact likelihood in Section~\ref{sec:likelihoods} reduces to one-dimensional integration. 
For estimation, we further need to evaluate the corresponding gradients repeatedly and accurately.  We first derive analytic expressions for the gradient of the log-likelihood in the known- and Poisson-arrival settings, and then develop a numerical procedure for evaluating these gradients.

\subsection{The Gradient of the Log-Likelihood}
In both the known-arrival and Poisson-arrival settings, by Theorems~\ref{thm:known-gamma} and~\ref{thm:poisson-gamma}, the likelihood takes the form \(L=CR\).
Our goal is to evaluate the derivative of $\log L = \log C+\log R$, with respect to the parameters.
The term $\log C$ and its gradient have closed forms.
As for the gradient of $\log R$, we use the known-arrival setting as an example to show how to obtain a tractable form.

Recall that
\begin{equation*}
R^{\mathrm K} = \frac{1}{(m+c)!} \int_0^\infty u^{m+c}e^{-u}  \prod_{j\in\cB} \overline F_{y_j}({\rho}_j u)\,du.
\end{equation*}
The parameters $\bv$ appear in $\rho_j$ inside $\overline F_{y_j}$.
Lemma~\ref{lem:differentiation} in Appendix~\ref{app:proofs} justifies differentiating under the integral. Differentiating $\log R^{\mathrm K}$ with respect to $v_j$ yields
\begin{align}
\partial_{v_j}\log R^{\mathrm K} &= \frac{1}{R^{\mathrm K}(m+c)!} \int_0^\infty u^{m+c}e^{-u} \partial_{v_j} \prod_{k\in\cB} \overline F_{y_k}({\rho}_k u)\,du \nonumber\\
&=
\frac{1}{R^{\mathrm K}(m+c)!} \int_0^\infty u^{m+c}e^{-u}  \prod_{k\in\cB} \overline F_{y_k}({\rho}_k u)\frac{\partial\log(\prod_{k\in\cB} \overline F_{y_k}({\rho}_k u))}{\partial v_j}\,du
\label{eq:score-R-derivative}
\end{align}
where we use \(f'(x)=f(x)(\log f(x))'\) in the second equality.
The logarithm helps us break down the product. 
For $k\in\cB$ and $j\in\cA$, we have
\[
\frac{\partial}{\partial v_j}\log\overline F_{y_k}(\rho_ku)
=\frac{\rho_ku\,\pi_{y_k-1}(\rho_ku)}{\overline F_{y_k}(\rho_ku)}
\left(\I{\{j=k\}}-\rho_j\I{\{j\in\cC\}}\right).
\]

Moreover, note that the remaining factors in the integrand in \eqref{eq:score-R-derivative} satisfy
\begin{equation*}
    \frac{1}{R^{\mathrm K}(m+c)!} \int_0^\infty u^{m+c}e^{-u}  \prod_{k\in\cB} \overline F_{y_k}({\rho}_k u)\,du = 1,
\end{equation*}
because of the definition of $R^{\mathrm K}$. 
This normalization defines a separate random variable $W$ with density
\begin{equation}
\label{eq:pdf-w}
f_W(u)=\frac{u^{m+c}e^{-u}\prod_{k\in\cB}\overline F_{y_k}(\rho_ku)}{R^{\mathrm K}(m+c)!},\qquad u>0,
\end{equation}
then \eqref{eq:score-R-derivative} can be compactly expressed as
\begin{equation*}
\frac{\partial\log R^{\mathrm K}}{\partial v_j}
=\E_W\left[\left.\frac{\partial}{\partial v_j}
\log\left\{\prod_{k\in\cB}\overline F_{y_k}(\rho_ku)\right\}\right|_{u=W}\right].
\end{equation*}
For $j\in\cC$, the explicit factor contributes $\partial_{v_j}\log C^{\mathrm K}=x_j-(m+c)\rho_j$; for $j\notin\cC$, that derivative is zero.  Combining these terms gives

\begin{theorem}[Gradient for known arrivals]
\label{thm:known-score}
In the known-arrival setting, for each product $j$ with positive initial inventory,
\begin{equation}
\frac{\partial\log L^{\mathrm K}}{\partial v_j}
=
\begin{cases}
\displaystyle
\E_{W}
\left[
\frac{\rho_jW\,\pi_{y_j-1}(\rho_jW)}
{\overline F_{y_j}(\rho_jW)}
\right],
& j\in\cB,\\[12pt]
\displaystyle
x_j-\rho_j
\left\{
c+m+
\sum_{k\in\cB}
\E_{W}
\left[
\frac{\rho_kW\,\pi_{y_k-1}(\rho_kW)}
{\overline F_{y_k}(\rho_kW)}
\right]
\right\},
& j\in\cC.
\end{cases}
\label{eq:known-score}
\end{equation}
Here, recall that
\(\cB\) and
\(\cC\) denote the sold-out products and the products that remain in stock, respectively,
\(c=\sum_{j\in\cC}x_j\),
\(m=T-s\), and
\({\rho}_j=\theta_j/(1+\sum_{k\in\cC}\theta_k)\).
The random variable \(W\) has the density given in \eqref{eq:pdf-w},
while \(\pi_k(\mu)\) and \(\overline F_k(\mu)\) denote the probability mass
function and tail probability of a Poisson random variable.
\end{theorem}

This result reflects how the parameters enter the likelihood.
For a sold-out product, the derivative with respect to $v_j$ is the expected
derivative of its log Poisson tail probability with respect to $v_j$, with the
integration variable held fixed.  
For a product that remains in stock, changing $v_j$ changes both the explicit factor and the common denominator $1+\sum_{k\in\cC}\theta_k$, and hence changes the tail argument of every sold-out product.  These effects produce the observed-sales term $x_j$ and the shared subtraction in \eqref{eq:known-score}.
Products with zero initial inventory do not enter the likelihood, so their corresponding log-likelihood derivatives are zero.

The representation also simplifies gradient computation.  
For a given observation of a period, all gradients use the same scalar random variable
\(W\).  Once the expectations associated
with the sold-out products are evaluated, their sum determines the gradient
of every product that remains in stock.  Thus, no separate numerical evaluation of the expectations is
carried out for each component of \(\bv\).  Section~\ref{subsec:score-computation}
uses this structure to evaluate the likelihood and its gradient
simultaneously.

Under Poisson arrivals, the additional factor $\Phi_s$ in Theorem~\ref{thm:poisson-gamma} contributes to the attraction gradients and to the derivative with respect to the log arrival rate. The same differentiation argument applies.
Define a separate random variable $\bar W$ whose density incorporates $\Phi_s$:
\begin{equation}
f_{\bar W}(u) = \frac{ u^ce^{-u} {\Phi_s(\lambda u/(1+\sum_{j\in\cC}\theta_j))} \prod_{k\in\cB}\overline F_{y_k}({\rho}_k u) }{ R^{\mathrm P}c! }, \qquad u>0,
\label{eq:pdf-w-poisson}
\end{equation}
In addition, define
\[
\psi_s(z)\coloneqq z\frac{d}{dz}\log\Phi_s(z),\quad z>0,\qquad \psi_s(0)=0,
\qquad
\overline m \coloneqq \E_{\bar W} \left[ \psi_s\left({\frac{\lambda\bar W}{1+\sum_{j\in\cC}\theta_j}}\right) \right].
\]
The quantity $\overline m$ replaces $m$ in the attraction gradients below. Appendix~\ref{app:proofs} provides the derivation.
\begin{theorem}[Gradient for Poisson arrivals]
\label{thm:poisson-score}
In the Poisson-arrival setting,
\begin{equation*}
\frac{\partial\log L^{\mathrm P}}{\partial v_j}
=
\begin{cases}
\displaystyle
\E_{\bar W}
\left[
\frac{\rho_j\bar W\,\pi_{y_j-1}(\rho_j\bar W)}
{\overline F_{y_j}(\rho_j\bar W)}
\right],
& j\in\cB,\\[12pt]
\displaystyle
x_j-\rho_j
\left\{ c+\overline m + \sum_{k\in\cB} \E_{\bar W} \left[ \frac{\rho_k\bar W\,\pi_{y_k-1}(\rho_k\bar W)} {\overline F_{y_k}(\rho_k\bar W)} \right] \right\},
& j\in\cC.
\end{cases}
\end{equation*}
In addition, for \(\eta=\log\lambda\),
\begin{equation*}
\frac{\partial\log L^{\mathrm P}}{\partial\eta}
=
\sum_{j=1}^Jx_j-\lambda+\overline m.
\end{equation*}
\end{theorem}
Both settings therefore require only one-dimensional expectations.  The next subsection describes their deterministic evaluation and the resulting optimization problem.

\subsection{Computation of the Gradient}
\label{subsec:score-computation}
The remaining computation has four components: approximating these expectations using numerical integration, evaluating the Poisson tail probability \(\overline F\), evaluating the functions \(\Phi_s\) and \(\psi_s\), and how to implement the gradient-based likelihood maximization.
Each component can be handled using standard numerical methods, while the specific structure of our likelihood allows these calculations to be implemented efficiently.  
We briefly describe the numerical methods used for the first three components below. 
For the last component, we use the L-BFGS-B method~\citep{liu1989limited} with a standard line search for step-size selection.

\paragraph{Generalized Gauss--Laguerre method for numerical integration.}
The expectation in the likelihood can be written generally as $\int_u u^\alpha e^{-u} f(u) du$.
This form is naturally suited to generalized Gauss--Laguerre quadrature \citep[Section~3.5(v)]{nist_dlmf_gauss_quadrature}.
For a chosen $Q$, the method uses $Q$ nodes and corresponding weights to approximate the integral by a finite weighted sum.
The quadrature approximation introduces numerical integration error, and increasing $Q$ provides a finer numerical approximation at a higher computational cost.  
Appendix~\ref{app:laguerre-evaluation} gives the implementation of the generalized Gauss--Laguerre method for our problem.

\paragraph{Evaluation of the Poisson tails.}
The Poisson tail \(\overline F_y(\mu)\) equals the regularized lower incomplete gamma function. We evaluate it using \texttt{torch.special.gammainc}~\citep{pytorch_gammainc}.
We jointly evaluate the Poisson tail probability \(\overline F_{y_{j,n}}(\mu)\) and the derivative \(\partial\log\overline F_{y_{j,n}}(\mu)/\partial\log\mu\) appearing in the gradient formulas, where $\mu=\rho_{j,n}u$ and $\rho_{j,n}=\theta_j/(1+\sum_{k\in\cC_n}\theta_k)$ for $j\in\cA_n$. Appendix~\ref{app:tail-evaluation} provides the numerical details for evaluating these quantities.

\paragraph{Evaluation of \(\Phi_s\) and \(\psi_s\).}
\label{subsubsec:phi-evaluation}
In the Poisson-arrival setting, both quantities involve an infinite sum. (See their definitions in \eqref{eq:phi_s-def} and after \eqref{eq:pdf-w-poisson}.)
The ratios of successive series terms eventually fall below one and continue to decrease, allowing the omitted tails to be bounded by geometric sums, then  it is sufficient to approximate them with the sum of a truncated sequence. Appendix~\ref{app:phi-evaluation} documents the details and provides an error analysis.

\paragraph{Computational complexity.}  
For observation \(n\), computing the total sales \(s_n=\sum_{j=1}^J x_{j,n}\), the sales \(c_n=\sum_{j\in\cC_n}x_{j,n}\) of the products that remain in stock, the common denominator \(V_n=1+\sum_{j\in\cC_n}\theta_j\), and the remaining closed-form terms in the gradient requires \(O(J)\) time.
The remaining work is performed over the \(Q\) nodes in the numerical integration.  
At each node, product-specific calculations are needed only for the sold-out products \(j\in\cB_n\), while all gradient components share the same normalized weights for approximating expectations under $W$ or $\bar W$.  Under Poisson arrivals, evaluating \(\Phi\) and \(\psi\) adds one scalar calculation per node.  Counting each scalar-function evaluation as one operation, the gradient evaluation requires
\begin{equation}
O\!\left(
NJ
+
Q\sum_{n=1}^N\bigl(1+|\cB_n|\bigr)
\right)
\le
O(NQJ).
\label{eq:quadrature-complexity}
\end{equation}
For a fixed \(Q\), this number of operations is linear in both \(N\) and \(J\). The arithmetic cost of each scalar evaluation also depends on the number of series terms required.
The computation is lighter when fewer products sell out, because fewer Poisson tails need to be evaluated.
For \(N=500\), \(J=100\), and \(Q=128\), the average time per gradient iteration is approximately \(0.038\) seconds under known arrivals and \(0.080\) seconds under Poisson arrivals on one NVIDIA RTX~5090 GPU.

\section{Theoretical Analysis}
\label{sec:estimation-theory}

The gradient expressions in Theorems~\ref{thm:known-score} and~\ref{thm:poisson-score} make gradient-based methods computationally efficient in each iteration. We next study the curvature of the log-likelihood and the identification of the model parameters, which provide conditions for the convergence of the gradient-based methods.

\subsection{Known Arrivals}
\label{subsec:known-theory}

We first consider the known-arrival setting.
Theorem~\ref{thm:known-gamma} factorizes the likelihood into two terms.
Because we maximize the log-likelihood, it suffices to establish the concavity of
$\log C^{\mathrm K}$ and $\log R^{\mathrm K}$. 
We focus on these two terms \emph{given the observation of a single selling period}.
We can show that the term $\log C^{\mathrm K}$ is (weakly) concave, while $\log R^{\mathrm K}$ is strictly concave when at least one product sells out. 
On the other hand, if no product sells out, then $\log C^{\mathrm K}$ is strictly concave. These  properties yield the following result.

\begin{theorem}[Global concavity in the known-arrival setting]
\label{thm:known-concavity}

For any known-arrival observation $(\by,\bx,T)$ with $T\ge1$, the sample log-likelihood $\log L^{\mathrm K}(\bv)$ is jointly strictly concave in the log attractions of products with positive initial inventory, provided at least one such product exists.
\end{theorem}
For a single-period observation, the log-likelihood is strictly increasing in $v_j$ whenever product $j$ sells out. Consequently, a finite maximizer does not exist if at least one product sells out in that period.
After aggregating the data from multiple periods, the sample log-likelihood is strictly concave in the parameters of products initially stocked in at least one period with $T_n\ge1$. Therefore, any finite stationary point is the unique global maximizer over these parameters.

Because identifiability concerns whether different parameter vectors generate different terminal-sales distributions, 
it is a weaker condition than global concavity.
Indeed, if the log-likelihood is strictly concave, then it leads to identifiability.
This observation leads to the following result.

\begin{corollary}[Identifiability in the known-arrival setting]
\label{cor:known-global-identification}
For any fixed $(\by,T)$ with $T\ge1$, the parameters $\theta_j$ of products with positive initial inventory are identifiable from the distribution of terminal sales $\bm X$.
\end{corollary}

\subsection{Poisson Arrivals}
\label{subsec:poisson-identification}

Unlike the known-arrival setting, the Poisson-arrival log-likelihood need not be globally concave in $(\eta,\bv)$ for a single observation. Lemma~\ref{lem:identified-nonconcavity} in Appendix~\ref{app:proofs} demonstrates this using a zero-sale observation, showing that nonconcavity can arise even without an observed stockout.

\subsubsection{Identifiability}
Despite this nonconcavity, the parameters are still identifiable from the terminal-sales distribution.
Under Poisson arrivals, the main identification challenge is separating the unknown arrival rate $\lambda$ from product attractions using only terminal sales. Before any stockout, product $j$ is purchased at rate $\lambda\theta_j/(1+\sum_{k=1}^J\theta_k)$. Hence, the pre-stockout sales dynamics depend on $\lambda$ and $\btheta$ only through these product-specific purchase rates, making the two components indistinguishable before stockouts occur.
For instance, when there is one product ($J=1$), we have the terminal sales $X_1\overset d=\min\{y_1,Z\}$, where
$Z\sim\Poi\!\left(\tfrac{\lambda\theta_1}{1+\theta_1}\right)$.
Thus, the terminal-sales distribution identifies only $\lambda\theta_1/(1+\theta_1)$ as a whole.

For $J\ge2$, stockout-induced substitution provides additional information. To separate the product purchase rates from the remaining parameter, we reparametrize:
\begin{equation}
\alpha_j=\frac{\lambda\theta_j}{1+\sum_{k=1}^J\theta_k},\qquad \alpha_+=\sum_{j=1}^J\alpha_j,\qquad \delta=\frac{\lambda}{1+\sum_{k=1}^J\theta_k}.
\label{eq:alpha-delta-parameterization}
\end{equation}
Here $\alpha_j$ is the purchase rate of product $j$ before any stockout, and $\delta$ is the corresponding no-purchase rate. The transformation is one-to-one: $\lambda=\alpha_++\delta$ and $\theta_j=\alpha_j/\delta$. Hence identifying $(\lambda,\btheta)$ is equivalent to identifying $(\bm\alpha,\delta)$. 
The zero-sale probability identifies the total pre-stockout purchase rate $\alpha_+$:
\begin{equation}
\PP(\bm X=\bzero\mid\by)=e^{-\alpha_+}.
\label{eq:poisson-zero-sale-id}
\end{equation}
For any product $j$ with $y_j\ge2$, selling exactly one unit of product $j$ and nothing else does not cause a stockout, so
\begin{equation}
\PP(X_j=1,\ X_k=0\ \text{for }k\neq j\mid\by)=\alpha_je^{-\alpha_+}.
\label{eq:alpha-intensities}
\end{equation}
Thus, $\alpha_j$ is identified for every product with at least two units of initial inventory.

If $y_j\ge2$ for every product, \eqref{eq:alpha-intensities} identifies the entire vector $\bm\alpha$. With $\bm\alpha$ fixed, Lemma~\ref{lem:scale-monotonicity} in Appendix~\ref{app:proofs} shows that expected total sales are strictly decreasing in $\delta$. Since this expectation is determined by the terminal-sales distribution, $\delta$ is also identified.

When $y_j=1$,
in Appendix~\ref{app:proofs} we show that, the probability of the event that only product $j$ is sold is strictly increasing in both $\alpha_j$ and $\delta$, holding $\alpha_+$ fixed. 
If two parameter vectors generate the same terminal-sales distribution and $\delta'>\delta$, equality of their one-sale probabilities requires $\alpha_j'<\alpha_j$ for every product with $y_j=1$, while \eqref{eq:alpha-intensities} fixes $\alpha_j$ for every product with $y_j\ge2$. This contradicts the equality of $\alpha_+$ implied by \eqref{eq:poisson-zero-sale-id}. Reversing the two parameter vectors rules out $\delta'<\delta$, so $\delta$ is identified, and the same strict monotonicity identifies the remaining components of $\bm\alpha$.

\begin{theorem}[Identifiability under Poisson arrivals]
\label{thm:poisson-global-identification}
Suppose $J\ge2$ and fix any $\by\in\Z_{++}^J$. Then $(\lambda,\bv)\in\R_{++}\times\R^J$ is identifiable from the conditional terminal-sales distribution.
\end{theorem}

Theorem~\ref{thm:poisson-global-identification} establishes identifiability without observed stockout times, variation in initial inventories, or the total customer arrivals. This differs from identifiability based on observed assortment variation. \citet{abdallah2021demand} study MNL estimation from period-level sales with observed assortments and external market-share information. Under a strong-connectivity condition, they establish identifiability of preference parameters but not of period-specific arrival rates. Here, the assortment changes induced by unobserved stockouts help identifying both the common Poisson arrival rate and all product attractions.

\subsubsection{Local Concavity}
\label{subsubsec:poisson-local-concavity}
As noted at the beginning of Section~\ref{subsec:poisson-identification}, the log-likelihood of a single period need not be concave under Poisson arrivals.
Unlike in the known-arrival setting, the sample log-likelihood therefore comes with no global concavity guarantee, and a stationary point reached by a gradient method need not be a maximizer.
We show that this difficulty disappears near the true value once enough selling periods are observed: there, the average log-likelihood is strongly concave with high probability.
Recall that a twice-differentiable function is called \emph{$\gamma$-strongly concave} on a convex set if its second derivative along every unit direction is at most $-\gamma$ on that set.

Strong concavity of the average log-likelihood on a neighborhood has three consequences for gradient-based methods.
First, the neighborhood contains at most one stationary point, which then maximizes the log-likelihood over the neighborhood.
A gradient method that converges to a point in the neighborhood has therefore found this maximizer rather than, say, a saddle point.
Second, the gradient bounds the distance to the maximizer over the neighborhood: at any point of the neighborhood, this distance is at most the norm of the gradient of the average log-likelihood divided by $\gamma$.
Within the neighborhood, a small gradient thus guarantees that the numerical solution is close to the maximizer, and numerical errors in the gradient, such as those from the quadrature in Section~\ref{subsec:score-computation}, lead to proportionally small errors in the solution.
Third, because the second derivatives are also bounded on the neighborhood (see Appendix~\ref{app:proofs}), gradient ascent with a backtracking line search converges linearly once its iterates are close enough to a stationary point in the neighborhood \citep[Section~9.3]{boyd2004convex}.

\begin{theorem}[Local strong concavity under Poisson arrivals]
\label{thm:poisson-local-concavity}
Let $\bbeta=(\eta,\bv)$ with $\eta=\log\lambda$, and let $\bbeta_0$ be its true value.
Suppose that $J\ge2$, that the initial inventories are nonrandom and take values in a finite set, that the terminal sales are independent across selling periods, and that the fraction of periods in which every product is initially stocked is bounded away from zero as $N\to\infty$.
Then there exist constants $r,\gamma>0$, independent of $N$, such that, with probability approaching $1$ as $N\to\infty$, the average log-likelihood $N^{-1}\sum_{n=1}^N\log L^{\mathrm P}(\lambda,\bv\mid\by_n,\bx_n)$ is $\gamma$-strongly concave in $\bbeta$ on the ball $\{\bbeta:\|\bbeta-\bbeta_0\|_2\le r\}$.
\end{theorem}

Because $r$ and $\gamma$ do not depend on $N$, the neighborhood and its curvature do not shrink as more selling periods are observed, so the guarantees above do not weaken as data accumulate.
The proof in Appendix~\ref{app:proofs} builds on the argument for Theorem~\ref{thm:poisson-global-identification}.
For a period in which every product is initially stocked, we can show the Fisher information is nonsingular at $\bbeta_0$, so its expected log-likelihood is strictly concave (Lemma~\ref{lem:poisson-population-concavity}).
Because the periods in which every product is initially stocked make up a positive fraction of the sample, the average log-likelihood inherits this curvature at $\bbeta_0$ as $N$ grows, and the concavity extends to a neighborhood of $\bbeta_0$.

\section{Numerical Study}
\label{sec:numerical-experiment}

In this section, we evaluate the accuracy and computational cost of our method and other benchmark methods, when available, in a large extensive numerical study.

\subsection{Experimental Setup}
\label{subsec:numerical_design}

\paragraph{Data generation.}
Each replication generates one dataset of $N$ selling periods for $J$ products.
Within each replication, we draw product attractions while fixing the no-purchase probability under the initial assortment at \(p_0=0.2\):
\begin{equation*}
\theta_j = \frac{1-p_0}{p_0} \frac{\exp(z_j)}{\sum_{\ell=1}^J\exp(z_{\ell})}, \qquad z_j\stackrel{\mathrm{iid}}{\sim}\mathcal N(0,1).
\label{eq:theta_softmax_normal}
\end{equation*}
Thus \(\sum_j\theta_j=4\). 
In each period, we generate the total number of customer arrivals.
We use a parameter $\kappa$ to control the number of customers relative to the number of products: we let \(T_n\equiv\lceil\kappa J\rceil\) in the known-arrival setting and
\(\widetilde T_n\sim\Poi(\lambda)\) with $\lambda=\kappa J$ in the Poisson-arrival setting.
To generate the initial inventory $y_{j,n}$, we use
\begin{equation}
    U_{j,n}\stackrel{\mathrm{iid}}{\sim}\operatorname{Uniform}(0,2),
\qquad
y_{j,n}
=
\max\left\{
1,
\left\lceil
c_s\kappa J
\frac{\theta_j}{1+\sum_{l=1}^J\theta_l}
U_{j,n}
\right\rceil
\right\}.
\label{eq:inventory_rule}
\end{equation}

The term \(\kappa J \theta_j (1+\sum_{l=1}^J \theta_l)^{-1}\) is the baseline expected demand for product \(j\) when all products remain available. The multipliers \(U_{j,n}\) create variation in stocking levels across selling periods, while \(c_s\) controls the inventory availability.

For each dataset, we measure the realized stockout rate by
$\frac{1}{NJ}\sum_{n=1}^N|\cB_n|$, the fraction of product--period observations ending in a stockout.
For a target stockout rate $\zeta$, we calibrate $c_s$ using independent pilot simulations and then generate the sample with the calibrated multiplier.
The details are provided in Appendix~\ref{app:stockout-calibration}.
The hyperparameters \((J,N,\kappa,\zeta)\) control data generation; each combination defines an experimental setting.

\paragraph{Experimental settings and benchmarks.}
The main experiment uses \(J\in\{10,50,100\}\), \(\kappa\in\{5,10\}\), \(\zeta\in\{0.25,0.50\}\), and \(N\in\{200,500\}\), with 10 replications per experimental setting. Within each replication, all methods use the same generated dataset.

Throughout the numerical study, we refer to our proposed likelihood-based estimator as \EMNL\unskip.
In the known-arrival setting, we use the following benchmark methods:
(1) \NMNL ignores stockout-induced assortment changes and fits the MNL model using the initial assortment throughout the selling period.
(2) \SAA~\citep{shao2020joint} approximates the likelihood by sampling latent stockout orders.
(3) For small experimental settings such as \(J=10\), we report \DAMC~\citep{wan2018demand}, a computationally intensive Bayesian method that explicitly samples the latent stockout trajectory.
Under Poisson arrivals, \SAA provides the closest likelihood-based benchmark and jointly estimates the arrival rate and choice parameters by sampling latent stockout histories. We also compare with \NMNL when it is supplied with the true arrival rate \(\lambda\), because it is otherwise unidentifiable, and with \EMNL under observed arrival counts to quantify the value of arrival information.

\paragraph{Computation and metrics.}
We impose a 600-second time limit per replication on all methods except \DAMC. Due to its computational cost, \DAMC is evaluated only for $J=10$ and is run without a time limit, as some of the larger instances require more than 600 seconds.
If a method runs over time, we report ``--'' in the table.
The experiments are run on one NVIDIA RTX~5090 GPU.

For each method, let $\widehat\theta_j$ denote its estimate of $\theta_j$, and write $\widehat{\btheta}=(\widehat\theta_1,\ldots,\widehat\theta_J)$. We report the root mean square of productwise relative errors, $\mathrm{rRMSE}_\theta$, and the relative $\ell_2$ error, $\mathrm{RelErr}_\theta$:
\begin{equation*}
\mathrm{rRMSE}_\theta
=
\left\{
\frac1J\sum_{j=1}^J
\left(
\frac{\widehat\theta_j-\theta_j}{\theta_j}
\right)^2
\right\}^{1/2},
\qquad
\mathrm{RelErr}_\theta
=
\frac{\|\widehat{\btheta}-\btheta\|_2}{\|\btheta\|_2}.
\label{eq:theta_metrics}
\end{equation*}
The first metric weights products equally when averaging squared relative errors; the second measures error relative to the norm of the attraction vector. Under Poisson arrivals, writing $\widehat\lambda$ for the estimated arrival rate, we additionally report the relative absolute error
\(\mathrm{RelErr}_\lambda=|\widehat\lambda-\lambda|/\lambda\).
Errors are computed within each replication and then averaged across replications.

\subsection{Known Arrivals}
\label{subsec:results_known}

\begin{table}[!t]
\centering
\scriptsize
\setlength{\tabcolsep}{1.5pt}
\resizebox{\textwidth}{!}{%
\begin{tabular}{rrr rrr rrr rrr rrr}
\toprule
\multicolumn{15}{r}{\textit{$\zeta=0.25$}}\\[-1pt]
$J$ & $\kappa$ & $N$
& \multicolumn{3}{c}{\EMNL}
& \multicolumn{3}{c}{\SAA}
& \multicolumn{3}{c}{\DAMC}
& \multicolumn{3}{c}{\NMNL}\\
& & &
$\mathrm{rRMSE}_\theta$ & $\mathrm{RelErr}_\theta$ & \makecell{Estimation\\time (s)}
& $\mathrm{rRMSE}_\theta$ & $\mathrm{RelErr}_\theta$ & \makecell{Estimation\\time (s)}
& $\mathrm{rRMSE}_\theta$ & $\mathrm{RelErr}_\theta$ & \makecell{Estimation\\time (s)}
& $\mathrm{rRMSE}_\theta$ & $\mathrm{RelErr}_\theta$ & \makecell{Estimation\\time (s)}\\
\midrule
10 & 5 & 200 & \textbf{0.0510} & \textbf{0.0335} & 0.75 & 0.0517 & 0.0336 & 0.95 & 0.0522 & 0.0338 & 158.5 & 0.1443 & 0.1693 & 0.30 \\
10 & 5 & 500 & \textbf{0.0309} & \textbf{0.0191} & 0.91 & 0.0314 & 0.0202 & 1.36 & 0.0313 & 0.0192 & 386.1 & 0.1402 & 0.1605 & 0.32 \\
10 & 10 & 200 & \textbf{0.0381} & \textbf{0.0264} & 0.77 & 0.0383 & 0.0269 & 0.93 & 0.0402 & 0.0267 & 152.0 & 0.1381 & 0.1513 & 0.32 \\
10 & 10 & 500 & 0.0269 & \textbf{0.0200} & 0.98 & 0.0276 & 0.0203 & 1.38 & \textbf{0.0266} & 0.0202 & 382.2 & 0.1377 & 0.1595 & 0.31 \\
50 & 5 & 200 & \textbf{0.0600} & \textbf{0.0251} & 1.00 & 0.0601 & 0.0256 & 1.58 & \multicolumn{3}{c}{--} & 0.1341 & 0.1389 & 0.34 \\
50 & 5 & 500 & \textbf{0.0397} & \textbf{0.0165} & 1.39 & 0.0402 & 0.0185 & 2.02 & \multicolumn{3}{c}{--} & 0.1318 & 0.1372 & 0.35 \\
50 & 10 & 200 & \textbf{0.0417} & \textbf{0.0192} & 1.03 & 0.0429 & 0.0232 & 1.47 & \multicolumn{3}{c}{--} & 0.1334 & 0.1430 & 0.33 \\
50 & 10 & 500 & \textbf{0.0270} & \textbf{0.0120} & 1.40 & 0.0278 & 0.0136 & 2.01 & \multicolumn{3}{c}{--} & 0.1279 & 0.1357 & 0.35 \\
100 & 5 & 200 & \textbf{0.0579} & \textbf{0.0252} & 1.10 & 0.0586 & 0.0276 & 1.88 & \multicolumn{3}{c}{--} & 0.1379 & 0.1345 & 0.34 \\
100 & 5 & 500 & \textbf{0.0360} & \textbf{0.0160} & 1.60 & 0.0364 & 0.0170 & 2.40 & \multicolumn{3}{c}{--} & 0.1311 & 0.1310 & 0.33 \\
100 & 10 & 200 & \textbf{0.0404} & \textbf{0.0165} & 1.16 & 0.0420 & 0.0199 & 2.18 & \multicolumn{3}{c}{--} & 0.1286 & 0.1280 & 0.37 \\
100 & 10 & 500 & \textbf{0.0251} & \textbf{0.0109} & 1.70 & 0.0262 & 0.0145 & 2.70 & \multicolumn{3}{c}{--} & 0.1251 & 0.1283 & 0.38 \\
\midrule
\multicolumn{15}{r}{\textit{$\zeta=0.50$}}\\[-1pt]
$J$ & $\kappa$ & $N$
& \multicolumn{3}{c}{\EMNL}
& \multicolumn{3}{c}{\SAA}
& \multicolumn{3}{c}{\DAMC}
& \multicolumn{3}{c}{\NMNL}\\
& & &
$\mathrm{rRMSE}_\theta$ & $\mathrm{RelErr}_\theta$ & \makecell{Estimation\\time (s)}
& $\mathrm{rRMSE}_\theta$ & $\mathrm{RelErr}_\theta$ & \makecell{Estimation\\time (s)}
& $\mathrm{rRMSE}_\theta$ & $\mathrm{RelErr}_\theta$ & \makecell{Estimation\\time (s)}
& $\mathrm{rRMSE}_\theta$ & $\mathrm{RelErr}_\theta$ & \makecell{Estimation\\time (s)}\\
\midrule
10 & 5 & 200 & \textbf{0.0536} & \textbf{0.0373} & 0.79 & 0.0552 & 0.0415 & 1.02 & 0.0539 & 0.0381 & 241.6 & 0.2834 & 0.3082 & 0.32 \\
10 & 5 & 500 & \textbf{0.0352} & 0.0221 & 0.95 & 0.0369 & 0.0266 & 1.48 & 0.0360 & \textbf{0.0209} & 586.8 & 0.2766 & 0.3001 & 0.31 \\
10 & 10 & 200 & 0.0402 & 0.0231 & 0.80 & 0.0441 & 0.0346 & 1.00 & \textbf{0.0398} & \textbf{0.0230} & 247.0 & 0.2979 & 0.3270 & 0.31 \\
10 & 10 & 500 & 0.0261 & 0.0152 & 1.00 & 0.0283 & 0.0293 & 1.55 & \textbf{0.0257} & \textbf{0.0149} & 624.2 & 0.2937 & 0.3257 & 0.31 \\
50 & 5 & 200 & \textbf{0.0564} & \textbf{0.0301} & 1.02 & 0.0597 & 0.0345 & 1.62 & \multicolumn{3}{c}{--} & 0.2723 & 0.2850 & 0.35 \\
50 & 5 & 500 & \textbf{0.0361} & \textbf{0.0173} & 1.41 & 0.0374 & 0.0269 & 2.13 & \multicolumn{3}{c}{--} & 0.2735 & 0.2908 & 0.34 \\
50 & 10 & 200 & \textbf{0.0461} & \textbf{0.0210} & 1.15 & 0.0518 & 0.0327 & 1.77 & \multicolumn{3}{c}{--} & 0.2648 & 0.2762 & 0.33 \\
50 & 10 & 500 & \textbf{0.0285} & \textbf{0.0153} & 1.60 & 0.0350 & 0.0296 & 2.39 & \multicolumn{3}{c}{--} & 0.2653 & 0.2818 & 0.32 \\
100 & 5 & 200 & \textbf{0.0600} & \textbf{0.0268} & 1.25 & 0.0636 & 0.0362 & 2.56 & \multicolumn{3}{c}{--} & 0.2587 & 0.2734 & 0.35 \\
100 & 5 & 500 & \textbf{0.0389} & \textbf{0.0169} & 1.85 & 0.0420 & 0.0266 & 3.04 & \multicolumn{3}{c}{--} & 0.2563 & 0.2707 & 0.36 \\
100 & 10 & 200 & \textbf{0.0443} & \textbf{0.0198} & 1.24 & 0.0534 & 0.0366 & 2.45 & \multicolumn{3}{c}{--} & 0.2606 & 0.2709 & 0.37 \\
100 & 10 & 500 & \textbf{0.0288} & \textbf{0.0125} & 1.87 & 0.0362 & 0.0263 & 3.18 & \multicolumn{3}{c}{--} & 0.2596 & 0.2705 & 0.40 \\
\bottomrule
\end{tabular}%
}
\caption{Mean estimation error and estimation time under known arrivals.}
\smallskip
\begin{minipage}{\textwidth}
\footnotesize
\textit{Notes.} Boldface denotes the smallest mean estimation error in each row; \DAMC is reported only for \(J=10\) because of its estimation time.
\end{minipage}
\label{tab:uniform_main_known}
\end{table}

Table~\ref{tab:uniform_main_known} reports estimation errors and estimation times under known arrivals.
Overall, our method \EMNL achieves the best accuracy in the most settings and is computationally efficient.
We summarize the patterns below.
First, the estimation error of \NMNL comes from ignoring the stockout substitution.
It doesn't diminish as the data size increases.
Second, when the Bayesian method \DAMC can compute a result within the time limit in the most settings, it is comparable to our method \EMNL. However, \DAMC  is not practical computationally even for instances of mild scales.
Third, in terms of scalability, \SAA is comparable to our method, with around 55\% more estimation time on average.
However, its estimation error is larger than ours especially for large $J$.
This is because \SAA approximates the likelihood by sampling latent stockout vectors, whereas \EMNL evaluates the exact one-dimensional representation by deterministic quadrature.
Fourth, the estimation error of \EMNL changes little between the two target stockout rates ($\zeta=0.25$ and $\zeta=0.50$).

\subsection{Poisson Arrivals}
\label{subsec:results_poisson}
Table~\ref{tab:uniform_main_poisson} reports estimation errors and estimation times under Poisson arrivals.
For Poisson arrivals, \SAA is the only benchmark that is applicable to this setting.
To enrich the comparison, we include \NMNL\unskip, which treats $\lambda$ as known and ignores the stockout-induced assortment changes.
The true value of $\lambda$ needs to be provided for \NMNL to be identifiable.
We also fit \EMNL after revealing the realized counts $\{\widetilde T_n\}_{n=1}^N$ from the same Poisson-generated data. Conditional on these counts, the fit uses the known-arrival likelihood. This comparison measures the effect of observing arrival counts on estimation error.

\begin{table}[htbp]
\centering
\scriptsize
\setlength{\tabcolsep}{3pt}
\resizebox{\textwidth}{!}{%
\begin{tabular}{rrr rrrr rrrr rr}
\toprule
\multicolumn{13}{l}{\textit{Panel A: Target stockout rate $\zeta=0.25$}}\\[-1pt]
$J$ & $\kappa$ & $N$ & \multicolumn{4}{c}{\EMNL} & \multicolumn{4}{c}{\SAA} & \makecell{\EMNL \\known arrivals} & \makecell{\NMNL \\true $\lambda$} \\
 & & & $\mathrm{rRMSE}_\theta$ & $\mathrm{RelErr}_\theta$ & $\mathrm{RelErr}_\lambda$ & \makecell{Estimation\\time (s)}
 & $\mathrm{rRMSE}_\theta$ & $\mathrm{RelErr}_\theta$ & $\mathrm{RelErr}_\lambda$ & \makecell{Estimation\\time (s)} & $\mathrm{rRMSE}_\theta$ & $\mathrm{rRMSE}_\theta$ \\
\midrule
10 & 5 & 200 & 0.3893 & \textbf{0.3808} & \textbf{0.0605} & 1.55 & \textbf{0.3883} & 0.3913 & 0.1743 & 14.95 & 0.0588 & 0.1476 \\
10 & 5 & 500 & \textbf{0.1862} & \textbf{0.1826} & \textbf{0.0350} & 1.62 & 0.4038 & 0.4057 & 0.1976 & 39.46 & 0.0342 & 0.1427 \\
10 & 10 & 200 & \textbf{0.2436} & \textbf{0.2435} & \textbf{0.0623} & 1.60 & 0.2808 & 0.2801 & 0.0930 & 12.78 & 0.0455 & 0.1473 \\
10 & 10 & 500 & \textbf{0.1038} & \textbf{0.1044} & \textbf{0.0230} & 1.62 & 0.2366 & 0.2350 & 0.0815 & 36.25 & 0.0231 & 0.1435 \\
50 & 5 & 200 & 0.4433 & 0.4399 & 0.0785 & 4.15 & \textbf{0.2593} & \textbf{0.2541} & \textbf{0.0740} & 30.35 & 0.0547 & 0.1332 \\
50 & 5 & 500 & \textbf{0.2774} & \textbf{0.2705} & \textbf{0.0497} & 4.08 & 0.3596 & 0.3593 & 0.1425 & 88.74 & 0.0347 & 0.1342 \\
50 & 10 & 200 & \textbf{0.2195} & \textbf{0.2110} & \textbf{0.0463} & 5.10 & 0.2391 & 0.2319 & 0.0827 & 34.22 & 0.0416 & 0.1366 \\
50 & 10 & 500 & \textbf{0.0959} & \textbf{0.0880} & \textbf{0.0223} & 4.92 & 0.1780 & 0.1752 & 0.0546 & 92.28 & 0.0248 & 0.1305 \\
100 & 5 & 200 & 0.3111 & 0.3015 & 0.0658 & 7.22 & \textbf{0.2003} & \textbf{0.1958} & \textbf{0.0633} & 41.19 & 0.0606 & 0.1315 \\
100 & 5 & 500 & 0.2577 & 0.2569 & \textbf{0.0565} & 8.12 & \textbf{0.2210} & \textbf{0.2198} & 0.0709 & 101.21 & 0.0376 & 0.1275 \\
100 & 10 & 200 & \textbf{0.1966} & \textbf{0.1920} & \textbf{0.0505} & 10.46 & 0.2851 & 0.2817 & 0.1096 & 43.68 & 0.0438 & 0.1340 \\
100 & 10 & 500 & \textbf{0.1351} & \textbf{0.1320} & \textbf{0.0313} & 11.06 & 0.3327 & 0.3316 & 0.1232 & 131.53 & 0.0288 & 0.1311 \\
\midrule
\multicolumn{13}{l}{\textit{Panel B: Target stockout rate $\zeta=0.50$}}\\[-1pt]
$J$ & $\kappa$ & $N$ & \multicolumn{4}{c}{\EMNL} & \multicolumn{4}{c}{\SAA} & \makecell{\EMNL\\known arrivals} & \makecell{\NMNL\\true $\lambda$} \\
 & & & $\mathrm{rRMSE}_\theta$ & $\mathrm{RelErr}_\theta$ & $\mathrm{RelErr}_\lambda$ & \makecell{Estimation\\time (s)}
 & $\mathrm{rRMSE}_\theta$ & $\mathrm{RelErr}_\theta$ & $\mathrm{RelErr}_\lambda$ & \makecell{Estimation\\time (s)} & $\mathrm{rRMSE}_\theta$ & $\mathrm{rRMSE}_\theta$ \\
\midrule
10 & 5 & 200 & \textbf{0.1880} & \textbf{0.1795} & \textbf{0.0560} & 1.49 & 0.3637 & 0.3714 & 0.1824 & 15.79 & 0.0614 & 0.3134 \\
10 & 5 & 500 & \textbf{0.1668} & \textbf{0.1596} & \textbf{0.0438} & 1.49 & 0.4220 & 0.4308 & 0.2186 & 33.92 & 0.0366 & 0.3086 \\
10 & 10 & 200 & \textbf{0.1372} & \textbf{0.1297} & \textbf{0.0290} & 1.92 & 0.2506 & 0.2463 & 0.1032 & 19.96 & 0.0428 & 0.3059 \\
10 & 10 & 500 & \textbf{0.1184} & \textbf{0.1110} & \textbf{0.0259} & 1.94 & 0.2866 & 0.2884 & 0.1168 & 38.34 & 0.0283 & 0.3106 \\
50 & 5 & 200 & \textbf{0.1568} & \textbf{0.1447} & \textbf{0.0435} & 4.72 & 0.3970 & 0.3881 & 0.2403 & 31.04 & 0.0683 & 0.2686 \\
50 & 5 & 500 & \textbf{0.0994} & \textbf{0.0873} & \textbf{0.0230} & 5.30 & 0.5314 & 0.5412 & 0.4141 & 81.11 & 0.0409 & 0.2645 \\
50 & 10 & 200 & \textbf{0.1125} & \textbf{0.0999} & \textbf{0.0240} & 6.31 & 0.3241 & 0.3293 & 0.1988 & 32.87 & 0.0480 & 0.2610 \\
50 & 10 & 500 & \textbf{0.0673} & \textbf{0.0597} & \textbf{0.0142} & 6.66 & 0.4957 & 0.5043 & 0.3291 & 106.07 & 0.0301 & 0.2663 \\
100 & 5 & 200 & \textbf{0.1810} & \textbf{0.1636} & \textbf{0.0330} & 9.53 & 0.4305 & 0.4324 & 0.3078 & 42.67 & 0.0669 & 0.2638 \\
100 & 5 & 500 & \textbf{0.0929} & \textbf{0.0782} & \textbf{0.0175} & 9.94 & 0.6583 & 0.6622 & 0.5944 & 113.81 & 0.0405 & 0.2582 \\
100 & 10 & 200 & \textbf{0.1198} & \textbf{0.1077} & \textbf{0.0298} & 10.91 & 0.4060 & 0.4124 & 0.2134 & 44.77 & 0.0431 & 0.2568 \\
100 & 10 & 500 & \textbf{0.0860} & \textbf{0.0787} & \textbf{0.0212} & 10.94 & 0.5686 & 0.5751 & 0.3791 & 132.87 & 0.0279 & 0.2557 \\
\bottomrule
\end{tabular}%
}
\caption{Poisson-arrival results averaged over ten replications.}
\smallskip
\begin{minipage}{\textwidth}
\footnotesize
\textit{Notes.} The last two columns use observed arrivals and the true arrival rate, respectively. We only highlight the error between \EMNL and \SAA.
\end{minipage}
\label{tab:uniform_main_poisson}
\end{table}

The advantage of \EMNL relative to \SAA is more pronounced under Poisson arrivals, for both estimation accuracy and computational efficiency.
In particular, the reduction of estimation time is more than 90\% and that of estimation error is more than 50\%.
When total arrivals are unobserved, \SAA must also sum over possible arrival counts while approximating the likelihood associated with each count.
\EMNL sums out the unobserved arrival count analytically and retains a one-dimensional likelihood representation.
Compared to \EMNL with known arrivals, the estimation error does increase significantly (more than 300\%). 
It highlights the challenge of not observing the total arrivals, and this challenge is fundamental.

\subsection{Large-Scale Experiments}
\label{subsec:results_large}

\begin{table}[!t]
\centering
\small
\setlength{\tabcolsep}{3pt}
\begin{tabular*}{\textwidth}{@{\extracolsep{\fill}}rr rrr rrr@{}}
\toprule
\multicolumn{8}{l}{\textit{Panel A: Known arrivals}}\\[2pt]
& &
\multicolumn{3}{c}{Target stockout rate $\zeta=0.25$} &
\multicolumn{3}{c}{Target stockout rate $\zeta=0.50$} \\
\cmidrule(lr){3-5}\cmidrule(lr){6-8}
$J$ & $N$
& $\mathrm{rRMSE}_\theta$
& $\mathrm{RelErr}_\theta$
& \makecell{Estimation\\time (s)}
& $\mathrm{rRMSE}_\theta$
& $\mathrm{RelErr}_\theta$
& \makecell{Estimation\\time (s)} \\
\midrule
500  & 500  & 0.0375 & 0.0144 & 2.96  & 0.0377 & 0.0166 & 4.18 \\
500  & 1000 & 0.0269 & 0.0102 & 5.05  & 0.0271 & 0.0118 & 7.20 \\
500  & 2000 & 0.0190 & 0.0075 & 9.10  & 0.0190 & 0.0084 & 13.59 \\
1000 & 500  & 0.0372 & 0.0152 & 8.24  & 0.0389 & 0.0160 & 10.51 \\
1000 & 1000 & 0.0263 & 0.0106 & 14.76 & 0.0277 & 0.0114 & 20.18 \\
1000 & 2000 & 0.0185 & 0.0073 & 27.92 & 0.0192 & 0.0079 & 38.48 \\
2000 & 500  & 0.0389 & 0.0137 & 13.88 & 0.0385 & 0.0155 & 17.55 \\
2000 & 1000 & 0.0273 & 0.0097 & 25.96 & 0.0272 & 0.0110 & 34.44 \\
2000 & 2000 & 0.0191 & 0.0069 & 48.77 & 0.0192 & 0.0078 & 66.90 \\
\bottomrule
\end{tabular*}
\par\medskip
\begin{tabular*}{\textwidth}{@{\extracolsep{\fill}}rr rrrr rrrr@{}}
\toprule
\multicolumn{10}{l}{\textit{Panel B: Poisson arrivals}}\\[2pt]
& &
\multicolumn{4}{c}{Target stockout rate $\zeta=0.25$} &
\multicolumn{4}{c}{Target stockout rate $\zeta=0.50$} \\
\cmidrule(lr){3-6}\cmidrule(lr){7-10}
$J$ & $N$
& $\mathrm{rRMSE}_\theta$
& $\mathrm{RelErr}_\theta$
& $\mathrm{RelErr}_\lambda$
& \makecell{Estimation\\time (s)}
& $\mathrm{rRMSE}_\theta$
& $\mathrm{RelErr}_\theta$
& $\mathrm{RelErr}_\lambda$
& \makecell{Estimation\\time (s)} \\
\midrule
500  & 500  & 0.1533 & 0.1415 & 0.0314 & 31.77  & 0.1465 & 0.1381 & 0.0305 & 47.49 \\
500  & 1000 & 0.1016 & 0.0920 & 0.0215 & 37.49  & 0.0962 & 0.0900 & 0.0203 & 60.23 \\
500  & 2000 & 0.0370 & 0.0284 & 0.0058 & 50.37  & 0.0549 & 0.0500 & 0.0116 & 93.30 \\
1000 & 500  & 0.0376 & 0.0145 & 0.0006 & 53.58  & 0.0442 & 0.0246 & 0.0047 & 81.83 \\
1000 & 1000 & 0.0273 & 0.0113 & 0.0007 & 63.80  & 0.0365 & 0.0246 & 0.0054 & 99.76 \\
1000 & 2000 & 0.0190 & 0.0076 & 0.0005 & 93.43  & 0.0279 & 0.0210 & 0.0049 & 207.74 \\
2000 & 500  & 0.0376 & 0.0148 & 0.0007 & 110.70 & 0.0441 & 0.0253 & 0.0048 & 149.96 \\
2000 & 1000 & 0.0267 & 0.0105 & 0.0006 & 142.24 & 0.0347 & 0.0227 & 0.0048 & 338.84 \\
2000 & 2000 & 0.0189 & 0.0075 & 0.0005 & 285.07 & 0.0277 & 0.0207 & 0.0048 & 565.57 \\
\bottomrule
\end{tabular*}
\caption{Mean estimation error and estimation time for \EMNL in the large-scale experiments.}
\label{tab:uniform_large}
\end{table}

Table~\ref{tab:uniform_large} reports results from the large-scale experiments using \(J,N\in\{500,1000,2000\}\), \(\kappa=5\), and \(\zeta\in\{0.25,0.50\}\), with 10 replications per experimental setting. We fit only \EMNL under both known and Poisson arrivals, as \SAA and \DAMC cannot be computed within the time limit.
Under known arrivals, estimation error is nearly unchanged as \(J\) increases at a fixed \(N\), whereas increasing \(N\) consistently reduces the error.  At \(J=N=2000\), mean rRMSE is approximately 0.019, with mean estimation times of 48.77 and 66.90 seconds for \(\zeta=0.25\) and \(0.50\), respectively.  Thus, the increase in parameter dimension does not materially degrade per-product accuracy in this design, and \EMNL remains computationally practical at the largest tested scale.

Under Poisson arrivals, mean rRMSE also decreases with \(N\).
Estimation time is higher at the larger target stockout rate \(\zeta=0.50\).  
But even at \(J=N=2000\), the average estimation time is 285.07 seconds for \(\zeta=0.25\) and 565.57 seconds for \(\zeta=0.50\). 
These results show that the one-dimensional likelihood expression makes \EMNL very robust and scalable.

\section{Concluding Remarks and Future Directions}
\label{sec:conclusion}

This paper develops maximum-likelihood estimation for MNL demand when only initial inventories and terminal product sales are observed, while customer choices, stockout times, and the assortments faced by individual customers remain latent.  
Because we can express the (log-)likelihood function and its gradient with a one-dimensional integral,
we are able to use standardized numerical recipes to maximize the likelihood efficiently, even for large-scale problems.

Several directions remain for future research.  
First, the continuous-time representation is made possible by the IIA property of the MNL choice model.
It is natural to ask whether other choice models, including observed covariates or customer heterogeneity, retain a tractable likelihood.
A one-dimensional representation is unlikely, but we may obtain a representation with a much lower dimension than the latent space of choice sequences.
Second, we have seen fruitful results in inventory management and demand estimation when applying the continuous-time view to dynamic stockout substitution, developed in \cite{zhou2025optimal}.
An exciting direction is to explore other operational levers that can be analyzed thanks to this viewpoint.

\bibliographystyle{informs2014}
\bibliography{reference}

\begin{thebibliography}{46}
\providecommand{\natexlab}[1]{#1}
\providecommand{\url}[1]{\texttt{#1}}
\providecommand{\urlprefix}{URL }

\bibitem[{Abdallah \protect\BIBand{} Vulcano(2021)}]{abdallah2021demand}
Abdallah T, Vulcano G (2021) Demand estimation under the multinomial logit
  model from sales transaction data. \emph{Manufacturing \& Service Operations
  Management} 23(5):1196--1216.

\bibitem[{Agrawal \protect\BIBand{} Smith(1996)}]{agrawal1996estimating}
Agrawal N, Smith SA (1996) Estimating negative binomial demand for retail
  inventory management with unobservable lost sales. \emph{Naval Research
  Logistics} 43(6):839--861.

\bibitem[{Akchen \protect\BIBand{} Caro(2026)}]{akchen2026size}
Akchen YC, Caro F (2026) On size substitution and its role in assortment and
  inventory planning. \emph{Manufacturing \& Service Operations Management}
  28(2):624--642.

\bibitem[{Anupindi et~al.(1998)Anupindi, Dada, \protect\BIBand{}
  Gupta}]{anupindi1998estimation}
Anupindi R, Dada M, Gupta S (1998) Estimation of consumer demand with stock-out
  based substitution: An application to vending machine products.
  \emph{Marketing Science} 17(4):406--423.

\bibitem[{Aouad \protect\BIBand{} Segev(2023)}]{aouad2023stability}
Aouad A, Segev D (2023) The stability of {MNL}-based demand under dynamic
  customer substitution and its algorithmic implications. \emph{Operations
  Research} 71(4):1216--1249.

\bibitem[{Boyd \protect\BIBand{} Vandenberghe(2004)}]{boyd2004convex}
Boyd S, Vandenberghe L (2004) \emph{Convex Optimization} (Cambridge, UK:
  Cambridge University Press).

\bibitem[{Chen \protect\BIBand{} Chao(2020)}]{chen2020dynamic}
Chen B, Chao X (2020) Dynamic inventory control with stockout substitution and
  demand learning. \emph{Management Science} 66(11):5108--5127.

\bibitem[{Chen et~al.(2026)Chen, Adulyasak, \protect\BIBand{}
  Cordeau}]{chen2026bayesian}
Chen X, Adulyasak Y, Cordeau JF (2026) Bayesian learning of demand and
  stockout-based substitution for risk-aware inventory optimization.
  \emph{Available at SSRN 7444520} .

\bibitem[{Cho et~al.(2024)Cho, Ferguson, Im, \protect\BIBand{}
  Pekg{\"u}n}]{cho2024robust}
Cho S, Ferguson M, Im J, Pekg{\"u}n P (2024) Robust demand estimation with
  customer choice-based models for sales transaction data. \emph{Production and
  Operations Management} 33(9):1801--1820.

\bibitem[{Cho et~al.(2023)Cho, Ferguson, Pekg{\"u}n, \protect\BIBand{}
  Vakhutinsky}]{cho2022estimating}
Cho S, Ferguson M, Pekg{\"u}n P, Vakhutinsky A (2023) Estimating personalized
  demand with unobserved no-purchases using a mixture model: An application in
  the hotel industry. \emph{Manufacturing \& Service Operations Management}
  25(4):1245--1262.

\bibitem[{Conlon \protect\BIBand{} Mortimer(2013)}]{conlon2013demand}
Conlon CT, Mortimer JH (2013) Demand estimation under incomplete product
  availability. \emph{American Economic Journal: Microeconomics} 5(4):1--30.

\bibitem[{Fisher \protect\BIBand{} Vaidyanathan(2014)}]{fisher2014demand}
Fisher M, Vaidyanathan R (2014) A demand estimation procedure for retail
  assortment optimization with results from implementations. \emph{Management
  Science} 60(10):2401--2415.

\bibitem[{Honhon et~al.(2010)Honhon, Gaur, \protect\BIBand{}
  Seshadri}]{honhon2010assortment}
Honhon D, Gaur V, Seshadri S (2010) Assortment planning and inventory decisions
  under stockout-based substitution. \emph{Operations Research}
  58(5):1364--1379.

\bibitem[{Hopp \protect\BIBand{} Xu(2008)}]{hopp2008static}
Hopp WJ, Xu X (2008) A static approximation for dynamic demand substitution
  with applications in a competitive market. \emph{Operations Research}
  56(3):630--645.

\bibitem[{Jain et~al.(2015)Jain, Rudi, \protect\BIBand{} Wang}]{jain2015demand}
Jain A, Rudi N, Wang T (2015) Demand estimation and ordering under censoring:
  Stock-out timing is (almost) all you need. \emph{Operations Research}
  63(1):134--150.

\bibitem[{Karabat{\i} et~al.(2009)Karabat{\i}, Tan, \protect\BIBand{}
  {\"O}zt{\"u}rk}]{karabati2009method}
Karabat{\i} S, Tan B, {\"O}zt{\"u}rk {\"O}C (2009) A method for estimating
  stock-out-based substitution rates by using point-of-sale data. \emph{IIE
  Transactions} 41(5):408--420.

\bibitem[{K{\"o}k \protect\BIBand{} Fisher(2007)}]{kok2007demand}
K{\"o}k AG, Fisher ML (2007) Demand estimation and assortment optimization
  under substitution: Methodology and application. \emph{Operations Research}
  55(6):1001--1021.

\bibitem[{Lee et~al.(2016)Lee, Gaur, Muthulingam, \protect\BIBand{}
  Swisher}]{lee2016stockout}
Lee J, Gaur V, Muthulingam S, Swisher GF (2016) Stockout-based substitution and
  inventory planning in textbook retailing. \emph{Manufacturing \& Service
  Operations Management} 18(1):104--121.

\bibitem[{Letham et~al.(2016)Letham, Letham, \protect\BIBand{}
  Rudin}]{letham2016bayesian}
Letham B, Letham LM, Rudin C (2016) Bayesian inference of arrival rate and
  substitution behavior from sales transaction data with stockouts.
  \emph{Proceedings of the 22nd ACM SIGKDD International Conference on
  Knowledge Discovery and Data Mining}, 1695--1704 (ACM).

\bibitem[{Li et~al.(2025)Li, Talluri, \protect\BIBand{}
  Tekin}]{li2025estimating}
Li A, Talluri K, Tekin M (2025) Estimating demand with unobserved no-purchases
  on revenue-managed data. \emph{Manufacturing \& Service Operations
  Management} 27(1):161--180.

\bibitem[{Liang et~al.(2021)Liang, Jasin, \protect\BIBand{}
  Uichanco}]{liang2021assortment}
Liang AJ, Jasin S, Uichanco J (2021) Assortment and inventory planning under
  dynamic substitution with {MNL} model: An {LP} approach and an asymptotically
  optimal policy. Technical report, University of Michigan, Ann Arbor, MI.

\bibitem[{Liang et~al.(2026)Liang, Mao, \protect\BIBand{}
  Wang}]{liang2026online}
Liang Y, Mao X, Wang S (2026) Online joint assortment--inventory optimization
  under {MNL} choices. \emph{Operations Research} 74(4):1909--1931.

\bibitem[{Liu \protect\BIBand{} Nocedal(1989)}]{liu1989limited}
Liu DC, Nocedal J (1989) On the limited memory {BFGS} method for large scale
  optimization. \emph{Mathematical Programming} 45(1--3):503--528.

\bibitem[{Liu et~al.(2026)Liu, Rusmevichientong, \protect\BIBand{}
  Wang}]{liu2026lessdata}
Liu Z, Rusmevichientong P, Wang M (2026) When less data is better: Eliminating
  stockout-induced bias in choice estimation. Available at SSRN 7106418.

\bibitem[{Mahajan \protect\BIBand{} van Ryzin(2001)}]{mahajan2001stocking}
Mahajan S, van Ryzin G (2001) Stocking retail assortments under dynamic
  consumer substitution. \emph{Operations Research} 49(3):334--351.

\bibitem[{Mersereau(2015)}]{mersereau2015demand}
Mersereau AJ (2015) Demand estimation from censored observations with inventory
  record inaccuracy. \emph{Manufacturing \& Service Operations Management}
  17(3):335--349.

\bibitem[{Musalem et~al.(2010)Musalem, Olivares, Bradlow, Terwiesch,
  \protect\BIBand{} Corsten}]{musalem2010structural}
Musalem A, Olivares M, Bradlow ET, Terwiesch C, Corsten D (2010) Structural
  estimation of the effect of out-of-stocks. \emph{Management Science}
  56(7):1180--1197.

\bibitem[{Nahmias(1994)}]{nahmias1994demand}
Nahmias S (1994) Demand estimation in lost sales inventory systems. \emph{Naval
  Research Logistics} 41(6):739--757.

\bibitem[{Newman et~al.(2014)Newman, Ferguson, Garrow, \protect\BIBand{}
  Jacobs}]{newman2014estimation}
Newman JP, Ferguson ME, Garrow LA, Jacobs TL (2014) Estimation of choice-based
  models using sales data from a single firm. \emph{Manufacturing \& Service
  Operations Management} 16(2):184--197.

\bibitem[{{NIST Digital Library of Mathematical
  Functions}(2026)}]{nist_dlmf_gauss_quadrature}
{NIST Digital Library of Mathematical Functions} (2026) Gauss quadrature.
  National Institute of Standards and Technology,
  \urlprefix\url{https://dlmf.nist.gov/3.5}, {S}ection 3.5(v), accessed
  September 14, 2026.

\bibitem[{{PyTorch Documentation}(2026)}]{pytorch_gammainc}
{PyTorch Documentation} (2026) \texttt{torch.special.gammainc}. PyTorch Special
  Functions Documentation,
  \urlprefix\url{https://docs.pytorch.org/docs/stable/special.html#torch.special.gammainc},
  accessed September 14, 2026.

\bibitem[{Shao \protect\BIBand{} Kleywegt(2020)}]{shao2020joint}
Shao H, Kleywegt AJ (2020) Joint estimation of discrete choice model and
  arrival rate with unobserved stock-out events. arXiv preprint
  arXiv:2003.02313.

\bibitem[{{\c S}im{\c s}ek \protect\BIBand{}
  Topaloglu(2018)}]{simsek2018estimation}
{\c S}im{\c s}ek AS, Topaloglu H (2018) Technical note---{An}
  expectation-maximization algorithm to estimate the parameters of the markov
  chain choice model. \emph{Operations Research} 66(3):748--760.

\bibitem[{Smith \protect\BIBand{} Agrawal(2000)}]{smith2000management}
Smith SA, Agrawal N (2000) Management of multi-item retail inventory systems
  with demand substitution. \emph{Operations Research} 48(1):50--64.

\bibitem[{Steeneck et~al.(2022)Steeneck, Eng-Larsson, \protect\BIBand{}
  Jauffred}]{steeneck2022estimating}
Steeneck D, Eng-Larsson F, Jauffred F (2022) Estimating lost sales for
  substitutable products with uncertain on-shelf availability.
  \emph{Manufacturing \& Service Operations Management} 24(3):1578--1594.

\bibitem[{Subramanian \protect\BIBand{} Harsha(2021)}]{subramanian2021demand}
Subramanian S, Harsha P (2021) Demand modeling in the presence of unobserved
  lost sales. \emph{Management Science} 67(6):3803--3833.

\bibitem[{Talluri \protect\BIBand{} Tekin(2025)}]{talluri2025estimation}
Talluri K, Tekin M (2025) Estimation using marginal competitor sales
  information. \emph{Journal of Operations Management} 71(5):588--610.

\bibitem[{Talluri \protect\BIBand{} van Ryzin(2004)}]{talluri2004revenue}
Talluri K, van Ryzin G (2004) Revenue management under a general discrete
  choice model of consumer behavior. \emph{Management Science} 50(1):15--33.

\bibitem[{van Ryzin \protect\BIBand{} Vulcano(2017)}]{vanryzin2017rank}
van Ryzin G, Vulcano G (2017) Technical note---{An} expectation-maximization
  method to estimate a rank-based choice model of demand. \emph{Operations
  Research} 65(2):396--407.

\bibitem[{Vulcano et~al.(2012)Vulcano, van Ryzin, \protect\BIBand{}
  Ratliff}]{vulcano2012estimating}
Vulcano G, van Ryzin G, Ratliff R (2012) Estimating primary demand for
  substitutable products from sales transaction data. \emph{Operations
  Research} 60(2):313--334.

\bibitem[{Wan et~al.(2018)Wan, Huang, Zhao, Deng, \protect\BIBand{}
  Fransoo}]{wan2018demand}
Wan M, Huang Y, Zhao L, Deng T, Fransoo JC (2018) Demand estimation under
  multi-store multi-product substitution in high density traditional retail.
  \emph{European Journal of Operational Research} 266(1):99--111.

\bibitem[{Wang \protect\BIBand{} Xu(2025)}]{wang2025joint}
Wang X, Xu W (2025) Joint estimation of the arrival rate and customer taste
  coefficients from censored transactional data. \emph{Production and
  Operations Management} 34(9):2853--2872.

\bibitem[{Zhang et~al.(2025)Zhang, Ma, \protect\BIBand{}
  Topaloglu}]{zhang2025leveraging}
Zhang J, Ma W, Topaloglu H (2025) Technical note---{Leveraging} the degree of
  dynamic substitution in assortment and inventory planning. \emph{Operations
  Research} 73(3):1248--1259.

\bibitem[{Zheng et~al.(2026)Zheng, Chen, Hedenstierna, \protect\BIBand{}
  Yang}]{zheng2026offline}
Zheng Y, Chen YF, Hedenstierna CPT, Yang J (2026) Offline learning and
  optimization for multi-product inventory management with stockout-based
  substitution. \emph{Production and Operations Management} 35(9):3272--3291.

\bibitem[{Zhou et~al.(2026)Zhou, Wang, \protect\BIBand{}
  Zhang}]{zhou2025optimal}
Zhou Z, Wang T, Zhang J (2026) Optimal inventory allocation for indifferent
  goods under dynamic substitution. \emph{Operations Research}
  74(4):2001--2012.

\bibitem[{Zhou \protect\BIBand{} Zhang(2026)}]{zhou2026optimal}
Zhou Z, Zhang J (2026) Approximations to stockout-based substitution under
  {MNL} model. Available at SSRN 6427181.

\end{thebibliography}
\clearpage

\begin{APPENDICES}

\renewcommand{\thesection}{\Alph{section}}
\renewcommand{\theHsection}{appendix.\arabic{section}}

\renewcommand{\theHsubsection}{appendix.\arabic{section}.\arabic{subsection}}
\renewcommand{\theHsubsubsection}{appendix.\arabic{section}.\arabic{subsection}.\arabic{subsubsection}}

\numberwithin{equation}{section}
\numberwithin{theorem}{section}
\numberwithin{proposition}{section}
\numberwithin{lemma}{section}
\numberwithin{definition}{section}
\numberwithin{figure}{section}
\numberwithin{table}{section}
\numberwithin{assumption}{section}
\numberwithin{corollary}{section}
\numberwithin{algorithm}{section}

\section{Proofs}
\label{app:proofs}

Throughout this appendix, we mostly consider a single selling period and suppress the index $n$.
We also write $V=1+\sum_{k\in\cC}\theta_k$, so that $\rho_j=\theta_j/V$ for $j\in\cA$.

We begin with Lemma~\ref{lem:clock-embedding}, which is due to \citet{zhou2025optimal}; we include a short proof for completeness because the proof of Lemma~\ref{lem:fixed-count-occupation} builds on it.
The idea is to show that the choices $(C_t)$ in the discrete-time system and the labels $(D_t)$ in the continuous-time system are generated by the same rule: given the past, the next choice follows the MNL probabilities over the products that have not sold out.

\begin{proof}[Proof of Lemma~\ref{lem:clock-embedding}]
We first check that each $D_t$ is well defined.
Since $N_{\mathrm{tot}}\ge N_0=\Pi_0$ and $\Pi_0(\tau)\to\infty$ almost surely, every $\sigma_t$ is finite almost surely.
Moreover, because independent Poisson processes almost surely have no common jump times, every jump of $N_{\mathrm{tot}}$ has size one, and exactly one of $N_0,\ldots,N_J$ jumps at each $\sigma_t$.

Next, fix $h\ge0$ and set $\sigma_0=0$.
The set of products that have not been sold out for the $(h+1)$st arrival,
\[
\widetilde S_{h+1}=\{j\in[J]:N_j(\sigma_h)<y_j\},
\]
is determined by $(D_1,\ldots,D_h)$.
By the strong Markov property at the stopping time $\sigma_h$, the processes $(\Pi_j(\sigma_h+\tau)-\Pi_j(\sigma_h))_{\tau\ge0}$, $j=0,\ldots,J$, are independent Poisson processes with rates $\theta_j$, independent of the history up to $\sigma_h$.
The $(h+1)$st choice is made by whichever of $\Pi_0$ and $\Pi_j$, $j\in\widetilde S_{h+1}$, jumps first after $\sigma_h$.
Since these first jump times are independent exponential random variables with rates $1$ and $(\theta_j)_{j\in\widetilde S_{h+1}}$,
\begin{equation}
\PP(D_{h+1}=j\mid D_1,\ldots,D_h)=
\begin{cases}
\dfrac{\theta_j}{1+\sum_{k\in\widetilde S_{h+1}}\theta_k}, & j\in\widetilde S_{h+1},\\[10pt]
\dfrac{1}{1+\sum_{k\in\widetilde S_{h+1}}\theta_k}, & j=0,\\[10pt]
0, & j\in[J]\setminus\widetilde S_{h+1}.
\end{cases}
\label{eq:app-next-label}
\end{equation}
In the discrete-time system, by \eqref{eq:inventory-update}, customer $h+1$ faces $S_{h+1}=\{j\in[J]:y_j-\sum_{i=1}^h\I{\{C_i=j\}}>0\}$, the same function of $(C_1,\ldots,C_h)$ as $\widetilde S_{h+1}$ is of $(D_1,\ldots,D_h)$.
Hence, by \eqref{eq:mnl}, $C_{h+1}$ given $(C_1,\ldots,C_h)$ has the same conditional distribution as $D_{h+1}$ given $(D_1,\ldots,D_h)$ in \eqref{eq:app-next-label}.
Since the joint distribution of the first $t$ choices is the product of these conditional distributions, \eqref{eq:embedding-law} follows.
\end{proof}

For the proof of Lemma~\ref{lem:fixed-count-occupation}, we first compute the expected time that the continuous-time system spends in the state $(m,\bx)$ in two ways: by integrating the probability $G_m(\tau;\btheta)$ of being in the state at time $\tau$, and by multiplying the probability of reaching the state by the expected holding time there.
We then show that reaching the state is equivalent to observing the terminal sales $\bx$ after $T=m+s$ customers, where $s=\sum_{j=1}^Jx_j$, which justifies the likelihood \eqref{eq:known-occupation}.

\begin{proof}[Proof of Lemma~\ref{lem:fixed-count-occupation}]
Write $\bm N(\tau)=(N_0(\tau),\ldots,N_J(\tau))$, and let $\tau_{\mathrm{in}}=\inf\{\tau\ge0:\bm N(\tau)=(m,\bx)\}$ and $\tau_{\mathrm{out}}=\inf\{\tau>\tau_{\mathrm{in}}:\bm N(\tau)\ne(m,\bx)\}$ be the times at which the continuous-time system enters and leaves $(m,\bx)$, with $\inf\varnothing=\infty$.
The event that $\bm N$ reaches $(m,\bx)$ is $H=\{\tau_{\mathrm{in}}<\infty\}$.

We first verify \eqref{eq:Gm}.
For fixed $\tau>0$, $N_0(\tau),\ldots,N_J(\tau)$ are independent, and $\PP(N_0(\tau)=m)=\pi_m(\tau)$.
For $j\in\cC$, $x_j<y_j$, so $N_j(\tau)=x_j$ if and only if $\Pi_j(\tau)=x_j$, which has probability $\pi_{x_j}(\theta_j\tau)$.
For $j\in\cB$, $x_j=y_j$, so $N_j(\tau)=x_j$ if and only if $\Pi_j(\tau)\ge y_j$, which has probability $\overline F_{y_j}(\theta_j\tau)$.
Multiplying these probabilities gives \eqref{eq:Gm}.

Next, let $O=\int_0^\infty\I{\{\bm N(\tau)=(m,\bx)\}}\,d\tau$ be the total time that $\bm N$ spends in $(m,\bx)$.
By Tonelli's theorem,
\begin{equation}
\E[O]=\int_0^\infty\PP\bigl(\bm N(\tau)=(m,\bx)\bigr)\,d\tau=\int_0^\infty G_m(\tau;\btheta)\,d\tau.
\label{eq:app-occupation-1}
\end{equation}
On the other hand, every coordinate of $\bm N$ is nondecreasing and right-continuous, so $\bm N$ never returns to $(m,\bx)$ after leaving it; thus $O=\tau_{\mathrm{out}}-\tau_{\mathrm{in}}$ on $H$, and $O=0$ on $H^c$.
In state $(m,\bx)$, only jumps of $\Pi_0$ and of $\Pi_j$, $j\in\cC$, change $\bm N$, because the other products are at their caps.
Hence, by the strong Markov property at $\tau_{\mathrm{in}}$, the holding time $\tau_{\mathrm{out}}-\tau_{\mathrm{in}}$ is, conditional on $H$, exponential with rate $V=1+\sum_{j\in\cC}\theta_j$, and
\begin{equation}
\E[O]=\PP(H)\,\E[\tau_{\mathrm{out}}-\tau_{\mathrm{in}}\mid H]=\frac{\PP(H)}{V}.
\label{eq:app-occupation-2}
\end{equation}
Equating \eqref{eq:app-occupation-1} and \eqref{eq:app-occupation-2} gives \eqref{eq:fixed-count-occupation}.

Finally, we show that $\PP(H)=\PP(\bm X=\bx\mid\by,T)$ with $T=m+s$.
Since every jump of $N_{\mathrm{tot}}$ has size one, $N_{\mathrm{tot}}(\tau)=T$ exactly when $\tau\in[\sigma_T,\sigma_{T+1})$, and $\bm N$ is constant on this interval.
As the coordinates of $(m,\bx)$ sum to $T$, it follows that $H=\{\bm N(\sigma_T)=(m,\bx)\}$, the event that the first $T$ choices $(D_1,\ldots,D_T)$ consist of $m$ no-purchase outcomes and $x_j$ purchases of each product $j$.
By Lemma~\ref{lem:clock-embedding}, $(D_1,\ldots,D_T)$ has the same distribution as $(C_1,\ldots,C_T)$, so $\PP(H)=\PP(\bm X=\bx\mid\by,T)$, which justifies the likelihood \eqref{eq:known-occupation}.
\end{proof}

We next prove the gradient formulas in Theorems~\ref{thm:known-score} and~\ref{thm:poisson-score}.
In both settings, the likelihood factorizes as $L=CR$.
The factor $C$ is explicit and can be differentiated directly, whereas the factor $R$ is an integral, which we differentiate under the integral sign.
We first establish two lemmas: Lemma~\ref{lem:scalar-functions} shows the properties of the Poisson tail $\overline F_y$ and of the functions $\Phi_s$ and $\psi_s$ that we need, and Lemma~\ref{lem:differentiation} uses these properties to justify the differentiation.

\begin{lemma}[Properties of the scalar functions]
\label{lem:scalar-functions}
\begin{itemize}
\item[(a)] For $y\ge1$ and $\mu>0$, $\overline F_y'(\mu)=\pi_{y-1}(\mu)$. Consequently, $d\log\overline F_y(\mu)/d\log\mu=h_y(\mu)$, where
\[
h_y(\mu)=\frac{\mu\pi_{y-1}(\mu)}{\overline F_y(\mu)}=\frac{y\pi_y(\mu)}{\overline F_y(\mu)}\in(0,y).
\]
\item[(b)] For $s\in\Z_+$ and $z\ge0$, $\Phi_s'(z)=\Phi_{s+1}(z)/(s+1)$,
\[
1\le\Phi_s(z)\le e^{2\sqrt z},
\qquad\text{and}\qquad
0\le\psi_s(z)\le\frac{z}{s+1}.
\]
\end{itemize}
\end{lemma}

\begin{proof}[Proof of Lemma~\ref{lem:scalar-functions}]
For part (a), $\overline F_y(\mu)=1-\sum_{\ell=0}^{y-1}\pi_\ell(\mu)$ and $\pi_\ell'=\pi_{\ell-1}-\pi_\ell$ with $\pi_{-1}\equiv0$, so the derivative telescopes to $\overline F_y'(\mu)=\pi_{y-1}(\mu)$, and $d\log\overline F_y(\mu)/d\log\mu=\mu\pi_{y-1}(\mu)/\overline F_y(\mu)$.
The second expression for $h_y$ follows from $\mu\pi_{y-1}(\mu)=y\pi_y(\mu)$, and $0<h_y(\mu)<y$ because $\overline F_y(\mu)=\sum_{\ell\ge y}\pi_\ell(\mu)>\pi_y(\mu)>0$.

For part (b), the power series \eqref{eq:phi_s-def} has an infinite radius of convergence, so it can be differentiated term by term:
\[
\Phi_s'(z)
=s!\sum_{m=1}^\infty\frac{z^{m-1}}{(m-1)!\,(s+m)!}
=s!\sum_{m=0}^\infty\frac{z^{m}}{m!\,(s+1+m)!}
=\frac{\Phi_{s+1}(z)}{s+1}.
\]
Since the coefficient of $z^m$ in $\Phi_{s+1}$ is $(s+1)/(s+1+m)\le1$ times that in $\Phi_s$, we have $\Phi_{s+1}\le\Phi_s$, and hence, for $z>0$ (the case $z=0$ being trivial),
\[
0\le\psi_s(z)=\frac{z\Phi_s'(z)}{\Phi_s(z)}=\frac{z}{s+1}\cdot\frac{\Phi_{s+1}(z)}{\Phi_s(z)}\le\frac{z}{s+1}.
\]
Iterating the comparison gives $\Phi_s(z)\le\Phi_0(z)=\sum_{m\ge0}z^m/(m!)^2$, and since $1/(m!)^2=\binom{2m}{m}/(2m)!\le4^m/(2m)!$,
\[
\Phi_0(z)\le\sum_{m=0}^\infty\frac{(2\sqrt z)^{2m}}{(2m)!}=\cosh(2\sqrt z)\le e^{2\sqrt z}.
\]
Finally, $\Phi_s(z)\ge1$ because the series has constant term one and nonnegative coefficients.
\end{proof}

The proofs below use the derivatives of $\log\rho_k$, $k\in\cB$, and of $\log z$, where $z=\lambda u/V$ is the argument of $\Phi_s$ in \eqref{eq:poisson-gamma} and $u>0$ is held fixed.
Since $\log\rho_k=v_k-\log V$, $\log z=\eta+\log u-\log V$, and $\partial\log V/\partial v_j=\rho_j\I{\{j\in\cC\}}$, we have, for $k\in\cB$ and $j\in[J]$,
\begin{equation}
\frac{\partial\log\rho_k}{\partial v_j}=\I{\{j=k\}}-\rho_j\I{\{j\in\cC\}},\qquad
\frac{\partial\log\rho_k}{\partial\eta}=0,\qquad
\frac{\partial\log z}{\partial v_j}=-\rho_j\I{\{j\in\cC\}},\qquad
\frac{\partial\log z}{\partial\eta}=1.
\label{eq:app-chain-rule}
\end{equation}
These derivatives do not depend on $u$ and lie in $[-1,1]$, because $\rho_j<1$ for $j\in\cC$ and $\cB\cap\cC=\varnothing$.

\begin{lemma}[Differentiation in the integral]
\label{lem:differentiation}
The functions $R^{\mathrm K}$ and $R^{\mathrm P}$ are continuously differentiable in $\bv$ and in $(\eta,\bv)$, respectively, and, for any coordinate $\vartheta$ of the corresponding parameter vector,
\begin{align}
\frac{\partial\log R^{\mathrm K}}{\partial\vartheta}
&=\E_W\left[\left.\frac{\partial}{\partial\vartheta}\log\left\{\prod_{k\in\cB}\overline F_{y_k}(\rho_ku)\right\}\right|_{u=W}\right],
\label{eq:app-dlogR-known}\\
\frac{\partial\log R^{\mathrm P}}{\partial\vartheta}
&=\E_{\bar W}\left[\left.\frac{\partial}{\partial\vartheta}\log\left\{\Phi_s\left(\frac{\lambda u}{V}\right)\prod_{k\in\cB}\overline F_{y_k}(\rho_ku)\right\}\right|_{u=\bar W}\right],
\label{eq:app-dlogR-poisson}
\end{align}
where $u$ is held fixed in the differentiation and $W$ and $\bar W$ have the densities \eqref{eq:pdf-w} and \eqref{eq:pdf-w-poisson}.
\end{lemma}

The idea behind Lemma~\ref{lem:differentiation} is simple.
If $R=\int_0^\infty f(u)\,du$ for a positive function $f$ that depends on the parameters, then, formally, $\partial_\vartheta R=\int_0^\infty f(u)\,\partial_\vartheta\log f(u)\,du$.
Dividing by $R$ turns the right-hand side into an expectation under the density $f/R$, which is how $W$ and $\bar W$ arise.
The proof only needs to justify the interchange of differentiation and integration.

\begin{proof}[Proof of Lemma~\ref{lem:differentiation}]
We give the proof for $R^{\mathrm P}$; for $R^{\mathrm K}$, replace $\Phi_s$ by one and $c$ by $m+c$.
By \eqref{eq:poisson-gamma}, $R^{\mathrm P}=\int_0^\infty f(u)\,du$ with the positive integrand
\[
f(u)=\frac{u^ce^{-u}}{c!}\,\Phi_s(z)\prod_{k\in\cB}\overline F_{y_k}(\rho_ku),\qquad z=\frac{\lambda u}{V}.
\]
By the chain rule, the definition of $\psi_s$, and Lemma~\ref{lem:scalar-functions}(a),
\begin{equation}
\frac{\partial\log f(u)}{\partial\vartheta}
=\psi_s(z)\frac{\partial\log z}{\partial\vartheta}
+\sum_{k\in\cB}h_{y_k}(\rho_ku)\frac{\partial\log\rho_k}{\partial\vartheta},
\label{eq:app-dlogf}
\end{equation}
so, by Lemma~\ref{lem:scalar-functions} and \eqref{eq:app-chain-rule}, $|\partial\log f(u)/\partial\vartheta|\le z+\sum_{k\in\cB}y_k$.

Fix a parameter value and a bounded open ball $\mathcal N$ around it, and let $\bar\lambda=\sup_{\mathcal N}\lambda<\infty$, so that $z\le\bar\lambda u$ on $\mathcal N$ because $V\ge1$.
Since $\Phi_s(z)\le e^{2\sqrt z}$ and $\overline F_{y_k}\le1$, for every parameter value in $\mathcal N$ and every $u>0$,
\[
\left|\frac{\partial f(u)}{\partial\vartheta}\right|
=f(u)\left|\frac{\partial\log f(u)}{\partial\vartheta}\right|
\le\frac{u^ce^{-u}}{c!}\,e^{2\sqrt{\bar\lambda u}}\left(\bar\lambda u+\sum_{k\in\cB}y_k\right)\eqqcolon g(u).
\]
Because $2\sqrt{\bar\lambda u}\le u/2+2\bar\lambda$, the function $g$ is bounded by a multiple of $u^c(1+u)e^{-u/2}$ and is therefore integrable.
By the mean value theorem, $g$ also bounds the difference quotients of $f(u)$ in $\vartheta$ on $\mathcal N$, so, by dominated convergence,
\[
\frac{\partial R^{\mathrm P}}{\partial\vartheta}
=\int_0^\infty\frac{\partial f(u)}{\partial\vartheta}\,du
=\int_0^\infty f(u)\,\frac{\partial\log f(u)}{\partial\vartheta}\,du,
\]
which, again by dominated convergence, is continuous on $\mathcal N$ because $\partial f(u)/\partial\vartheta$ is continuous in the parameters.
Dividing by $R^{\mathrm P}>0$ and noting that $f/R^{\mathrm P}$ is the density \eqref{eq:pdf-w-poisson} of $\bar W$ gives \eqref{eq:app-dlogR-poisson}.
For $R^{\mathrm K}$, the dominating function $\bigl(\sum_{k\in\cB}y_k\bigr)u^{m+c}e^{-u}/(m+c)!$ works for all parameter values.
\end{proof}

\begin{proof}[Proof of Theorem~\ref{thm:known-score}]
By \eqref{eq:known-prefactor} and $\log\rho_k=v_k-\log V$, we have $\log C^{\mathrm K}=\text{constant}+\sum_{k\in\cC}x_kv_k-(m+c)\log V$.
By Lemmas~\ref{lem:scalar-functions}(a) and~\ref{lem:differentiation}, and since the derivatives in \eqref{eq:app-chain-rule} do not depend on $u$,
\begin{equation}
\frac{\partial\log R^{\mathrm K}}{\partial v_j}
=\sum_{k\in\cB}\E_W\bigl[h_{y_k}(\rho_kW)\bigr]\frac{\partial\log\rho_k}{\partial v_j}.
\label{eq:app-known-R-derivative}
\end{equation}
If $j\in\cB$, then $\log C^{\mathrm K}$ does not depend on $v_j$, and $\partial\log\rho_k/\partial v_j=\I{\{j=k\}}$ for $k\in\cB$, so $\partial\log L^{\mathrm K}/\partial v_j=\E_W[h_{y_j}(\rho_jW)]$, which is the first case of \eqref{eq:known-score}.
If $j\in\cC$, then $\partial\log C^{\mathrm K}/\partial v_j=x_j-(m+c)\rho_j$ and $\partial\log\rho_k/\partial v_j=-\rho_j$ for every $k\in\cB$, so adding $\partial\log R^{\mathrm K}/\partial v_j=-\rho_j\sum_{k\in\cB}\E_W[h_{y_k}(\rho_kW)]$ gives the second case.
\end{proof}

Because $h_y>0$, the first case of \eqref{eq:known-score} also shows that the log-likelihood of a single period is strictly increasing in $v_j$ whenever product $j$ sells out in that period, as noted in Section~\ref{subsec:known-theory}.

\begin{proof}[Proof of Theorem~\ref{thm:poisson-score}]
The proof parallels that of Theorem~\ref{thm:known-score}.
By \eqref{eq:poisson-prefactor}, $\log C^{\mathrm P}=\text{constant}-e^{\eta}+s\eta+\sum_{k\in\cC}x_kv_k-c\log V$.
Substituting \eqref{eq:app-dlogf} into \eqref{eq:app-dlogR-poisson} gives, for any coordinate $\vartheta$ of $(\eta,\bv)$,
\begin{equation}
\frac{\partial\log R^{\mathrm P}}{\partial\vartheta}
=\overline m\,\frac{\partial\log z}{\partial\vartheta}
+\sum_{k\in\cB}\E_{\bar W}\bigl[h_{y_k}(\rho_k\bar W)\bigr]\frac{\partial\log\rho_k}{\partial\vartheta},
\label{eq:app-poisson-R-derivative}
\end{equation}
where $\overline m=\E_{\bar W}[\psi_s(\lambda\bar W/V)]$.
If $j\in\cB$, neither $C^{\mathrm P}$ nor $z$ depends on $v_j$, so $\partial\log L^{\mathrm P}/\partial v_j=\E_{\bar W}[h_{y_j}(\rho_j\bar W)]$ as in the known-arrival setting.
If $j\in\cC$, then $\partial\log C^{\mathrm P}/\partial v_j=x_j-c\rho_j$ and $\partial\log z/\partial v_j=\partial\log\rho_k/\partial v_j=-\rho_j$ for every $k\in\cB$, so
\[
\frac{\partial\log L^{\mathrm P}}{\partial v_j}=x_j-\rho_j\left\{c+\overline m+\sum_{k\in\cB}\E_{\bar W}\bigl[h_{y_k}(\rho_k\bar W)\bigr]\right\}.
\]
Products outside $\cA$ do not enter the likelihood.
Finally, $\partial\log C^{\mathrm P}/\partial\eta=s-\lambda$, $\partial\log z/\partial\eta=1$, and $\partial\log\rho_k/\partial\eta=0$, so $\partial\log L^{\mathrm P}/\partial\eta=s-\lambda+\overline m$.
\end{proof}

We now turn to Theorem~\ref{thm:known-concavity}. Since $\log L^{\mathrm K}=\log C^{\mathrm K}+\log R^{\mathrm K}$, we establish joint strict concavity by considering the two terms separately. The explicit term $\log C^{\mathrm K}$ is concave and becomes strictly
concave when no product sells out, in which case $R^{\mathrm K}=1$. When at least one product sells out, we establish strict concavity of $\log R^{\mathrm K}$ in two steps: we first show that its integrand is jointly strictly log-concave in the active log attractions and the
integration variable $u$, and then apply the following lemma to show that integrating out $u$ preserves strict log-concavity.

\begin{lemma}[Integration preserves log-concavity]
\label{lem:logconcave-integral}
Let $f:\R^J\times(0,\infty)\to(0,\infty)$ be continuous and jointly log-concave, and suppose that $F(\bv)=\int_0^\infty f(\bv,u)\,du<\infty$ for every $\bv\in\R^J$.
Then $\log F$ is concave on $\R^J$. If $f$ is jointly strictly log-concave, then $\log F$ is strictly concave.
\end{lemma}

\begin{proof}[Proof of Lemma~\ref{lem:logconcave-integral}]
Fix $\bv_0,\bv_1\in\R^J$ and $t\in(0,1)$, and let $\bv_t=(1-t)\bv_0+t\bv_1$; we show that $F(\bv_t)\ge F(\bv_0)^{1-t}F(\bv_1)^t$.
The idea is to pair the points that split $F(\bv_0)$ and $F(\bv_1)$ at the same fraction.
For $i\in\{0,1\}$, the function $K_i(w)=\int_0^wf(\bv_i,u)\,du$ is continuously differentiable on $(0,\infty)$ with $K_i'(w)=f(\bv_i,w)>0$ and increases from $0$ to $F(\bv_i)$.
Hence, for each $a\in(0,1)$, there is a unique $u_i(a)>0$ with $K_i(u_i(a))=aF(\bv_i)$, and, by the inverse function theorem, $u_i$ is a continuously differentiable increasing bijection from $(0,1)$ onto $(0,\infty)$ with
\begin{equation}
f\bigl(\bv_i,u_i(a)\bigr)\,u_i'(a)=F(\bv_i),\qquad i=0,1.
\label{eq:app-transport}
\end{equation}
The convex combination $u_t=(1-t)u_0+tu_1$ is again such a bijection, so the change of variables $u=u_t(a)$ gives
\begin{align*}
F(\bv_t)
&=\int_0^1 f\bigl(\bv_t,u_t(a)\bigr)\,u_t'(a)\,da\\
&\ge\int_0^1 f\bigl(\bv_0,u_0(a)\bigr)^{1-t}f\bigl(\bv_1,u_1(a)\bigr)^{t}\,u_0'(a)^{1-t}u_1'(a)^{t}\,da\\
&=\int_0^1 F(\bv_0)^{1-t}F(\bv_1)^t\,da=F(\bv_0)^{1-t}F(\bv_1)^t.
\end{align*}
The inequality uses the log-concavity of $f$ at $(\bv_t,u_t(a))=(1-t)(\bv_0,u_0(a))+t(\bv_1,u_1(a))$ and the weighted arithmetic--geometric mean inequality $u_t'(a)\ge u_0'(a)^{1-t}u_1'(a)^t$; the second equality follows from \eqref{eq:app-transport}.
If $f$ is jointly strictly log-concave and $\bv_0\ne\bv_1$, the log-concavity inequality is strict for every $a\in(0,1)$. Hence $F(\bv_t)>F(\bv_0)^{1-t}F(\bv_1)^t$.
\end{proof}

\begin{proof}[Proof of Theorem~\ref{thm:known-concavity}]
We work with the active log attractions $(v_j)_{j\in\cA}$, since products outside $\cA$ do not enter the likelihood. By \eqref{eq:known-prefactor},
\[
\log C^{\mathrm K}=\text{constant}+\sum_{j\in\cC}x_jv_j-(m+c)\log\left(1+\sum_{j\in\cC}e^{v_j}\right),
\]
which is concave because the log-sum-exp function is convex and $m+c\ge0$.

If $\cB=\varnothing$, then $\cC=\cA$, $m+c=T\ge1$, and $R^{\mathrm K}=1$. Recall that $\rho_j=\theta_j/V=e^{v_j}/(1+\sum_{k\in\cA}e^{v_k})$ in this case. The Hessian entries of $\log C^{\mathrm K}$ are
\[
\frac{\partial^2\log C^{\mathrm K}}{\partial v_j\partial v_k}
=-T\left(\I\{j=k\}\rho_j-\rho_j\rho_k\right),
\qquad j,k\in\cA.
\]
Since $\rho_j>0$ and $\sum_{k\in\cA}\rho_k<1$, the diagonal entries are negative and satisfy
\[
-\frac{\partial^2\log C^{\mathrm K}}{\partial v_j^2}
-\sum_{k\in\cA\setminus\{j\}}
\left|\frac{\partial^2\log C^{\mathrm K}}{\partial v_j\partial v_k}\right|
=T\rho_j\left(1-\sum_{k\in\cA}\rho_k\right)>0.
\]
Thus the Hessian is symmetric, strictly diagonally dominant, and negative definite. Hence $\log L^{\mathrm K}=\log C^{\mathrm K}$ is jointly strictly concave.

Then we consider $\cB\ne\varnothing$. By \eqref{eq:known-gamma}, $R^{\mathrm K}(\bv)=\int_0^\infty f(\bv,u)\,du\le1$ with the positive and continuous integrand
\[
f(\bv,u)=\frac{u^{m+c}e^{-u}}{(m+c)!}\prod_{j\in\cB}\overline F_{y_j}\bigl(\rho_j(\bv)u\bigr),
\]
so, by Lemma~\ref{lem:logconcave-integral}, it suffices to show that $f$ is  jointly strictly log-concave in $(\bv,u)$.
The key property is that the probability that a Poisson random variable reaches a given level is log-concave in the logarithm of its mean: for $y\ge1$, the function $q_y(w)=\log\overline F_y(e^w)$ is strictly increasing and strictly concave.
To see this, let $\mu=e^w$.
By Lemma~\ref{lem:scalar-functions}(a), $q_y'(w)=h_y(\mu)>0$, and differentiating $\log h_y(\mu)=\log y+\log\pi_y(\mu)-\log\overline F_y(\mu)$ with respect to $w$, using $d\log\pi_y(\mu)/d\log\mu=y-\mu$, gives
\[
q_y''(w)=h_y(\mu)\bigl\{y-\mu-h_y(\mu)\bigr\}.
\]
Hence $q_y$ is strictly concave if $h_y(\mu)>y-\mu$, which holds immediately when $\mu\ge y$.
When $0<\mu<y$,
\[
\frac{\overline F_y(\mu)}{\pi_y(\mu)}
=\sum_{\ell=0}^\infty\frac{\mu^\ell}{(y+1)(y+2)\cdots(y+\ell)}
<\sum_{\ell=0}^\infty\left(\frac{\mu}{y}\right)^\ell
=\frac{y}{y-\mu},
\]
so $h_y(\mu)=y\pi_y(\mu)/\overline F_y(\mu)>y-\mu$.

Now,
\[
\log f(\bv,u)=\text{constant}+(m+c)\log u-u+\sum_{j\in\cB}q_{y_j}\bigl(\log\rho_j(\bv)+\log u\bigr).
\]
Fix two distinct points $(\bv_0,u_0)$ and $(\bv_1,u_1)$, and let $(\bv_t,u_t)=(1-t)(\bv_0,u_0)+t(\bv_1,u_1)$ for $t\in(0,1)$. For each $j\in\cB$, joint concavity of $\log\rho_j(\bv)+\log u$ and strict monotonicity and concavity of $q_{y_j}$ give
\[
\begin{aligned}
&q_{y_j}\bigl(\log\rho_j(\bv_t)+\log u_t\bigr)\\
&\quad\ge q_{y_j}\bigl((1-t)(\log\rho_j(\bv_0)+\log u_0)+t(\log\rho_j(\bv_1)+\log u_1)\bigr)\\
&\quad\ge(1-t)q_{y_j}\bigl(\log\rho_j(\bv_0)+\log u_0\bigr)+tq_{y_j}\bigl(\log\rho_j(\bv_1)+\log u_1\bigr).
\end{aligned}
\]
The first inequality can hold with equality only if $u_0=u_1$ and $\bv_0,\bv_1$ agree on $\cC$, because $\log u$ and the negative log-sum-exp term are strictly concave. In that case, since the two points are distinct, they must differ in $v_j$ for some $j\in\cB$, making the second inequality strict by the strict concavity of $q_{y_j}$.

Since $(m+c)\log u-u$ is concave, it follows that
\[
\log f(\bv_t,u_t)>(1-t)\log f(\bv_0,u_0)+t\log f(\bv_1,u_1).
\]
Thus $\log f$ is jointly strictly concave. By Lemma~\ref{lem:logconcave-integral}, $\log R^{\mathrm K}$ is strictly concave, and adding the concave term $\log C^{\mathrm K}$ proves that $\log L^{\mathrm K}$ is jointly strictly concave.
\end{proof}

We then prove the Poisson-arrival results in Section~\ref{subsec:poisson-identification}. The sales-dynamics and monotonicity lemmas support both the identifiability result and the local strong concavity result. For these arguments, we use the parameters $(\bm\alpha,\delta)$ defined in \eqref{eq:alpha-delta-parameterization}, with $\lambda=\alpha_++\delta$ and $\theta_j=\alpha_j/\delta$. The identification arguments concern a fixed positive inventory vector; the local concavity result also allows periods in which some products are not initially stocked.

\begin{lemma}[Sales dynamics under Poisson arrivals]
\label{lem:sales-dynamics}
Represent a selling period by the interval $[0,1]$. Under Poisson arrivals, the terminal sales have the distribution of the sales by time one in the continuous-time Markov chain in which, whenever the set $A$ of products in stock is nonempty, the next sale occurs at rate
\begin{equation}
\Lambda_A(\delta)=\alpha_A\frac{\alpha_++\delta}{\alpha_A+\delta},\qquad\text{where }\alpha_A=\sum_{k\in A}\alpha_k,
\label{eq:app-sales-rate}
\end{equation}
and is of product $j\in A$ with probability $\alpha_j/\alpha_A$; sales stop once $A=\varnothing$. Moreover, $\Lambda_{[J]}(\delta)=\alpha_+$, and $\Lambda_A(\delta)$ is strictly decreasing in $\delta$ for every nonempty proper subset $A$ of $[J]$.
\end{lemma}

\begin{proof}[Proof of Lemma~\ref{lem:sales-dynamics}]
Let customers arrive on $[0,1]$ according to a Poisson process with rate $\lambda$ and choose according to \eqref{eq:mnl}; the number of customers is then $\Poi(\lambda)$, and given this number, their choices are as in Section~\ref{sec:data}. While the set of products in stock is $A$, each arriving customer buys product $j\in A$ with probability $\theta_j/(1+\sum_{k\in A}\theta_k)$, independently of the past, so, by Poisson thinning, product $j$ sells at rate
\[
\frac{\lambda\theta_j}{1+\sum_{k\in A}\theta_k}
=\frac{(\alpha_++\delta)\,\alpha_j/\delta}{1+\alpha_A/\delta}
=\alpha_j\frac{\alpha_++\delta}{\alpha_A+\delta},
\]
where we used $\lambda=\alpha_++\delta$ and $\theta_k=\alpha_k/\delta$. Summing over $j\in A$ gives the total rate $\Lambda_A(\delta)$, and the next sale is of product $j$ with probability $\alpha_j/\alpha_A$. Finally, $\Lambda_{[J]}(\delta)=\alpha_+$, and $\Lambda_A'(\delta)=\alpha_A(\alpha_A-\alpha_+)/(\alpha_A+\delta)^2<0$ whenever $0<\alpha_A<\alpha_+$, that is, whenever $A$ is a nonempty proper subset of $[J]$.
\end{proof}

Lemma~\ref{lem:sales-dynamics} clarifies the role of $\delta$. Before the first stockout, product $j$ sells at rate $\alpha_j$, which does not depend on $\delta$. After a stockout, $\delta$ determines how much of the demand for the sold-out products is redirected to the remaining products: as $\delta\to0$, $\Lambda_A(\delta)\to\alpha_+$, so all of this demand is redirected, whereas as $\delta\to\infty$, $\Lambda_A(\delta)\to\alpha_A$, so none of it is. Hence information about $\delta$ comes only from the post-stockout sales dynamics, including the absence of further sales. In particular, when $J=1$, nothing remains in stock after the stockout, so the product sells at rate $\alpha_1$ until it sells out, and $X_1\overset{d}{=}\min\{y_1,Z\}$ with $Z\sim\Poi(\alpha_1)$. This distribution depends on $(\lambda,\theta_1)$ only through $\alpha_1=\lambda\theta_1/(1+\theta_1)$ and not on $\delta$, which proves the claim for $J=1$ in Section~\ref{subsec:poisson-identification}. For $J\ge2$, the next lemma shows that, with $\bm\alpha$ fixed, a larger $\delta$ leads to fewer sales.

\begin{lemma}[Monotonicity in $\delta$]
\label{lem:scale-monotonicity}
Fix $\bm\alpha\in\R_{++}^J$ and $\by\ge\bm1$, let $\bm X(\delta)$ denote the terminal sales under $(\bm\alpha,\delta)$, and let $\mu_{\by}(\bm\alpha,\delta)=\E[X_+(\delta)]$ be the expected total sales, where $X_+(\delta)=\sum_jX_j(\delta)$.
\begin{itemize}
\item[(a)] For $0<\delta_1<\delta_2$, $\bm X(\delta_1)$ and $\bm X(\delta_2)$ can be coupled so that $\bm X(\delta_1)\ge\bm X(\delta_2)$ componentwise.
\item[(b)] If $J\ge2$, then $\mu_{\by}(\bm\alpha,\delta)$ is continuously differentiable and $\partial_\delta\mu_{\by}(\bm\alpha,\delta)<0$ for every $\delta>0$.
\end{itemize}
\end{lemma}

\begin{proof}[Proof of Lemma~\ref{lem:scale-monotonicity}]
For part (a), we construct the chain of Lemma~\ref{lem:sales-dynamics} for all $\delta$ simultaneously. Let $y_+=\sum_jy_j$, and draw the products sold, $j_1,\ldots,j_{y_+}$, by choosing $j_{h+1}=j$ with probability $\alpha_j/\alpha_{A_h}$ for $j\in A_h$, where $A_h$ is the set of products still in stock after the first $h$ sales. Independently, draw $\xi_1,\ldots,\xi_{y_+}\stackrel{\mathrm{iid}}{\sim}\operatorname{Exp}(1)$, and, for each $\delta>0$, let the $q$th sale occur at time
\begin{equation}
\tau_q(\delta)=\sum_{h=0}^{q-1}\frac{\xi_{h+1}}{\Lambda_{A_h}(\delta)},\qquad q=1,\ldots,y_+.
\label{eq:app-sale-times}
\end{equation}
By Lemma~\ref{lem:sales-dynamics}, the sales by time one, $X_j(\delta)=\sum_q\I{\{j_q=j,\ \tau_q(\delta)\le1\}}$, have the distribution of the terminal sales under $(\bm\alpha,\delta)$. Only the sale times depend on $\delta$, and each $\tau_q(\delta)$ is nondecreasing in $\delta$ because each $\Lambda_{A_h}(\delta)$ is nonincreasing. Since the sales completed by time one are the first $X_+(\delta)$ sales of the common sequence, it follows that $X_+(\delta_2)\le X_+(\delta_1)$ and $\bm X(\delta_2)\le\bm X(\delta_1)$.

For part (b), let $J\ge2$ and $0<\delta_1<\delta_2$, and fix a product $k$. On the event $\mathcal E_k=\{j_1=\cdots=j_{y_k}=k\}$, product $k$ sells out at the $y_k$th sale while the other products remain in stock, so $\tau_{y_k}=\sum_{h=1}^{y_k}\xi_h/\alpha_+$ does not depend on $\delta$, and the $(y_k+1)$st sale, which exists because $J\ge2$, follows after the time $\xi_{y_k+1}/\Lambda_{-k}(\delta)$, where
\[
\Lambda_{-k}(\delta)\coloneqq\Lambda_{[J]\setminus\{k\}}(\delta)=(\alpha_+-\alpha_k)\frac{\alpha_++\delta}{\alpha_+-\alpha_k+\delta}
\]
is strictly decreasing in $\delta$. The nonnegative integer $X_+(\delta_1)-X_+(\delta_2)$ is positive on $\mathcal E_k\cap\{\tau_{y_k+1}(\delta_1)\le1<\tau_{y_k+1}(\delta_2)\}$, and $\tau_{y_k+1}(\delta_1)\le\tau_{y_k+1}(\delta_2)$, so
\begin{equation}
\mu_{\by}(\bm\alpha,\delta_1)-\mu_{\by}(\bm\alpha,\delta_2)
\ge\PP\bigl(\mathcal E_k,\ \tau_{y_k+1}(\delta_1)\le1\bigr)-\PP\bigl(\mathcal E_k,\ \tau_{y_k+1}(\delta_2)\le1\bigr).
\label{eq:app-mu-gap}
\end{equation}
The event $\mathcal E_k$ has probability $(\alpha_k/\alpha_+)^{y_k}$ and is independent of $(\xi_h)$, and $\tau_{y_k}$ has density $\alpha_+^{y_k}u^{y_k-1}e^{-\alpha_+u}/(y_k-1)!$, so conditioning on $\tau_{y_k}=u$ gives
\begin{equation}
\PP\bigl(\mathcal E_k,\ \tau_{y_k+1}(\delta)\le1\bigr)=\frac{\alpha_k^{y_k}}{(y_k-1)!}\int_0^1u^{y_k-1}e^{-\alpha_+u}\Bigl\{1-e^{-\Lambda_{-k}(\delta)(1-u)}\Bigr\}\,du,
\label{eq:app-Fk}
\end{equation}
whose derivative with respect to $\delta$ is
\[
\frac{\alpha_k^{y_k}\Lambda_{-k}'(\delta)}{(y_k-1)!}\int_0^1u^{y_k-1}(1-u)e^{-\alpha_+u-\Lambda_{-k}(\delta)(1-u)}\,du<0.
\]
The finite-state sales process has transition rates that are smooth in $\delta>0$, so $\mu_{\by}$ is continuously differentiable. Dividing \eqref{eq:app-mu-gap} by $\delta_2-\delta_1$ and letting $\delta_2\downarrow\delta_1=\delta$ gives
\[
\partial_\delta\mu_{\by}(\bm\alpha,\delta)\le\frac{\partial}{\partial\delta}\PP\bigl(\mathcal E_k,\tau_{y_k+1}(\delta)\le1\bigr)<0.
\]
This proves part (b).
\end{proof}

We are now ready to prove Theorem~\ref{thm:poisson-global-identification}. The proof first identifies $\alpha_+$, as well as $\alpha_j$ for every product with $y_j\ge2$, from the probabilities of selling nothing and of selling exactly one unit. It then identifies $\delta$ in one of two ways. If every product has $y_j\ge2$, we use Lemma~\ref{lem:scale-monotonicity}. Otherwise, some product has $y_j=1$; selling one unit of it causes a stockout, so its one-sale probability depends on $\delta$.

\begin{proof}[Proof of Theorem~\ref{thm:poisson-global-identification}]
Let $p_{\bzero}=\PP(\bm X=\bzero\mid\by)$ and $p_j=\PP(\bm X=\bm e_j\mid\by)$, where $\bm e_j$ is the $j$th unit vector. By Lemma~\ref{lem:sales-dynamics}, while all products are in stock, sales occur at rate $\alpha_+$ and each sale is of product $j$ with probability $\alpha_j/\alpha_+$, so $p_{\bzero}=e^{-\alpha_+}$ identifies $\alpha_+$. If $y_j\ge2$, then $\{\bm X=\bm e_j\}$ is the event that the first sale, at some time $u\in[0,1]$, is of product $j$ and that no further sale occurs, at the unchanged rate $\alpha_+$, during $(u,1]$; hence $p_j=\int_0^1\alpha_je^{-\alpha_+u}e^{-\alpha_+(1-u)}\,du=\alpha_je^{-\alpha_+}$, which identifies $\alpha_j$.

If $y_j\ge2$ for every product, then $\bm\alpha$ is identified, and so is $\delta$, because the expected total sales $\mu_{\by}(\bm\alpha,\delta)$ are strictly decreasing in $\delta$ by Lemma~\ref{lem:scale-monotonicity}(b).

Otherwise, consider a product $j$ with $y_j=1$, and write $a=\alpha_j$. Once product $j$ is sold, it is sold out, and the other products sell at rate $\Lambda_{-j}=(\alpha_+-a)(\alpha_++\delta)/(\alpha_+-a+\delta)$, so, as above,
\begin{equation}
p_j=\varphi(a,\delta)\coloneqq a\int_0^1e^{-\alpha_+u}\,e^{-\Lambda_{-j}(1-u)}\,du,
\label{eq:app-unit-one-sale}
\end{equation}
where, given $\alpha_+$, the function $\varphi$ is the same for every product with unit inventory. Since $0<a<\alpha_+$ because $J\ge2$,
\[
\frac{\partial\Lambda_{-j}}{\partial a}=-\frac{\delta(\alpha_++\delta)}{(\alpha_+-a+\delta)^2}<0,\qquad
\frac{\partial\Lambda_{-j}}{\partial\delta}=-\frac{a(\alpha_+-a)}{(\alpha_+-a+\delta)^2}<0,
\]
so $\varphi$ is strictly increasing in both $a$ and $\delta$; intuitively, after product $j$ sells out, the probability that nothing else is sold is larger when fewer customers substitute, that is, when $\delta$ is larger. Now suppose that $(\bm\alpha',\delta')$ induces the same distribution of $\bm X$ as $(\bm\alpha,\delta)$, so that $\alpha_+'=\alpha_+$ and $\alpha_k'=\alpha_k$ whenever $y_k\ge2$. If $\delta'>\delta$, then, for every product $j$ with $y_j=1$, $\varphi(\alpha_j',\delta')=p_j=\varphi(\alpha_j,\delta)<\varphi(\alpha_j,\delta')$, so $\alpha_j'<\alpha_j$, and summing over all products gives $\alpha_+'<\alpha_+$, a contradiction. By symmetry, $\delta'<\delta$ is also impossible, so $\delta'=\delta$, and then $\alpha_j'=\alpha_j$ for every product with $y_j=1$. In both cases, $(\bm\alpha,\delta)$, and hence $(\lambda,\bv)$, is identified.
\end{proof}

We next turn to Theorem~\ref{thm:poisson-local-concavity}.
Its proof rests on the curvature of the expected log-likelihood of a single period at the true parameter, which we establish first.
Recall that $\bbeta=(\eta,\bv)$ with $\eta=\log\lambda$ and that $\bbeta_0$ is its true value.
We write $L^{\mathrm P}(\bbeta\mid\by,\bx)$ for the likelihood \eqref{eq:poisson-mixture} evaluated at $(\lambda,\bv)=(e^\eta,\bv)$.
For a fixed initial inventory $\by\in\Z_+^J$, let $p_{\bbeta}(\bx\mid\by)=L^{\mathrm P}(\bbeta\mid\by,\bx)$ denote the probability of the terminal sales $\bx$, and define the expected log-likelihood
\[
\mathscr L_{\by}^{\mathrm P}(\bbeta)
=\E_{\bbeta_0}\bigl[\log L^{\mathrm P}(\bbeta\mid\by,\bm X)\bigm|\by\bigr]
=\sum_{\bx}p_{\bbeta_0}(\bx\mid\by)\log p_{\bbeta}(\bx\mid\by),
\]
where the sum is over the finitely many feasible sales vectors, that is, those with $0\le x_j\le y_j$ for every product $j$.
Each feasible $\bx$ has positive probability; for example, exactly $\sum_jx_j$ customers may arrive and buy these units.
Moreover, $p_{\bbeta}(\bx\mid\by)$ is infinitely differentiable in $\bbeta$.
Indeed, by Lemma~\ref{lem:sales-dynamics} and its proof, it is the time-one transition probability from $\bzero$ to $\bx$ of the continuous-time Markov chain on the feasible sales vectors that moves from $\bx'$ to $\bx'+\bm e_j$ at rate $\lambda\theta_j/(1+\sum_{k:x_k'<y_k}\theta_k)$ whenever $x_j'<y_j$, where $\bm e_j$ is the $j$th unit vector.
It is therefore an entry of the exponential of a finite generator matrix whose entries are smooth in $\bbeta$, and the matrix exponential is a real-analytic function of the matrix entries.

\begin{lemma}[Expected log-likelihood curvature]
\label{lem:poisson-population-concavity}
\begin{itemize}
\item[(a)] For every $\by\in\Z_+^J$, $\nabla^2\mathscr L_{\by}^{\mathrm P}(\bbeta_0)\preceq0$.
\item[(b)] If $J\ge2$ and $\by\in\Z_{++}^J$, then $\nabla^2\mathscr L_{\by}^{\mathrm P}(\bbeta_0)\prec0$.
\end{itemize}
\end{lemma}

In statistical terms, $-\nabla^2\mathscr L_{\by}^{\mathrm P}(\bbeta_0)$ is the Fisher information of a period with initial inventory $\by$, and part (b) states that it is nonsingular.
To prove the lemma, we express $\bm d^\top\nabla^2\mathscr L_{\by}^{\mathrm P}(\bbeta_0)\bm d$ as a negative weighted sum of squared directional derivatives of terminal-sales probabilities.
We then apply the argument for Theorem~\ref{thm:poisson-global-identification} to first order to show that, for every $\bm d\ne\bzero$, at least one of these derivatives is nonzero at $\bbeta_0$.

\begin{proof}[Proof of Lemma~\ref{lem:poisson-population-concavity}]
We first prove part (a).
Fix $\by\in\Z_+^J$.
Since the sum defining $\mathscr L_{\by}^{\mathrm P}$ is finite and its terms are smooth, it can be differentiated term by term, with the weights $p_{\bbeta_0}(\bx\mid\by)$ held fixed.
Hence, for every direction $\bm d\in\R^{J+1}$,
\[
\begin{aligned}
\bm d^\top\nabla^2\mathscr L_{\by}^{\mathrm P}(\bbeta_0)\bm d
&=\sum_{\bx}\bm d^\top\nabla^2p_{\bbeta_0}(\bx\mid\by)\bm d
-\sum_{\bx}\frac{\{\bm d^\top\nabla p_{\bbeta_0}(\bx\mid\by)\}^2}{p_{\bbeta_0}(\bx\mid\by)}\\
&=-\sum_{\bx}\frac{\{\bm d^\top\nabla p_{\bbeta_0}(\bx\mid\by)\}^2}{p_{\bbeta_0}(\bx\mid\by)}
\le0.
\end{aligned}
\]
The first equality evaluates $\nabla^2\log p=\nabla^2p/p-\nabla p\,\nabla p^\top/p^2$ at $\bbeta_0$, where the weight $p_{\bbeta_0}(\bx\mid\by)$ cancels one factor of $p$.
The second equality holds because $\sum_{\bx}\nabla^2p_{\bbeta}(\bx\mid\by)=\bm0$, which follows by differentiating $\sum_{\bx}p_{\bbeta}(\bx\mid\by)=1$ twice.
This proves part (a).

We next prove part (b), so let $J\ge2$ and $\by\in\Z_{++}^J$.
We express the zero- and one-sale probabilities through the pre-stockout purchase rates $\alpha_j$, their sum $\alpha_+$ and the no-purchase rate $\delta$ in \eqref{eq:alpha-delta-parameterization}, which satisfy $\lambda=\alpha_++\delta$ and $\theta_j=\alpha_j/\delta$.

Fix a nonzero direction $\bm d$ and suppose, for a contradiction, that $\bm d^\top\nabla^2\mathscr L_{\by}^{\mathrm P}(\bbeta_0)\bm d=0$. Every denominator in the curvature identity is positive, so every squared term must be zero. Consider the path $\bbeta(t)=\bbeta_0+t\bm d$, and use a dot to denote differentiation at $t=0$. The zero-curvature assumption therefore gives
\[
\dot p_{\bx}
=\left.\frac{d}{dt}p_{\bbeta(t)}(\bx\mid\by)\right|_{t=0}
=\bm d^\top\nabla p_{\bbeta_0}(\bx\mid\by)
=0
\]
for every feasible $\bx$. The quantities $\dot\alpha_j$, $\dot\delta$, and $\dot\alpha_+$ denote derivatives along this same path. We will show that they must all be zero, and then use the inverse parameter relations to obtain a contradiction.

The zero-sale outcome is feasible, so its probability $p_{\bzero}$ also has zero derivative. All products stay in stock until the first sale, which occurs at rate $\alpha_+$, so $p_{\bzero}=e^{-\alpha_+}$, and hence
\[
0=\dot p_{\bzero}=-e^{-\alpha_+}\dot\alpha_+.
\]
The exponential factor is positive, so $\dot\alpha_+=0$. Next, let $p_j$ denote the probability of selling exactly one unit of product $j$ and no other products. When $y_j\ge2$, this event causes no stockout, and $p_j=\alpha_je^{-\alpha_+}$. Therefore,
\[
0=\dot p_j=e^{-\alpha_+}(\dot\alpha_j-\alpha_j\dot\alpha_+)
=e^{-\alpha_+}\dot\alpha_j,
\]
which gives $\dot\alpha_j=0$ for every product with at least two units of inventory.

If every product has at least two units, all purchase-rate derivatives are now zero. It remains to determine $\dot\delta$. Expected total sales are given by $\mu_{\by}(\bm\alpha,\delta)=\sum_{\bx}(\sum_jx_j)p_{\bx}$. Since every $\dot p_{\bx}=0$, this expectation also has zero derivative. Moreover, $\mu_{\by}$ is smooth in $(\bm\alpha,\delta)$, because each $p_{\bx}$ is smooth in $\bbeta$ and $\bbeta$ is a smooth function of $(\bm\alpha,\delta)$. Since all $\dot\alpha_j=0$, the chain rule gives
\[
0=\sum_{\bx}\left(\sum_jx_j\right)\dot p_{\bx}
=\dot\mu_{\by}
=\partial_\delta\mu_{\by}\,\dot\delta.
\]
Lemma~\ref{lem:scale-monotonicity}(b) gives $\partial_\delta\mu_{\by}<0$, and therefore $\dot\delta=0$.

Suppose instead that at least one product has unit inventory. For such a product, its one-sale event depletes its inventory, so the probability also depends on the purchase rates after that stockout. Equation~\eqref{eq:app-unit-one-sale} writes this probability as $p_j=\varphi(\alpha_j,\delta)$ when $\alpha_+$ is held fixed. Differentiating \eqref{eq:app-unit-one-sale} under the integral sign, which is justified because the integrand is smooth on the compact interval $[0,1]$, gives the partial derivatives with respect to the first and second arguments,
\[
\begin{aligned}
\varphi_a(a,\delta)&=\int_0^1e^{-\alpha_+u-\Lambda_{-j}(1-u)}\Bigl\{1-a(1-u)\frac{\partial\Lambda_{-j}}{\partial a}\Bigr\}\,du,\\
\varphi_\delta(a,\delta)&=-a\,\frac{\partial\Lambda_{-j}}{\partial\delta}\int_0^1(1-u)\,e^{-\alpha_+u-\Lambda_{-j}(1-u)}\,du.
\end{aligned}
\]
Both are strictly positive, because $\partial\Lambda_{-j}/\partial a<0$ and $\partial\Lambda_{-j}/\partial\delta<0$ by the proof of Theorem~\ref{thm:poisson-global-identification}. Along our parameter path, $\dot\alpha_+=0$, so changes in $\alpha_+$ contribute no term to the first derivative of $p_j$. Consequently,
\[
0=\dot p_j=\varphi_a(\alpha_j,\delta)\dot\alpha_j+\varphi_\delta(\alpha_j,\delta)\dot\delta.
\]
If $\dot\delta>0$, this equality requires $\dot\alpha_j<0$ for every unit-inventory product. If $\dot\delta<0$, it requires $\dot\alpha_j>0$ for every such product. Both possibilities contradict $\dot\alpha_+=\sum_j\dot\alpha_j=0$, because the purchase-rate derivatives of all other products are already zero. Thus $\dot\delta=0$, and the displayed equality then gives $\dot\alpha_j=0$ for each unit-inventory product.

In both cases, $\dot\alpha_j=0$ for every product and $\dot\delta=0$. To return to the original parameters, use
\[
\eta=\log(\alpha_++\delta),\qquad
v_j=\log\alpha_j-\log\delta.
\]
Differentiating along the same path gives
\[
\dot\eta=\frac{\dot\alpha_++\dot\delta}{\alpha_++\delta}=0,\qquad
\dot v_j=\frac{\dot\alpha_j}{\alpha_j}-\frac{\dot\delta}{\delta}=0,\quad j\in[J].
\]
Thus $\dot{\bbeta}=\bzero$. However, the definition $\bbeta(t)=\bbeta_0+t\bm d$ gives $\dot{\bbeta}=\bm d\ne\bzero$, a contradiction. The expected directional curvature is therefore strictly negative in every nonzero direction, which proves part (b).
\end{proof}

We prove Theorem~\ref{thm:poisson-local-concavity} in three steps. First, we use the preceding lemma and the positive frequency of periods in which all products are initially stocked to establish a uniform negative bound on the average expected Hessian. Second, we show that the sample Hessian at the true parameter concentrates around its expectation. Finally, continuity of the Hessians extends this negative curvature to a fixed neighborhood of the true parameter.

\begin{proof}[Proof of Theorem~\ref{thm:poisson-local-concavity}]
For period $n$, write $L_n^{\mathrm P}(\bbeta)=L^{\mathrm P}(\bbeta\mid\by_n,\bx_n)$, so that the average log-likelihood in the theorem is $\mathcal L_N^{\mathrm P}(\bbeta)=N^{-1}\sum_{n=1}^N\log L_n^{\mathrm P}(\bbeta)$.
Since the second derivative of $\mathcal L_N^{\mathrm P}$ at $\bbeta$ along a unit direction $\bm d$ is $\bm d^\top\nabla^2\mathcal L_N^{\mathrm P}(\bbeta)\bm d$, we need to find $r,\gamma>0$, independent of $N$, such that, with probability approaching one, $\nabla^2\mathcal L_N^{\mathrm P}(\bbeta)\preceq-\gamma I_{J+1}$ at every point of the ball $\{\bbeta:\|\bbeta-\bbeta_0\|_2\le r\}$.
We first obtain a negative bound for the expected Hessian at $\bbeta_0$. Let $\mathcal Y$ denote the finite support of the initial inventories, and write $H_{\by}=\nabla^2\mathscr L_{\by}^{\mathrm P}(\bbeta_0)$, which equals $\E_{\bbeta_0}[\nabla^2_{\bbeta}\log L^{\mathrm P}(\bbeta_0\mid\by,\bm X)\mid\by]$ because the expectation is a finite sum. By Lemma~\ref{lem:poisson-population-concavity}, $H_{\by}\preceq0$ for every $\by$, and, since $J\ge2$, $H_{\by}\prec0$ whenever $\by\in\Z_{++}^J$. The assumed positive proportion of periods stocking every product ensures that $\mathcal Y\cap\Z_{++}^J$ is nonempty. Since this set is finite and each $-H_{\by}$ with $\by\in\mathcal Y\cap\Z_{++}^J$ is positive definite, the smallest eigenvalue among these matrices, denoted by $\gamma_0$, is positive. Thus $H_{\by}\preceq-\gamma_0 I_{J+1}$ for every $\by\in\mathcal Y\cap\Z_{++}^J$.

By assumption, there exists $\epsilon_0>0$ such that at least $\epsilon_0N$ periods stock every product for all sufficiently large $N$. Set $\gamma=\epsilon_0\gamma_0/2$. Then, for all sufficiently large $N$,
\[
\begin{aligned}
\frac1N\sum_{n=1}^NH_{\by_n}
&\preceq\frac1N\sum_{\substack{1\le n\le N\\\by_n\in\Z_{++}^J}}H_{\by_n}\\
&\preceq-\gamma_0\frac{\#\{n\le N:\by_n\in\Z_{++}^J\}}{N}I_{J+1}\\
&\preceq-\epsilon_0\gamma_0 I_{J+1}
=-2\gamma I_{J+1}.
\end{aligned}
\]
The first inequality omits the remaining periods, whose expected Hessians are negative semidefinite; the second uses $H_{\by}\preceq-\gamma_0I_{J+1}$ for every $\by\in\mathcal Y\cap\Z_{++}^J$; and the third uses the assumed proportion of such periods.
We will bound the sampling fluctuation and the change in the Hessian over a neighborhood by $\gamma/2$ each, leaving a negative bound of $-\gamma I_{J+1}$.

We next control the sampling fluctuation at $\bbeta_0$. A feasible inventory--sales pair $(\by,\bx)$ consists of an initial inventory vector $\by\in\mathcal Y$ and a terminal-sales vector $\bx\in\Z_+^J$ satisfying $0\le x_j\le y_j$ for every product $j$. For each $\by$, there are $\prod_{j=1}^J(y_j+1)$ such sales vectors. Since $\mathcal Y$ is finite, the total number of feasible pairs is also finite.
Let $\|\cdot\|_{\mathrm{op}}$ denote the matrix operator norm induced by the Euclidean norm. For each feasible pair, the likelihood is positive and twice continuously differentiable in a neighborhood of $\bbeta_0$. Its logarithm therefore has a Hessian with finite entries at $\bbeta_0$, and this matrix has a finite operator norm. Taking the maximum over the finitely many feasible pairs gives
\[
M:=\max_{\substack{\by\in\mathcal Y,\;\bx\in\Z_+^J\\0\le\bx\le\by}}\left\|\nabla^2_{\bbeta}\log L^{\mathrm P}(\bbeta_0\mid\by,\bx)\right\|_{\mathrm{op}}<\infty.
\]
The constant $M$ depends only on the fixed parameter $\bbeta_0$ and the inventory support $\mathcal Y$, and is independent of $N$ and the realized sample.

Define
\[
\Delta_n:=\nabla^2_{\bbeta}\log L_n^{\mathrm P}(\bbeta_0)-H_{\by_n}.
\]
Because the inventories are nonrandom and the terminal sales are independent across periods, the matrices $\Delta_n$ are independent and have zero means. Since the squared Frobenius norm is the sum of the squared matrix entries, the variance identity gives
\[
\begin{aligned}
\E_{\bbeta_0}\|\Delta_n\|_{\mathrm F}^2
&=\E_{\bbeta_0}\left\|\nabla^2_{\bbeta}\log L_n^{\mathrm P}(\bbeta_0)\right\|_{\mathrm F}^2-\|H_{\by_n}\|_{\mathrm F}^2\\
&\le(J+1)M^2,
\end{aligned}
\]
where the last inequality uses $\|\cdot\|_{\mathrm F}^2\le(J+1)\|\cdot\|_{\mathrm{op}}^2$ for $(J+1)\times(J+1)$ matrices. Independence and zero means eliminate the cross terms when expanding the squared Frobenius norm of the sum. Therefore,
\[
\E_{\bbeta_0}\left\|\frac1N\sum_{n=1}^N\Delta_n\right\|_{\mathrm{op}}^2
\le\E_{\bbeta_0}\left\|\frac1N\sum_{n=1}^N\Delta_n\right\|_{\mathrm F}^2
=\frac1{N^2}\sum_{n=1}^N\E_{\bbeta_0}\|\Delta_n\|_{\mathrm F}^2
\le\frac{(J+1)M^2}{N}.
\]
Markov's inequality applied to the squared operator norm now yields
\[
\PP_{\bbeta_0}\left(\left\|\frac1N\sum_{n=1}^N\Delta_n\right\|_{\mathrm{op}}>\frac{\gamma}{2}\right)
\le\frac{4(J+1)M^2}{N\gamma^2}\longrightarrow0.
\]
A symmetric matrix is bounded above by its operator norm times the identity. Thus, on the event where this norm is at most $\gamma/2$, for all sufficiently large $N$,
\[
\begin{aligned}
\nabla^2_{\bbeta}\mathcal L_N^{\mathrm P}(\bbeta_0)
&=\frac1N\sum_{n=1}^NH_{\by_n}+\frac1N\sum_{n=1}^N\Delta_n\\
&\preceq-2\gamma I_{J+1}+\frac{\gamma}{2}I_{J+1}
=-\frac{3\gamma}{2}I_{J+1}.
\end{aligned}
\]

It remains to extend this bound to a fixed neighborhood of $\bbeta_0$. For each feasible inventory--sales pair, continuity of the log-likelihood Hessian at $\bbeta_0$ gives a positive radius within which its change in operator norm is at most $\gamma/2$. Taking the minimum of these radii over the finitely many pairs gives a common $r>0$, independent of $N$ and the realized sample, such that
\[
\max_{\substack{\by\in\mathcal Y,\;\bx\in\Z_+^J\\0\le\bx\le\by}}
\sup_{\|\bbeta-\bbeta_0\|_2\le r}
\left\|\nabla^2_{\bbeta}\log L^{\mathrm P}(\bbeta\mid\by,\bx)-\nabla^2_{\bbeta}\log L^{\mathrm P}(\bbeta_0\mid\by,\bx)\right\|_{\mathrm{op}}
\le\frac{\gamma}{2}.
\]
Averaging over periods and applying the triangle inequality yields, for every sample,
\[
\sup_{\|\bbeta-\bbeta_0\|_2\le r}
\left\|\nabla^2_{\bbeta}\mathcal L_N^{\mathrm P}(\bbeta)-\nabla^2_{\bbeta}\mathcal L_N^{\mathrm P}(\bbeta_0)\right\|_{\mathrm{op}}
\le\frac{\gamma}{2}.
\]
Consequently, on the same event used above, every $\bbeta$ with $\|\bbeta-\bbeta_0\|_2\le r$ satisfies
\[
\begin{aligned}
\nabla^2_{\bbeta}\mathcal L_N^{\mathrm P}(\bbeta)
&\preceq\nabla^2_{\bbeta}\mathcal L_N^{\mathrm P}(\bbeta_0)+\frac{\gamma}{2}I_{J+1}\\
&\preceq-\frac{3\gamma}{2}I_{J+1}+\frac{\gamma}{2}I_{J+1}
=-\gamma I_{J+1}.
\end{aligned}
\]
This event has probability approaching one, and both $r$ and $\gamma$ are independent of $N$, which completes the proof.
\end{proof}

The proof also yields the two properties used in the discussion of gradient-based methods in Section~\ref{subsubsec:poisson-local-concavity}.
First, the Hessian of $\mathcal L_N^{\mathrm P}$ is bounded on the ball for every sample: the Hessian of each period's log-likelihood lies within $\gamma/2$ of its value at $\bbeta_0$, whose operator norm is at most $M$, so $\|\nabla^2_{\bbeta}\mathcal L_N^{\mathrm P}(\bbeta)\|_{\mathrm{op}}\le M+\gamma/2$ whenever $\|\bbeta-\bbeta_0\|_2\le r$.
Second, suppose that $\mathcal L_N^{\mathrm P}$ is $\gamma$-strongly concave on the ball, and let $\widehat\bbeta$ be its maximizer over the ball, which exists because the ball is compact.
Fix $\bbeta$ in the ball and consider the derivative of $t\mapsto\mathcal L_N^{\mathrm P}(\widehat\bbeta+t(\bbeta-\widehat\bbeta))$ on $[0,1]$.
At $t=0$, this derivative equals $\nabla\mathcal L_N^{\mathrm P}(\widehat\bbeta)^\top(\bbeta-\widehat\bbeta)\le0$, because $\widehat\bbeta$ maximizes $\mathcal L_N^{\mathrm P}$ over the convex ball.
Its own derivative is at most $-\gamma\|\bbeta-\widehat\bbeta\|_2^2$ on $[0,1]$, because the segment from $\widehat\bbeta$ to $\bbeta$ lies in the ball.
Integrating over $[0,1]$ gives $\nabla\mathcal L_N^{\mathrm P}(\bbeta)^\top(\bbeta-\widehat\bbeta)\le-\gamma\|\bbeta-\widehat\bbeta\|_2^2$, and the Cauchy--Schwarz inequality yields $\|\bbeta-\widehat\bbeta\|_2\le\|\nabla\mathcal L_N^{\mathrm P}(\bbeta)\|_2/\gamma$.

Finally, we verify the nonconcavity claim using the zero-sale observation, for which the likelihood is explicit.

\begin{lemma}[Nonconcavity despite identifiability]
\label{lem:identified-nonconcavity}
Under the conditions of Theorem~\ref{thm:poisson-global-identification}, consider the zero-sale observation $\bx=\bzero$, and let $\bv=\bar\bv+r\bm1$ for a fixed $\bar\bv\in\R^J$, where $r\in\R$ and $\bm1$ is the all-ones vector. Then the Hessian of $\log L^{\mathrm P}$ with respect to $(\eta,r)$ is indefinite at every $(\eta,r)$, so the log-likelihood of a single period is not concave in $(\eta,\bv)$.
\end{lemma}

\begin{proof}[Proof of Lemma~\ref{lem:identified-nonconcavity}]
By the proof of Theorem~\ref{thm:poisson-global-identification}, $\log L^{\mathrm P}=-\alpha_+=-e^{\eta}b(r)$, where $b(r)=\Theta(r)/\{1+\Theta(r)\}$, with $\Theta(r)=\sum_je^{\bar v_j+r}$, is the probability that a customer buys some product when all products are in stock. Since $\Theta'=\Theta$, we have $b'=b(1-b)$ and $b''=b(1-b)(1-2b)$, so the Hessian with respect to $(\eta,r)$ is
\[
-e^{\eta}
\begin{pmatrix}
b & b(1-b)\\
b(1-b) & b(1-b)(1-2b)
\end{pmatrix},
\]
with determinant $e^{2\eta}\{b^2(1-b)(1-2b)-b^2(1-b)^2\}=-e^{2\eta}b^3(1-b)<0$. A symmetric $2\times2$ matrix with a negative determinant is indefinite, and since the restriction of a concave function to an affine set is concave, $\log L^{\mathrm P}$ is not concave in $(\eta,\bv)$.
\end{proof}

\section{Computational Methods for Gradient Evaluation}
\label{app:score-computation}

This appendix describes how we evaluate the likelihood and its gradient numerically, as summarized in Section~\ref{subsec:score-computation}.
By Theorems~\ref{thm:known-gamma}--\ref{thm:poisson-score}, all the required quantities are built from one-dimensional expectations with respect to a Gamma distribution.
Appendix~\ref{app:laguerre-evaluation} approximates these expectations by generalized Gauss--Laguerre quadrature.
Appendices~\ref{app:tail-evaluation} and~\ref{app:phi-evaluation} explain how we evaluate the scalar functions at the quadrature nodes, namely $\overline F_y$ and $h_y$, and $\Phi_s$ and $\psi_s$, and bound the resulting truncation errors.

\subsection{Generalized Gauss--Laguerre Evaluation}
\label{app:laguerre-evaluation}

Both $R^{\mathrm K}$ and $R^{\mathrm P}$, as well as the expectations under $W$ and $\bar W$, are built from integrals of the form $\int_0^\infty u^\alpha e^{-u}f(u)\,du/\alpha!$ for a nonnegative integer $\alpha$.
Generalized Gauss--Laguerre quadrature is designed for integrals with the weight $u^\alpha e^{-u}$.
Let $(u_q^{(\alpha)},w_q^{(\alpha)})_{q=1}^Q$ denote the nodes and weights of this quadrature rule, and set $\bar w_q^{(\alpha)}=w_q^{(\alpha)}/\alpha!$ to account for the normalizing constant of the Gamma density. Then
\begin{equation}
\frac1{\alpha!}\int_0^\infty u^\alpha e^{-u}f(u)\,du
\approx\sum_{q=1}^Q\bar w_q^{(\alpha)}f(u_q^{(\alpha)}).
\label{eq:laguerre-rule}
\end{equation}
For observation $n$, let $L_n^r=C_n^rR_n^r$ denote the likelihood factorization in setting $r\in\{\mathrm K,\mathrm P\}$. The exponent of $u$ in the quadrature weight is $\alpha_n^{\mathrm K}=m_n+c_n$, where $m_n=T_n-s_n$, under known arrivals and $\alpha_n^{\mathrm P}=c_n$ under Poisson arrivals. The corresponding Gamma shape parameter is one plus this exponent. Fix $r$ and write $u_{nq}=u_q^{(\alpha_n^r)}$ and $\bar w_{nq}=\bar w_q^{(\alpha_n^r)}$. When $\cB_n=\varnothing$, we use the closed-form likelihood and gradient, setting $\widehat R_{n,Q}^r=R_n^r$ and $\widehat L_{n,Q}^r=L_n^r$. The quadrature formulas below apply when $\cB_n\ne\varnothing$.

To write both settings together, define
\[
\Xi_{\mathrm K,n}(u)=1,\qquad
\Xi_{\mathrm P,n}(u)=\Phi_{s_n}\left(\frac{\lambda u}{1+\sum_{j\in\cC_n}\theta_j}\right).
\]
Treating scalar evaluations at the nodes as exact, form the positive unnormalized weights
\begin{equation}
\xi_{nq}^r=\bar w_{nq}\Xi_{r,n}(u_{nq})
\prod_{j\in\cB_n}\overline F_{y_{j,n}}(\rho_{j,n}u_{nq}).
\label{eq:finite-node-masses}
\end{equation}
Their sum approximates the Gamma expectation, giving $\widehat R_{n,Q}^r=\sum_{q=1}^Q\xi_{nq}^r$ and $\widehat L_{n,Q}^r=C_n^r\widehat R_{n,Q}^r$.
Normalizing these weights gives
\begin{equation*}
\widetilde w_{nq}^r=\frac{\xi_{nq}^r}{\sum_{\ell=1}^Q\xi_{n\ell}^r}.
\end{equation*}
The unnormalized weights and their sum are evaluated on the log scale. The normalized weights $\widetilde w_{nq}^r$ approximate the distributions of $W$ and $\bar W$ used in Theorems~\ref{thm:known-score} and~\ref{thm:poisson-score}. In particular, the expectation for sold-out product $j$ is approximated by $\sum_{q=1}^Q\widetilde w_{nq}^r h_{y_{j,n}}(\rho_{j,n}u_{nq})$, and under Poisson arrivals,
\begin{equation*}
\widehat{\overline m}_{n,Q}
=\sum_{q=1}^Q\widetilde w_{nq}^{\mathrm P}
\psi_{s_n}\left(\frac{\lambda u_{nq}}{1+\sum_{j\in\cC_n}\theta_j}\right).
\end{equation*}
Substituting these quantities into the two gradient theorems gives the gradient computed by quadrature.

Let $\nabla$ denote the gradient with respect to $\bv$ for $r=\mathrm K$ and with respect to $(\eta,\bv^\top)^\top$ for $r=\mathrm P$. The nodes and the weights $\bar w_{nq}$ depend only on the exponent $\alpha_n^r$, which is determined by the observation, and remain fixed as the parameters vary. Differentiating the finite sum therefore gives
\begin{equation}
\nabla\log\widehat L_{n,Q}^r
=\nabla\log C_n^r
+\sum_{q=1}^Q\widetilde w_{nq}^r\nabla\log\left\{
\Xi_{r,n}(u_{nq})\prod_{j\in\cB_n}\overline F_{y_{j,n}}(\rho_{j,n}u_{nq})\right\}.
\label{eq:finite-node-score}
\end{equation}
Thus, with exact scalar evaluations, the gradient computed by quadrature is exactly the gradient of the approximate log-likelihood $\log\widehat L_{n,Q}^r$.
In other words, the optimizer works with a gradient that is consistent with the objective it maximizes.

\paragraph{Choice of the Gamma shape parameter.}
Under Poisson arrivals, the reported implementation increases the Gamma shape parameter by a fixed nonnegative integer to move the quadrature nodes toward the region contributing most to the integral. For any fixed integer $d\ge0$,
\begin{equation}
\frac1{c_n!}\int_0^\infty u^{c_n}e^{-u}f(u)\,du
=\frac{(c_n+d)!}{c_n!}
\E_{U\sim\GammaDist(c_n+d+1,1)}[U^{-d}f(U)].
\label{eq:shifted-gamma-reference}
\end{equation}
The factor $U^{-d}$ and the factorial ratio exactly correct the change of Gamma density. The corresponding nodes and coefficients are
\begin{equation*}
 u_{nq}^{(d)}=u_q^{(c_n+d)},\qquad
 \omega_{nq}^{(d)}=\bar w_q^{(c_n+d)}\frac{(c_n+d)!}{c_n!}
 (u_{nq}^{(d)})^{-d}.
\end{equation*}
Replacing $u_{nq}$ and $\bar w_{nq}$ by these quantities throughout the quadrature formulas above gives the quadrature rule with the increased Gamma shape parameter. Because $d$ is chosen before optimization and held fixed, the resulting nodes and coefficients remain parameter-independent, so \eqref{eq:finite-node-score} still holds. The coefficients $\omega_{nq}^{(d)}$ are positive but need not sum to one; normalized weights are obtained only after forming the unnormalized weights $\xi_{nq}^r$ and dividing each by their sum.

\subsection{Evaluation of Poisson Tails and Their Derivatives}
\label{app:tail-evaluation}

For $y\ge1$ and $\mu>0$, the Poisson tail equals the regularized lower incomplete gamma function,
$\overline F_y(\mu)=\{(y-1)!\}^{-1}\int_0^\mu t^{y-1}e^{-t}\,dt$.
When this evaluation is numerically unstable for a small tail, we use the positive series
\begin{equation}
\overline F_y(\mu)=\pi_y(\mu)\sum_{\ell=0}^\infty a_\ell,\qquad
 a_0=1,\qquad a_{\ell+1}=a_\ell\frac{\mu}{y+\ell+1}.
\label{eq:tail-series-app}
\end{equation}
This series follows from $\overline F_y(\mu)=\sum_{\ell\ge0}\pi_{y+\ell}(\mu)$, because $a_\ell=\pi_{y+\ell}(\mu)/\pi_y(\mu)$.
Since $\mu\pi_{y-1}(\mu)=y\pi_y(\mu)$, the same terms give
$h_y(\mu)=y/\sum_{\ell=0}^\infty a_\ell$ without dividing two small probabilities.

The adjacent-term ratios decrease with $\ell$. Once $r_L=\mu/(y+L+1)<1$, we have $a_{L+k}\le a_Lr_L^k$ for $k\ge1$, and hence
\[
0\le\sum_{\ell>L}a_\ell\le B_L:=\frac{a_Lr_L}{1-r_L}.
\]
Writing $S_L=\sum_{\ell=0}^La_\ell$, the exact sum lies between $S_L$ and $S_L+B_L$. Taking logarithms and reciprocals gives
\[
\begin{aligned}
0&\le\log\overline F_y(\mu)-\log\{\pi_y(\mu)S_L\}
\le\log(1+B_L/S_L),\\
0&\le\frac{y}{S_L}-h_y(\mu)\le\frac{yB_L}{S_L^2}.
\end{aligned}
\]
Thus one recurrence supplies both scalar quantities and explicit truncation bounds.

\subsection{Stable Evaluation of
\texorpdfstring{\(\Phi_s\) and \(\psi_s\)}{Phi and psi}}
\label{app:phi-evaluation}

For $s\in\Z_+$ and $z>0$, write
\begin{equation}
 t_\ell=\frac{s!z^\ell}{\ell!(s+\ell)!},\qquad
 \Phi_s(z)=\sum_{\ell=0}^\infty t_\ell,\qquad
 \psi_s(z)=\frac{\sum_{\ell=0}^\infty\ell t_\ell}{\sum_{\ell=0}^\infty t_\ell}.
\label{eq:phi-positive-series-app}
\end{equation}
The ratio $t_{\ell+1}/t_\ell=z/\{(\ell+1)(s+\ell+1)\}$ decreases with $\ell$, so the terms are nondecreasing up to their maximum and nonincreasing thereafter. We retain an interval $\ell_-,\ldots,\ell_+$ containing all indices attaining the maximum, including both indices when two adjacent terms tie, and use the same interval for both quantities. Define $S_0=\sum_{\ell=\ell_-}^{\ell_+}t_\ell$ and $S_1=\sum_{\ell=\ell_-}^{\ell_+}\ell t_\ell$.

To bound the omitted terms with $\ell>\ell_+$, let $r_+=z/\{(\ell_++1)(s+\ell_++1)\}<1$. Since $t_{\ell_++k}\le t_{\ell_+}r_+^k$ for $k\ge1$, summing the geometric bounds for the omitted terms, both unweighted and weighted by their indices, gives
\[
 B_{0,\mathrm R}=\frac{t_{\ell_+}r_+}{1-r_+},\qquad
 B_{1,\mathrm R}=t_{\ell_+}\left\{
 \frac{\ell_+r_+}{1-r_+}+\frac{r_+}{(1-r_+)^2}\right\}.
\]
For the omitted terms with $\ell<\ell_-$, when $\ell_->0$, let $r_-=\ell_-(s+\ell_-)/z<1$. The bound $t_{\ell_--k}\le t_{\ell_-}r_-^k$ for $1\le k\le\ell_-$ gives
\[
 B_{0,\mathrm L}=\frac{t_{\ell_-}r_-}{1-r_-},\qquad
 B_{1,\mathrm L}=(\ell_--1)B_{0,\mathrm L},
\]
because each of these omitted indices is at most $\ell_--1$. When $\ell_-=0$, no terms are omitted below the retained interval, so both corresponding bounds are zero.

Let $B_i=B_{i,\mathrm L}+B_{i,\mathrm R}$ for $i=0,1$. The omitted denominator and numerator terms in \eqref{eq:phi-positive-series-app} lie in $[0,B_0]$ and $[0,B_1]$, respectively. Taking logarithms, and subtracting the exact and truncated ratios using $\Phi_s(z)\ge S_0$, yields
\begin{equation}
\begin{aligned}
0&\le\log\Phi_s(z)-\log S_0
\le\log\left(1+\frac{B_0}{S_0}\right),\\
\left|\psi_s(z)-\frac{S_1}{S_0}\right|
&\le\frac{B_1+(S_1/S_0)B_0}{S_0}.
\end{aligned}
\label{eq:phi-psi-truncation-certificate}
\end{equation}
Thus the approximations are $\widehat\Phi_s(z)=S_0$ and $\widehat\psi_s(z)=S_1/S_0$.

The reported implementation constructs the interval either by a forward recurrence from $\ell=0$ or by expanding around a term attaining the maximum, with relative truncation tolerance $B_0/S_0\le\varepsilon$, where~${\varepsilon=2\times10^{-15}}$.
Here $B_0/S_0$ is an upper bound on the omitted sum divided by the retained sum. The implementation evaluates the sums on the log scale, computing $\log\widehat\Phi_s(z)$ as the logarithm of the sum of the retained terms and $\widehat\psi_s(z)$ as the ratio $S_1/S_0$. At most $10{,}000$ terms are retained. The stated truncation bound is met when this tolerance is satisfied; reaching the term cap alone does not ensure it. No separate tolerance is imposed on $\psi_s$, whose error is bounded by \eqref{eq:phi-psi-truncation-certificate}. At $z=0$, set $\widehat\Phi_s(0)=\Phi_s(0)=1$ and $\widehat\psi_s(0)=\psi_s(0)=0$.

For a fixed retained interval, $z\,d\log S_0/dz=S_1/S_0$. This derivative identity applies while that interval remains fixed; the gradient identity in \eqref{eq:finite-node-score} assumes exact scalar evaluations.

\section{Calibration of the Target Stockout Rate}
\label{app:stockout-calibration}

The stockout rate is a key factor in the numerical study: it determines how often the assortment changes within a selling period, and hence how much stockout-induced substitution the data contain.
To compare settings with different numbers of products and customers at the same stockout frequency, we fix the target stockout rate $\zeta$ and adjust the inventory multiplier $c_s$ in \eqref{eq:inventory_rule} accordingly (Section~\ref{subsec:numerical_design}).
Because the stockout rate implied by a given $c_s$ depends on the attractions, the arrivals, and the substitution within each period, it is not available in closed form, so we calibrate $c_s$ by simulation, separately for each $(J,\kappa)$ and $\zeta\in\{0.25,0.50\}$.
The calibration relies on the fact that, in a common simulation, larger inventories cannot lead to more stockouts.
We first establish this fact and then describe the search.

\begin{lemma}[Monotonicity of stockouts in inventory]
\label{lem:stockout-monotone}
Let $\by'\ge\by\ge\bm1$ be two inventory vectors.
Generate the two sales processes from the continuous-time representation in Section~\ref{sec:likelihoods}, using the same Poisson processes $\Pi_0,\ldots,\Pi_J$ and the same number of customers $T$.
Then every product that sells out under $\by'$ also sells out under $\by$.
\end{lemma}

\begin{proof}[Proof of Lemma~\ref{lem:stockout-monotone}]
We use superscripts to indicate the initial inventory.
Since $\min\{y_j',\Pi_j(\tau)\}\ge\min\{y_j,\Pi_j(\tau)\}$ for every $j$ and $\tau$, \eqref{eq:capped-counts} gives
\[
N_{\mathrm{tot}}^{\by'}(\tau)
=\Pi_0(\tau)+\sum_j\min\{y_j',\Pi_j(\tau)\}
\ge\Pi_0(\tau)+\sum_j\min\{y_j,\Pi_j(\tau)\}
=N_{\mathrm{tot}}^{\by}(\tau).
\]
In other words, choices accumulate at least as fast under $\by'$, so the $T$th choice occurs no later: $\sigma_T^{\by'}\le\sigma_T^{\by}$.
If product $j$ sells out under $\by'$, then $\Pi_j(\sigma_T^{\by'})\ge y_j'$.
Since $\Pi_j$ is nondecreasing,
\[
\Pi_j(\sigma_T^{\by})\ge\Pi_j(\sigma_T^{\by'})\ge y_j'\ge y_j,
\]
so product $j$ also sells out under $\by$.
By Lemma~\ref{lem:clock-embedding}, each of the two processes has the correct distribution of terminal sales, so the construction is a valid coupling.
\end{proof}

The same conclusion holds in the Poisson-arrival setting after conditioning on a common number of arrivals.
We apply Lemma~\ref{lem:stockout-monotone} across candidate multipliers by reusing the inventory draws $U_{j,n}$ from \eqref{eq:inventory_rule}, the arrival counts, and the underlying Poisson processes.
For each fixed draw $U_{j,n}$, the generated inventory $y_{j,n}(c_s)$ is nondecreasing in $c_s$.
Let $x_{j,n}(c_s)$ denote the corresponding terminal sales.
For a pilot sample containing $N_{\mathrm{pilot}}$ periods, Lemma~\ref{lem:stockout-monotone} implies that the realized stockout rate
\[
\widehat\zeta_{\mathrm{pilot}}(c_s)
=\frac1{N_{\mathrm{pilot}}J}\sum_{n=1}^{N_{\mathrm{pilot}}}\sum_{j=1}^J
\I{\{x_{j,n}(c_s)=y_{j,n}(c_s)\}}
\]
is nonincreasing in $c_s$, and so is its expectation.

The search is performed on the log scale. Because the ceiling operation in \eqref{eq:inventory_rule} makes the pilot stockout rate a step function, bisection narrows the search interval; among the evaluated multipliers, we select the one whose pilot rate is closest to $\zeta$.

Each candidate uses three pilot samples of 200 periods in the known-arrival setting and five pilot samples of 500 periods in the Poisson-arrival setting. For targets $0.25$ and $0.50$, the resulting main-sample stockout rates are $0.2480$ and $0.5000$ under known arrivals and $0.2499$ and $0.4969$ under Poisson arrivals. The main samples are generated independently of the pilot samples.

\end{APPENDICES}

\end{document}